%% file: survey.tex
\documentclass[journal,final]{IEEEtran}

\usepackage{dblfloatfix}
\usepackage{multirow}
\usepackage{diagbox}
\usepackage{graphicx}
\usepackage{mathtools, algorithm}
\usepackage{amssymb}
\usepackage{subfigure}
\usepackage{amsmath}
\usepackage{todonotes}
\usepackage{bbm}

\usepackage{algorithm}
\usepackage{algorithmicx}
\usepackage{algpseudocode}
\usepackage{amsthm}
\usepackage{color}
\usepackage{tabularx,booktabs}
\usepackage{bbding}
\usepackage[bottom]{footmisc}

\theoremstyle{definition}

\usepackage{tabularx}
\newcolumntype{Y}{>{\centering\arraybackslash}X}

\usepackage{multirow}
\usepackage[normalem]{ulem}
\useunder{\uline}{\ul}{}
\usepackage{longtable}
\usepackage{tabu}

\usepackage{threeparttable}

\usepackage[hyphens]{url}

\usepackage{silence}
\usepackage{textcomp}

\usepackage{tikz}
\newcommand*\emptycirc[1][1ex]{\tikz\draw (0,0) circle (#1);} 
\newcommand*\halfcirc[1][1ex]{%
  \begin{tikzpicture}
  \draw[fill] (0,0)-- (90:#1) arc (90:270:#1) -- cycle ;
  \draw (0,0) circle (#1);
  \end{tikzpicture}}
\newcommand*\fullcirc[1][1ex]{\tikz\fill (0,0) circle (#1);} 

\usepackage[colorlinks=false,citecolor=black,urlcolor=black,bookmarks=false,hypertexnames=true]{hyperref} 

\usepackage{tikz}
\usetikzlibrary{calc,trees,positioning,arrows,chains,shapes.geometric,%
    decorations.pathreplacing,decorations.pathmorphing,shapes,%
    matrix,shapes.symbols,backgrounds}
\tikzset{
>=stealth',
  blockchain/.style={rectangle, rounded corners, fill=black!15, draw=black, very thick},
  both/.style={rectangle, rounded corners, fill=black!35, draw=black, very thick},
  auctionrelated/.style={rectangle, rounded corners, draw=black, very thick},
  punktchain/.style={rectangle, rounded corners, draw=black, very thick,text width=8.2em, minimum height=3em, text centered, on chain},
}         

\usepackage{xargs}                      %

\newcommandx{\improve}[2][1=]{\todo[linecolor=green,backgroundcolor=green!25,bordercolor=black,#1]{#2}}
\newcommandx{\change}[2][1=]{\todo[linecolor=red,backgroundcolor=red!25,bordercolor=black,#1]{#2}}
\newcommandx{\unsure}[2][1=]{\todo[linecolor=yellow,backgroundcolor=yellow!25,bordercolor=black,#1]{#2}}
\newcommandx{\info}[2][1=]{\todo[linecolor=blue,backgroundcolor=blue!25,bordercolor=black,#1]{#2}}
\newcommandx{\thiswillnotshow}[2][1=]{\todo[disable,#1]{#2}}

\IEEEoverridecommandlockouts

\makeatletter
\def\LT@makecaption#1#2#3{%
\LT@mcol\LT@cols c{\hbox to\z@{\hss\parbox[t]\LTcapwidth{%
\footnotesize\bgroup\par\centering\@IEEEtabletopskipstrut{\normalfont\footnotesize #2}\\{\normalfont\footnotesize\scshape #3}\par\addvspace{0.5\baselineskip}\egroup\endgraf%
\@IEEEtablecaptionsepspace}%
\hss}}}
\makeatother

\begin{document}
\title{Integration of Blockchain and Auction Models: A Survey, Some Applications, and Challenges}

\author{Zeshun Shi,~\IEEEmembership{Member,~IEEE,}
        Cees de Laat,~\IEEEmembership{Member,~IEEE,}
        Paola Grosso,~\IEEEmembership{Member,~IEEE,} \\
        and~Zhiming Zhao,~\IEEEmembership{Senior Member,~IEEE}%
\IEEEcompsocitemizethanks{
\IEEEcompsocthanksitem 
\IEEEcompsocthanksitem The authors are with the Informatics Institute, University of Amsterdam, Amsterdam, 1098 XH, the Netherlands. E-mails: \{z.shi2, delaat, p.grosso, z.zhao\}@uva.nl
\IEEEcompsocthanksitem Zhiming Zhao (z.zhao@uva.nl) is the corresponding author.}

}

\maketitle

\begin{abstract}
In recent years, blockchain has gained widespread attention as an emerging technology for decentralization, transparency, and immutability in advancing online activities over public networks.
As an essential market process, auctions have been well studied and applied in many business fields due to their efficiency and contributions to fair trade.
Complementary features between blockchain and auction models trigger a great potential for research and innovation.
On the one hand, the decentralized nature of blockchain can provide a trustworthy, secure, and cost-effective mechanism to manage the auction process; 
on the other hand, auction models can be utilized to design incentive and consensus protocols in blockchain architectures.
These opportunities have attracted enormous research and innovation activities in both academia and industry; however, there is a lack of an in-depth review of existing solutions and achievements.  
In this paper, we conduct a comprehensive state-of-the-art survey of these two research topics.
We review the existing solutions for integrating blockchain and auction models, with some application-oriented taxonomies generated. Additionally, we highlight some open research challenges and future directions towards integrated blockchain-auction models.

\end{abstract}

\begin{IEEEkeywords}
Blockchain, auction models, decentralized applications, incentive mechanisms.
\end{IEEEkeywords}
\IEEEpeerreviewmaketitle

\section{Introduction}
\label{introduction}
Over the past decade, the world has witnessed the success of blockchain as a novel technology to build decentralized systems. In general, blockchain is a decentralized ledger technology that incorporates cryptography, peer-to-peer (P2P) networks, and consensus mechanisms. 
The ledger is maintained by all nodes participating in the system and is decentralized, tamper-proof, transparent, and secure \cite{bashir2018mastering}. In 2008, Satoshi Nakamoto first introduced blockchain as the foundation technology for a cryptocurrency named Bitcoin \cite{nakamoto2008bitcoin}. After that, with smart contracts bringing programmability to the blockchain, it is now widely believed that blockchain can be applied to build decentralized systems in various application scenarios, e.g., transportation and logistics, agriculture and food, energy and utilities, healthcare, and life sciences \cite{da2019application}. According to MarketsandMarkets \cite{BlockchainMarket_2020}, the worldwide blockchain market is predicted to expand to \$39.7 billion
and cover specific applications across more than 15 industries.

An auction is a process of buying and selling goods or services. This process involves offering items for bidding, waiting for bids to be accepted, and then selling goods to the highest bidder under the supervision of an auctioneer \cite{krishna2009auction}. Typically, auctions tend to be centrally organized and offline. Due to their fairness properties, auctions are widely used in trading activities for artworks, cars, radio spectra, online advertisements \cite{doi:10.1146/annurev-economics-080218-025818}. 
In the field of economics, auction theory has become one of the most successful and active branches \cite{klemperer1999auction}. Hundreds of auction models have been designed to serve different auction scenarios.
A case in point is the spectrum auction that the Federal Communications Commission (FCC) has been conducting since 1994 \cite{milgrom2000putting}. Since then, spectrum auctions have contributed more than \$200 billion of revenue to the U.S. government. The two designers of the FCC auction were awarded the Nobel Prize in 2020 for their improvements to auction theory and the invention of new auction formats \cite{nobel_2020}.

Potential research and innovation opportunities across both blockchain and auction models have emerged recently \cite{nguyen_blockchain-based_2020}.
On the one hand, traditional centralized auctions usually require a third-party auctioneer or auction house to manage the entire auction process, which is expensive due to high commission fees. They also suffer from a single point of failure, and auctioneers can potentially be malicious in some cases \cite{wu_cream_2019}. In this context, blockchain has emerged as a decentralized platform to support trustworthy online auction applications. In 2018, for the first time in the world, multi-million dollar artworks by Andy Warhol were tokenized and auctioned successfully using the Ethereum blockchain \cite{ethereum_2021,andy_2018}. 
It is also reported that major auction houses (e.g., Sotheby's and Christie's) are actively working on applying blockchain in secure and trusted auction use cases \cite{Christie_2018}. Thus, we can foresee that this mechanism of bidding for ownership of items with blockchain could become the future trend. On the other hand, peers in the blockchain can use auctions to handle dynamic relationships.
For instance, auction theory can be leveraged to model the transaction fee market of blockchain platforms. The transaction fee mechanism of the Ethereum blockchain has been a first-price auction since its inception; each transaction has an associated transaction fee (bid), which is paid by its submitter to the miner for priority processing \cite{roughgarden_transaction_2020}. Auctions are also found, as the literature indicates, in other blockchain activities such as miner selection \cite{amin_secured_2020} and block reward allocation \cite{liu_auction_2019}. 

The opportunities of applying blockchain in auctions or enhancing blockchain using auctions have attracted many research and innovation activities; however, there is a lack of surveys to systemically review those different technical developments and achievements, and to identify the important open challenges. 
In this paper, we attempt to answer the following questions through a systematic literature survey: 1) What are the characteristics of existing blockchain technologies and auction models? 2) How can blockchain technologies and auction models enhance each other? 3) What blockchain-based auction applications have been published, and how can these applications be classified? 
4) What auction-based solutions have been proposed for enhancing blockchains?
5) What open challenges can we identify in the integration between blockchain and auction models?

\subsection{Contributions}
In this survey, we draw a comprehensive research landscape of the integration between blockchain and auction models to answer the above-mentioned research questions.
Both aspects of the integration, namely blockchain-based auction models and auction-enhanced blockchain technologies, are carefully reviewed.
The main contributions of this paper can be summarized as follows:
\begin{itemize}

\item Review existing blockchain technologies and auction models, and provide a conceptual schema to analyze research and innovation opportunities from their integration. 

\item Systematically review the blockchain-based auction applications, and auction-based solutions to enhance blockchain technologies. 

\item Provide a taxonomy to classify the existing applications and solutions in the integration between blockchain technologies and auction models. 

\item Identify open research challenges from the reviewed models, and provide guidance to design applications that require integration between blockchain and auction models.

\end{itemize}

\subsection{Related Works}
\label{relatedwork}

\begin{figure}[!t]
    \centering
    \input{pics/Related}
    \caption{Summary of existing related survey studies, categorized according to the year of publication and their focus. Here the value of N represents the number of surveys in each time interval.}
    \label{related}
\end{figure}
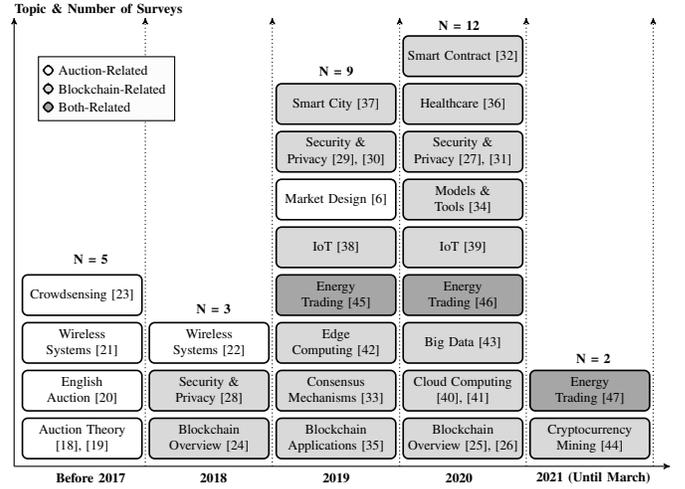

During the past years, auction-based theories and models have attracted extensive attention from many researchers. Most surveys on auction-related topics we can find were published before 2017 in the field of economics. 
Those surveys mainly concern the introduction and comparison of different auction models \cite{klemperer2004auctions,jain2006classification,wahaballa2015taxonomy}, market design \cite{doi:10.1146/annurev-economics-080218-025818}, as well as the application of auctions in specialized areas such as wireless systems \cite{zhang2012auction,habiba2018auction} and crowdsensing \cite{zhang2015incentives}.
The investigation efforts of blockchain, on the other hand, are relatively new. Despite the fact that blockchain is a newly emerged technology, almost every aspect of blockchain has been extensively studied in the literature. These surveys cover topics including blockchain overview \cite{zheng2018blockchain,butijn2020blockchains,kolb2020core}, security \& privacy \cite{hassan2020differential,conti2018survey,zhang2019security,lee2019systematic,peng2020privacy}, smart contract \cite{Hu_2021}, consensus mechanism \cite{wang2019survey}, models \& tools \cite{huang2021survey}, and various blockchain-based applications \cite{casino2019systematic} such as healthcare \cite{de2020survey}, smart city \cite{xie2019survey}, Internet of Things (IoT) \cite{ferrag2018blockchain,lao2020survey}, cloud/edge computing \cite{gai2020blockchain,10.1145/3403954,yang2019integrated}, big data \cite{deepa2020survey}, and cryptocurrency \cite{hacioglu2021crafting}. The summary of these survey topics over the publication years is shown in Fig. \ref{related}. Overall, both the publication number and the research diversity have increased significantly in the last few years.

A Blockchain can provide a decentralized environment to support auction activities, thereby improving the security and trustworthiness of auctions. On the other hand, previous research has suggested the application prospect of using auction models to optimize blockchain workflows, e.g., transaction fee mechanism design, miner selection, and block reward distribution. However, although there are so many studies on blockchain and auction models respectively, the issue of combining the two has rarely been addressed in previous survey works.
The studies most relevant to our research are three survey papers working on blockchain-based energy trading solutions, where auction models are partially discussed \cite{wang2019energy,oprea2020local,hassan2021optimizing}. 
The authors in these studies only focus on one specific application field and do not offer the comprehensiveness of this work. Besides, no research work has so far summarized how to use auction models to optimize blockchain technologies.

In summary, most of the existing surveys discussed the two topics separately. 
There is no general survey on the current landscape of integrated blockchain-auction models. 
Therefore, the purpose of this survey is to summarize previous publications and to complement existing research on the integration of blockchain and auction models. To the best of our knowledge, this paper is the first comprehensive survey to fill these gaps.

\subsection{Organization}
\begin{figure}[!t]
    \centering
    \includegraphics[width=\linewidth]{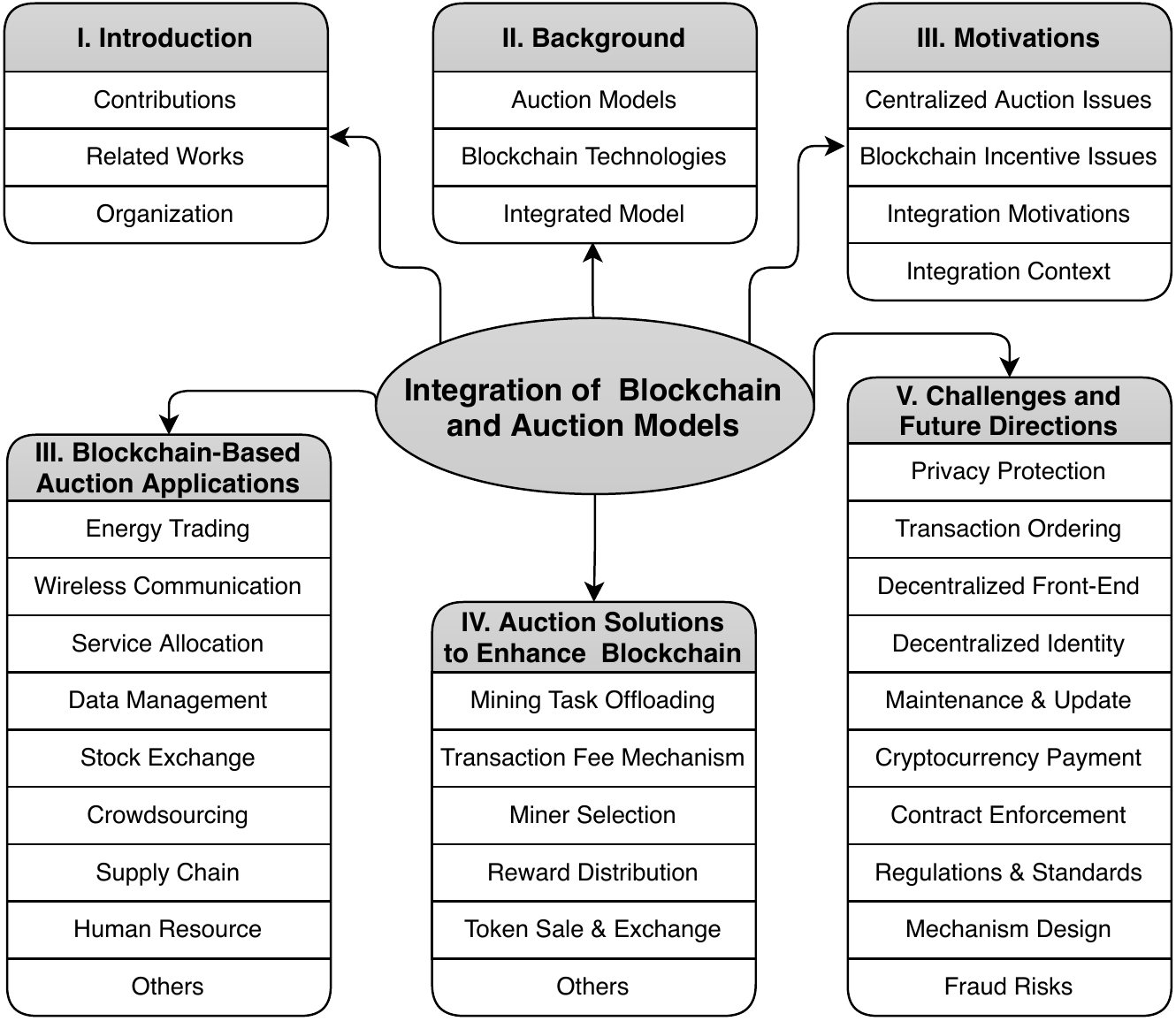}
    \caption{Road map and organization of this survey.}
    \label{roadmap}
\end{figure}

Fig.~\ref{roadmap} illustrates the road map and organization of this paper. As shown in the figure, the remainder of this survey is organized as follows. Firstly, some preliminary knowledge of auction models and blockchain technologies, as well as the blockchain-auction integration model architecture are presented in Section~\ref{background}. Then, Section~\ref{motivation} introduces the motivations and considerations for the integration. Section~\ref{blockchain4auction} reviews blockchain-based auction applications, including a survey on auction models and blockchain technologies used in different application fields. Section~\ref{auction4blockchain} explores several aspects of using auction models to enhance blockchain technologies. Section~\ref{challenge} highlights and summarizes the current research challenges and solutions. Finally, the survey is concluded in Section~\ref{conslusion}. The acronyms used in this paper are listed in Table \ref{tab:ACRONYMS} for easy reference.

\input{table/Table_Abbreviation}

\section{Background Knowledge}
\label{background}
In this section, we begin with a brief overview of different auction models and blockchain technologies. 
We then proceed to discuss the opportunities and considerations behind combining them.

\subsection{Auction Models}

\input{table/Table_Auction3}

An auction is a sale activity in which potential buyers make competitive bids for objects or services \cite{klemperer2004auctions}. There are usually several fundamental elements in an auction: 1) a seller who owns and wants to sell the objects; 2) one or several bidders who want to buy the objects via the auction; 3) the auction objects traded between the seller and the buyer(s); and 4) an auctioneer who works as an intermediary agent to host and control the auction process. 

Auction models can be classified from different dimensions, e.g., the bidding process, the number of items, the roles of buyers/sellers, and the bidding participants \cite{habiba2018auction}. In the rest of this section, we review auction models that are frequently used in the blockchain-related literature. A comparison of those auction models is also shown in Table \ref{tab:auctions}.

\subsubsection{Open-Outcry Auction vs. Sealed-Bid Auction} From the perspective of the bidding process, an auction model can be either open-outcry or sealed-bid. In an open-outcry auction, a bidder's bidding activities are transparent and visible to all bidders. Whereas in a sealed-bid auction, bidders submit their bids to the auctioneer privately, and the bids are only known by the auctioneer until the auction ends. Typical open-outcry auctions and sealed-bid auctions are summarized as follows:

\begin{itemize}
    \item \textit{English Auction} (also called open-outcry ascending-price auction). In an English auction, the price begins low and rises as buyers submit their bids until only one bidder is left and no higher bids are obtained within the specified time span. The whole process of requesting bids is open and transparent. It can be very competitive, with pressure rising as bidders' offers increase. Since the auctioneer would try to get the best price for the seller, an English auction is expected to benefit the seller.
    An English auction can be profitable for sellers, but they often pose problems for bidders. In addition, it requires iterative communications and adjustments, which can sometimes be a bit difficult and costly.
    \item \textit{Dutch Auction} (also called open-outcry descending-price auction or clock auction). In a Dutch auction, the auctioneer starts by announcing a high asking bid and then keeps lowering this bid until a buyer is willing to accept it. 
    This auction is often used to sell goods that must be sold quickly (e.g., fresh produce). For example, such auctions are very common in the Dutch flower sales market.
    In some cases, Dutch auctions may result in inappropriate bidding, which may be caused by a lack of sufficient information among bidders.
    \item \textit{First-Price Sealed-Bid (FPSB) Auction} (also called blind auction). In an FPSB auction, all bidders submit sealed bids to the auctioneer simultaneously, and the highest bidder wins and pays his/her bid. Other bidders' bids will not be revealed during the auction until a winner is determined. Therefore, bidders do not compete openly with each other, but they can collect information about their competitors' bids before submitting their own.
    Since bidders could not see the bids of other participants, they could not adjust their bids accordingly. In addition, bidders are vulnerable to the winner's curse.
    \item \textit{Vickrey Auction} (also called second-price sealed-bid auction). It is similar to an FPSB auction but with a different payment mechanism. 
    After all bidders submit sealed bids to the auctioneer, the highest bidder still wins but only pays the second-highest bid. 
    In Vickrey auctions, truthful bidding is the dominant strategy \cite{klemperer1999auction}. One concern with this type of auction is that it has been well studied in theory but not very popular in practice.
\end{itemize}

\subsubsection{Single-Item Auction vs. Multi-Item Auction} From the perspective of the number of items, an auction model can be single-item or multi-item. The above-mentioned four auction models are the main types of auctions where a single item is sold \cite{easley2010networks}. However, in some situations, selling multiple items at the same time is a more efficient way. Multi-item auctions can be further subdivided into two cases: an auction is said to be homogeneous if all items offered in the auction are identical; otherwise, it is considered heterogeneous.

\begin{itemize}
    \item \textit{Combinatorial Auction} (also called multi-lot auction). This is a popular auction in which heterogeneous items are sold at the same time. Bidders can place bids on combinations (or ``packages") of items. 
    It is suitable to auction scenarios where bidders have non-additive valuations for bundled items.
    Despite allowing more expression for bidders, combinatorial auctions present computational and mechanism design challenges compared to traditional auctions. For example, the winner determination problem is often a computationally intensive NP-hard problem.
    \item \textit{Multi-unit Auction}. This is an auction in which several homogeneous items are sold. Based on the different payments for each unit, it can be further divided into two types, i.e., pay-as-bid auction (or discriminatory price auction) and uniform price auction (or clearing price auction) \cite{wittwer2018pay}. In the former, bidders pay their bids for each unit they won. Whereas in the latter, all winning bidders pay the same price regardless of their actual bid.
    It should be noted that the incentive of the multi-unit auction may cause bidders to bid less than their true value, resulting in inefficient allocations.
\end{itemize}

\subsubsection{Forward Auction vs. Reverse Auction}
An auction model can be either forward or reverse in terms of the roles of buyers/sellers.
A forward auction is also called a seller-determined auction, in which one seller sells products to multiple potential buyers (bidders).
The auction models discussed so far are all forward ones.
In a reverse auction, however, the roles of buyers and sellers are swapped: sellers need to bid and compete for the opportunity to sell their products.

\begin{itemize}
    \item \textit{Reverse Auction} (also called buyer-determined auction or procurement auction). In a reverse auction, one buyer needs to trade with multiple potential sellers. The buyer first makes a request for the required goods or services. Then sellers place bids for the goods or services they are willing to deliver. A reverse auction is highly suitable for procurement activities proposed by governments, companies, and organizations since it motivates sellers' competition.
    One of the main disadvantages of a reverse auction is that it does not require bidders to provide information about the specific costs involved in the contract. This can lead a buyer to choose a seller who appears to bid the lowest price but offers inferior products or poor customer service.
\end{itemize}

\subsubsection{Single-Sided Auction vs. Double Auction} 
In terms of the participants in the bidding process, an auction model can be single-sided or double-sided. The single-sided approach has been widely implemented in traditional auctions (e.g., forward and reverse auctions). However, in some cases, they cannot accommodate additional sellers/bidders in a large-scale situation. The double auction is an extension of the conventional auction, which adopts the many-to-many strategy to generate multiple winning bidders in each round \cite{lee2020preserving}. 

\begin{itemize}
    \item \textit{Double Auction} (also called double-sided auction). In this auction, multiple sellers and buyers submit their bids/offers, respectively. The market institution (auctioneer) then chooses a price that clears the market. Many different market clearing mechanisms already exist, including average mechanism, VCG (Vickrey-Clarke-Groves) mechanism, trade reduction mechanism, and McAfee's mechanism \cite{Babaioff_2004}. In reality, a double auction is suitable for marketplaces with multiple sellers and buyers, e.g., stock exchanges.
    The double auction mechanism is challenging to handle the auction of heterogeneous items with multiple attributes due to the substantial execution time and cost required.
\end{itemize}

\subsubsection{Others}
Some other emerging auction models found in the literature are listed as follows:
\begin{itemize}
    \item \textit{All-Pay Auction}. Every bidder must pay regardless of whether he/she wins or not. The auction is awarded to the highest bidder as in a conventional auction. It is popular among governments and central banks. However, overbidding is a common behavior in the auction process and can result in winner's curse.
    \item \textit{Multi-Attribute Auction}. The bids could have multiple attributes (e.g., service time and quality) other than price. In this case, a scoring mechanism is needed to calculate the total bidding value. 
    It is suitable when the auction needs to consider multiple attributes (e.g., service allocation).
    One challenge in multi-attribute auctions is designing a reasonable scoring mechanism to determine which bid is the best. Unfortunately, this cannot be addressed by simply comparing different attributes.
    \item \textit{Sponsored Search Auction} (also called keyword auction). It is specially designed for search advertising scenarios. In this auction, $n$ advertisers (bidders) compete for the assignment of $k$ advertisement slots/positions. Each bidder submits a bid, then the highest bidder gets the first slot (with his/her bid), the second-highest bidder gets the second slot, and so forth. Based on the winner's different payment strategies, it can be further divided into generalized first-price (GFP) auction, generalized second-price (GSP) auction, and VCG auction.
    There are some trade-offs among them: GFP auctions are easy to use but less stable; GSP auctions incorporate the advantages of Vickrey auctions but do not support truthful bids; VCG is a truthful auction and is relatively stable, but users may find it difficult to understand and use in reality.
\end{itemize}

\subsection{Blockchain Technologies}
\label{background_blockchain}
Introduced by Satoshi Nakamoto in 2008, blockchain was initially used as the underlying technology for Bitcoin. It records transactions among distributed participants as identical copies through a decentralized ledger, which is represented as a chain of blocks. Based on the consensus among distributed participants, new blocks are generated and attached to the chain using a cryptographic algorithm. In this process, a blockchain builds trust among its distributed users by virtue of the immutability and security of the ledger.

\subsubsection{Blockchain Architecture}%
Blockchain researchers and practitioners often model blockchain systems using a layered architecture, and abstract typical blockchain technologies and functional components as six bottom-up layers: data, network, consensus, incentive, contract, and application layer \cite{zhang2019security}.
The three layers at the bottom are usually considered a blockchain's basic elements, while the upper three layers are the extended elements.

\begin{itemize}
    \item \textit{Data Layer}. This layer defines the schema, data structure, and storage of all the data information on the blockchain. As the name suggests, a blockchain uses the ``chained blocks" data structure as its backbone. Each block consists of several transactions, with useful information (e.g., version, hash, nonce, timestamp, and Merkle root) contained in the block header. The blocks are chained to each other via cryptographic algorithms (e.g., asymmetric encryption, digital signature, hashing algorithm), making the data layer constitute a tamper-proof database for the blockchain. In this regard, Bitcoin uses double iterative SHA-256 as the hash function, while Ethereum uses KECCAK-256. ECDSA (Elliptic Curve Digital Signature Algorithm) is the transaction signature algorithm used by both Bitcoin and Ethereum.
    \item \textit{Network Layer}. 
    This layer models protocols for connecting blockchain nodes and validating data transferred across them. Blockchain nodes are typically connected using a P2P paradigm, where the network is maintained by all peer nodes together, and no single agent can control the whole system. Based on the type of underlying P2P network (e.g., whether it is structured or unstructured), different blockchain platforms may use different communication protocols. Bitcoin, for example, uses a gossip-based protocol to select peers and exchange states. When new transactions are generated on a node, they are first propagated to the neighboring nodes for validation. If the data structure and syntax are valid, they are saved for further processing; otherwise, they are simply rejected. Ethereum, on the other hand, relies on the Kademlia distributed hash table (DHT) protocol to manage communication in its P2P network. This is different from the unstructured P2P network used by Bitcoin \cite{wang2021ethna}.
    \item \textit{Consensus Layer}. This layer is the foundation and core of a blockchain system. It defines protocols and algorithms for decentralized nodes to reach a consensus on the update of the blockchain. The most common and successful consensus algorithm is Proof of Work (PoW). Other alternatives like Proof of Stake (PoS), Delegated Proof of Stake (DPoS), Practical Byzantine Fault Tolerance (PBFT), Proof of Elapsed Time (PoET), Proof of Authority (PoA), and Raft, have also been widely discussed recently \cite{wang2019survey}. These algorithms will be discussed in detail in the next section.
    \item \textit{Incentive Layer}. This layer 
    provides incentive mechanisms for a blockchain to motivate participants to validate the data and maintain the whole system. Incentive mechanisms are typically based on block rewards and transaction fees. For example, the issuance mechanism of the Bitcoin blockchain guarantees that successful miners are rewarded with 6.25 Bitcoins when a new valid block is mined. At the same time, the transaction fees associated with each transaction can be allocated to the corresponding miners. This layer is essential in permissionless blockchains. Whereas in a permissioned blockchain, the incentive mechanism is often optional since the participants are selected organizations \cite{kolb2020core}.
    \item \textit{Contract Layer}. This layer defines decentralized programming paradigms in a blockchain, which was initially promoted by the Ethereum smart contract technology. A smart contract is a tamper-proof and self-executing program running on the blockchain, which enables a much broader range of application innovations in addition to cryptocurrencies. The concept of smart contracts has also extended to other blockchain platforms, e.g., chaincodes \cite{Chaincode_2017} and transaction processors \cite{Transaction_2017} are smart contracts offered by Hyperledger Fabric and Sawtooth, respectively. 
    \item \textit{Application Layer}. This layer defines application programming interfaces (APIs) and programming models for developing specific applications. Blockchain was once well known for its cryptocurrency application (e.g., Bitcoin). Now with the popularity of smart contract technology,
    blockchain-based applications, namely decentralized applications (DApps), are showing huge market potential in many industrial sectors \cite{BlockchainMarket_2020}.
\end{itemize}

\subsubsection{Consensus Algorithms}
Consensus algorithms lie at the heart of blockchain technology. Considering the decentralized nodes involved in the blockchain network and the potential instability of communication, the design of consensus algorithms is full of challenges. Since the invention of the blockchain, new consensus mechanisms have been created continuously. Some of them are the improvements on PoW, while others are the traditional distributed fault-tolerant algorithms. This section introduces some of the consensus algorithms commonly used by popular blockchain platforms.
\begin{itemize}
    \item \textit{Proof of Work (PoW)}. This is the most famous and successful blockchain consensus algorithm. In PoW, miners need to earn bookkeeping rights by demonstrating the amount of work they contribute. The process of proving workload is to solve a puzzle (also known as mining), and the miner who solves the puzzle faster has priority for bookkeeping \cite{wang2019survey}. The advantage of PoW is the high level of decentralization it can provide. PoW is also considered to be the most secure blockchain consensus mechanism to date. The disadvantage is that it can cause energy waste because mining requires a lot of computational resources. In addition, it limits the performance of the blockchain network. Bitcoin and Ethereum use PoW as the underlying consensus algorithm.
    \item \textit{Proof of Stake (PoS)}. In the PoS consensus, whoever has more stakes (i.e., tokens) gets the right to produce blocks. A fundamental assumption of POS is that the stake owners prefer to maintain the consistency and security of the blockchain system \cite{pos}. It has the prominent advantage of being more efficient than PoW. However, the security needs to be further validated due to the low level of decentralization. Examples of industry-leading PoS blockchains include Cardano and Avalanche. Ethereum, originally designed as a PoW blockchain, is also being upgraded to a PoS version called Ethereum 2.0.
    \item \textit{Delegated Proof of Stake (DPoS)}. This is a voting-based consensus algorithm; token holders vote for a certain number of representatives (based on the tokens held in their hands) to be responsible for producing new blocks and maintaining the network. As a variant of PoS, DPoS optimizes the traditional PoS using a voting mechanism. However, it can suffer from low enthusiasm for voting and concentration of power. Blockchain projects that use DPoS include EOS and Lisk.
    \item \textit{Proof of Authority (PoA)}. This consensus algorithm aims to unify the state of the blockchain by electing authoritative validators with good reputations \cite{de2018pbft}. There are many similarities between PoA and PoS, for example, they both do not require mining and therefore have good performance. The disadvantage of PoA is the low level of decentralization it caused. This consensus algorithm typically serves test networks and private blockchains. For example, Ethereum Kovan testnet and the private Ethereum version on Azure Blockchain Workbench are both based on the PoA protocol.
    \item \textit{Practical Byzantine Fault Tolerance (PBFT)}. The idea of Byzantine Fault Tolerance (BFT) was first proposed in the 1980s and there are many implementations of the algorithm. Among them, PBFT is the most famous one, which provides (n-1)/3 fault tolerance while guaranteeing system liveness and safety \cite{de2018pbft}. It has the advantage of dealing with the inefficiency of the original BFT algorithm and tolerating malicious peers, while the drawbacks are limited scalability and high latency. In the current blockchain community, Hyperledger Sawtooth supports a pluggable PBFT consensus protocol. Hyperledger Fabric claims to support PBFT but is not yet fully implemented.
    \item \textit{Proof of Elapsed Time (PoET)}. PoET is a type of consensus that uses a trusted execution environment (TEE) to improve the efficiency of the current PoW protocol. It uses the randomly generated elapsed time to determine the right for bookkeeping. PoET provides an excellent solution to the random miner selection problem, but has the disadvantage of being necessarily dependent on dedicated hardware for security. PoET is primarily promoted and used in Hyperledger Sawtooth \cite{poet}.
    \item \textit{Raft}. Raft is a distributed consensus algorithm that functions similarly to Paxos \cite{howard2020paxos}. Compared to Paxos, it is easier to understand and implement in real systems. In Raft, each node has three states: follower, candidate, and leader. The Leader is selected for bookkeeping in a continuous iterative voting process. Raft has the advantage of low algorithm complexity and easy implementation. However, it only supports crash fault tolerance and cannot solve the problem of malicious nodes. Raft is the consensus algorithm mainly used by Hyperledger Fabric and Oracle Blockchain.
    \item \textit{Others}. In addition, researchers have identified dozens of new consensus algorithms. Other commonly used consensus algorithms include tangle-based solutions, which are widely used in directed acyclic graph (DAG) blockchains (e.g., IOTA). In addition, some platforms adopt customized consensus solutions. For example, the Corda blockchain achieves consensus by confirming the validity and uniqueness of transactions \cite{r3consensus}.
\end{itemize}
\subsubsection{Blockchain Types}
In general, there are three types of blockchain networks: permissionless, permissioned, and hybrid blockchain. 
This section provides a brief summary of them.
A more detailed comparison is shown in Table \ref{blockchaintypes}.

\begin{itemize}
    \item \textit{Permissionless Blockchain}. In a permissionless or public blockchain (e.g., Bitcoin or Ethereum), anyone can join the network by submitting or validating transactions. 
    To address the lack of trust among anonymous players, a consensus mechanism is often used to determine who gets the right to package transactions and produce new blocks in a given round. PoW is a good illustration of such a consensus algorithm and has been validated with the popularity of blockchain. However, it has been criticized for being inefficient and consuming too much energy in order to reach a consensus. It is widely believed that in a PoW-based permissionless blockchain, the waste of energy is inevitable in order to establish trust among strangers without any prior knowledge of each other.
    \item \textit{Permissioned Blockchain}. A permissioned blockchain is operated as a closed ecosystem that can only be accessed by users with permissions. A user can only view the ledger or validate new transactions after being approved by the authority of the blockchain. In this way, malicious or crashed nodes can be identified through more energy-efficient consensus algorithms such as PBFT, PoET, and Raft. 
    The ability of assigning specific network permissions to users and the enhanced performance give permissioned blockchains a great potential for wider industrial application. 
    Hyperledger is one of the most successful blockchain communities and has incubated several permissioned blockchain platforms such as Fabric and Sawtooth \cite{Hyperledger}.
    However, there are also some arguments that the ``partially decentralized" nature of the permissioned blockchain may lead to compromises in trust \cite{wang2019survey}.
    \item \textit{Hybrid Blockchain}. It aims to combine the strengths of both permissionless and permissioned blockchains 
    and to customize the degree of decentralization based on specific application needs.
    A hybrid blockchain enables highly regulated organizations to have greater flexibility and control over which data is kept private versus shared on a public ledger \cite{hybrid_2018}.
    A typical example is the Aergo platform, which consists of a public chain network using the DPoS consensus and several customized sidechains dedicated to specific applications based on leader-based PoA consensus mechanisms \cite{Aergo_2021}.
\end{itemize}

\input{table/Table_Blockchain}

\subsubsection{Suitability for Auctions}
To select the most suitable blockchain technologies for an auction application, users need to consider some basic questions. For example, does the blockchain have to provide a cryptocurrency to support auction payments? Is the auction designed to be implemented on a private or public network? In addition, some specific business requirements for the auction model need to be considered, such as user scenarios, security, privacy, and scalability. A permissionless Ethereum blockchain is focused on providing a universal platform for various transactions and applications. It has the advantage of being easy to use, secure, and having a wide user base. Therefore, it is suitable for open-outcry auctions and double auctions where a large number of bidders are required. However, its full decentralization and transparency come at the cost of performance and privacy. Therefore, it is more suitable for single-item auctions instead of multi-item auction models that require complex on-chain computation. On the other hand, due to privacy, regulatory, and scalability concerns, enterprises may prefer to use permissioned blockchains rather than permissionless ones to enable auctions. Hyperledger Fabric, for example, provides high throughput to help with on-chain winner determination calculations for some complex auctions (e.g., VCG auctions). However, the disadvantages of using it for auctions are also obvious; it is not equipped with a stable cryptocurrency. Besides, as a permissioned blockchain, it faces greater challenges in terms of data security and immutability.
It should be noted that the choice of blockchain platform should be flexible for different auction scenarios. Most existing blockchain platforms are quite extensible and can be improved for different application requirements. For example, Ethereum has designed an alternative privacy deployment version to address the issues in permissionless deployment. Hyperledger Fabric could add an extra token component to solve the problem of not having native tokens, as the system is based on a highly modular design. Furthermore, it is also possible to use a hybrid blockchain to incorporate the advantages of both permissionless and permissioned blockchains. However, the usability of such a model for auctions still requires further validation.

\begin{figure*}[!t]
\centering
\includegraphics[width=\linewidth]{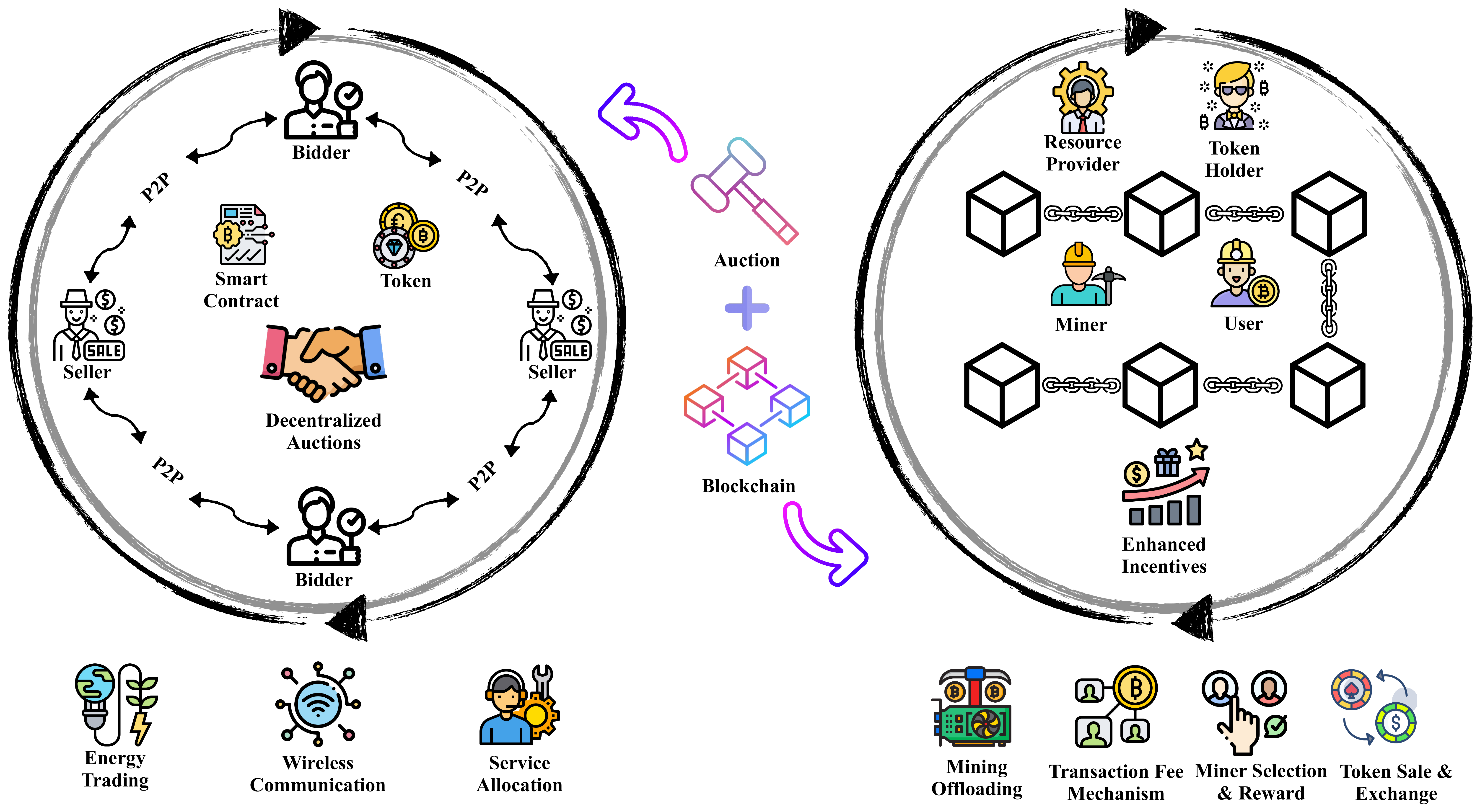}
\caption{Architecture of the integrated blockchain-auction model.}
\label{integrationarchitecture}
\end{figure*}

\subsection{Integrated Blockchain-Auction Model}
With the support of an extensive literature review, we propose a conceptual blockchain-auction integration architecture model, as shown in Fig. \ref{integrationarchitecture}. This architecture consists of two components forming an organic system. First, in the middle of the architecture are blockchain and auction technologies respectively. When blockchain is applied to optimize the auction model, the decentralized auction scenario on the left side of the architecture is demonstrated. In this case, the blockchain serves as the underlying infrastructure to support direct P2P transactions between buyers and sellers without the need for a trusted third party (TTP). The programmability of smart contracts provides customizable and automated execution of a large number of auction models, with tokens further optimizing the auction payment. The most common decentralized auction applications include energy trading, wireless communication, and service allocation, which will be discussed in Section \ref{blockchain4auction}. Next, when the auction model is applied to optimize the blockchain technology, the enhanced incentives for the blockchain are shown on the right side of the architecture. Blockchain maintenance requires enough users, active miners, and sufficient computational and network resources. Therefore, the auction model can be leveraged to provide suitable incentives to stakeholders (including regular blockchain users, miners, token holders, and external resource providers) and maintain the blockchain's economic stability and operation. In general, auction models can enhance blockchain incentives in the following scenarios: mining task offloading, transaction fee mechanism design, miner selection \& reward distribution, and token sale \& exchange. This will be discussed in detail in Section \ref{auction4blockchain}.

\section{Motivations and Considerations for the Integration}
\label{motivation}
The integration between blockchain technologies and auction models can promote innovations on both sides. 
On the one hand, blockchain can be used to enable a decentralized auction system and improve the trustworthiness of centralized auctions. On the other hand, auction models can be leveraged to motivate decentralized peer nodes as a kind of price incentive mechanism to enhance blockchain technology. 
In this section, we discuss the research and innovation opportunities brought by the integration between blockchain and auction models. 

\subsection{Current Issues of Centralized Auctions}
Auctions can be centralized or decentralized, and they differ in how they achieve their auction goals. Centralized auction applications are operated and owned by a single company (e.g., eBay) and run on a centralized server cluster. The developer can retain full control over the auction application in a centralized auction platform. As a result, centralized auction applications can typically handle higher traffic volumes. More importantly, centralized auctions offer low-cost hosting, fast runtime, easy development, and a tightly controlled user experience. All of these factors have made centralized online auctions a huge success over the past few decades. However, these advantages also come at some serious costs. In general, centralized auctions have the following issues and challenges.

\subsubsection{Centralized Auctions Are Inflexible}
One problem with centralized auctions is inflexibility. With the current centralized model of online auctions, each platform has several fixed auction formats, rules, policies, and user groups. As a result, both auction buyers and sellers are at risk of vendor lock-in, a situation where the cost of switching to a different vendor is so high that the customer is essentially stuck with the original vendor. Another reason for inflexibility is that platforms need time and money to develop new auction formats to match new technologies and user needs. The limitations of auction formats mean that dynamic user needs cannot be satisfied. As a result, auctions in the current marketplace are usually arranged and operated in an inefficient and inflexible manner.
Besides, centralized auctions are subject to censorship from central authorities. The content of the auction is subject to the laws and regulations of the country in which it is held, as well as the platform's own rules and policies \cite{ahlqvist2022survey}.

\subsubsection{Centralized Auctions Are Opaque and Untrustworthy}
Centralized auctions operate in an opaque manner (i.e., a black box). Large auction companies and service providers are by default regarded as trusted parties that can potentially maintain, control, and manage user data, access, and activity. While this can be beneficial for users, it can potentially be used as a source of control to enforce surveillance or lead to abuse of trustworthiness \cite{zarrin2021blockchain}. For example, system administrators can obtain sensitive data in a private auction easily and help some bidders to win the auction.

\subsubsection{Centralized Auctions Pose Security and Privacy Risks}
Centralized auction platforms collect user data collectively and store it in a certain number of servers to support the hosting of various types of services and applications \cite{zarrin2021blockchain}. Unfortunately, this exposes vulnerabilities and user data to cybercriminals, leading to serious security and privacy concerns. A prime example is eBay's report in 2014 that hackers had infiltrated their systems and stolen the passwords of 145 million users. In addition to account passwords, hackers obtained user private information such as names, email addresses, dates of birth, physical addresses, and phone numbers \cite{ebaybreak}. These security incidents may cause negative influence and huge financial losses on auction users.

\subsubsection{Centralized Auctions Suffer From Single Points of Failure}
Centralized auctions based on the client-server model are prone to single points of failure. These failures can cause the entire auction system to stop functioning due to network or system problems. If the centralized server goes down, the auction application will go offline, and users may not be able to use the application in a timely manner until the error is fixed. For auction use cases that require high availability and reliability (such as luxury jewelry and art auctions), a single point of failure is highly undesirable, which can cause severe property damage to users.

\subsubsection{Centralized Auctions May Trigger Huge Expenses}
Centralized auctions tend to have higher commission costs. Online auction platforms such as eBay and eBid take a percentage of the final sale price from users to compensate for data processing and marketing costs. For example, eBay's auction commission fee is 12.9\% of total sales. In contrast, eBid is cheaper but also requires a base fee of 5\% of total sales. When the value of the auctioned item increases, the cost of a centralized auction will increase significantly. This will discourage the widespread adoption of auction applications by regular users \cite{shi2022bayesian}.

\subsection{Current Issues of Blockchain Incentives}
The successful operation of Bitcoin has well demonstrated the strong stability and security of the blockchain system. However, many issues still need to be overcome for the wide adoption of the blockchain, e.g., privacy protection, scalability, interoperability, and regulation issues. In this section, we specifically discuss the issues presented in blockchain that can be potentially improved and solved by auction-based incentives.
From an incentive perspective, blockchain faces a fundamental challenge; it must motivate users to continuously join and maintain the system while preventing some users from colluding and gaining disproportionate control. There are a number of design flaws in the operation of existing blockchain incentives.

\subsubsection{Lack of Optimal Incentives to Offload Mining Tasks}
When node devices want to run PoW-based blockchain applications, they have to spend significant computational resources to solve the PoW cryptographic puzzle. Resource-constrained miner nodes are usually unable to complete block computation in a short time, which limits the overall performance of the blockchain \cite{liu_smart_2020}. Mining task offloading is an instinctive approach to solve this problem. However, solutions that rely solely on traditional offloading algorithms ignore the dynamically changing needs of miners and various offloading devices in balancing risk and reward, making offloading inefficient.

\subsubsection{Lack of an Ideal Transaction Fee Mechanism}
Some popular permissionless blockchain platforms (e.g., Bitcoin and Ethereum) use a built-in transaction fee mechanism; users attach a transaction fee when submitting a new transaction, and miners prioritize processing the transactions with the highest fee. This mechanism incentivizes miners to participate in transaction confirmation, thus ensuring the continuous operation of the blockchain. However, it is difficult for users to estimate how much they need to pay in order to have their transactions accepted on the blockchain \cite{ferreira2021dynamic}. In addition, this mechanism can lead to huge fluctuations in transaction fees as the volume of blockchain transactions explodes. 

\subsubsection{Lack of Miner Selection Mechanisms to Build a Reliable Blockchain Consensus}
The appointment of miners is crucial to the proper functioning of the blockchain. If miners violate the rules, it may result in the loss or tampering of transaction data. Severe cases may even lead to errors in the entire blockchain network. Some blockchain consensus algorithms require the selection of miners to process and verify transactions. For example, the PoS consensus is often criticized for selecting nodes with more stakes as miners, leading to centralization of control and unfairness \cite{saad_e-pos_2021}. Therefore, it is essential to ensure randomness and fairness in the miner selection process.

\subsubsection{Lack of Effective Mechanisms for Trading Tokens}
The practice of buying and selling cryptocurrencies or tokens to earn profits is known as cryptocurrency trading. There are many popular platforms available for cryptocurrency issuance and trading. However, the current blockchain community still lacks a fair and efficient trading strategy that allows buyers and sellers to match demand in a short period of time. Auctions are a great way to exchange tokens for fiat currency or other tokens in such cases.

\subsection{Motivations for the Two-Way Integration}

Blockchain technologies effectively eliminate intermediaries, thereby reducing transaction costs and ensuring trust among auction stakeholders \cite{hawlitschek2018limits}. In general, blockchain technologies can enhance auction models from the following aspects: 

\begin{itemize}
    \item \textit{Immutability of the Auction Transaction.} Every transaction executed on the blockchain is public, verifiable, and immutable. This means that the blockchain can be leveraged as an audit certificate device that prevents participants from cheating during the auction. The winning bidders can also use the blockchain as a transaction proof \cite{braghin2018designing}.
    \item \textit{Automation of the Auction Process.} A smart contract automates the auction process on the blockchain. Almost all auction logic can be predefined in smart contracts to facilitate the exchange of goods or services as well as the token payment.
    \item \textit{Decentralization of the Auction Management.} There is no need for a specific third-party auctioneer, which ensures trustworthiness and greatly reduces the auction cost. By contrast, traditional centralized auctions can be very expensive and subject to cheating auctioneers; auction houses typically charge 8-20\% of the hammer price as a commission \cite{Commissionsfee}.
    \item \textit{Flexibility in the Auction Payment.} Cryptocurrencies embedded in the blockchain can improve the security and flexibility of auction payments. At the same time, a decentralized payment scheme obviates the need for financial intermediaries, making transactions more convenient and less costly.
\end{itemize}

\begin{figure}[ht]
\centering
\includegraphics[width=\linewidth]{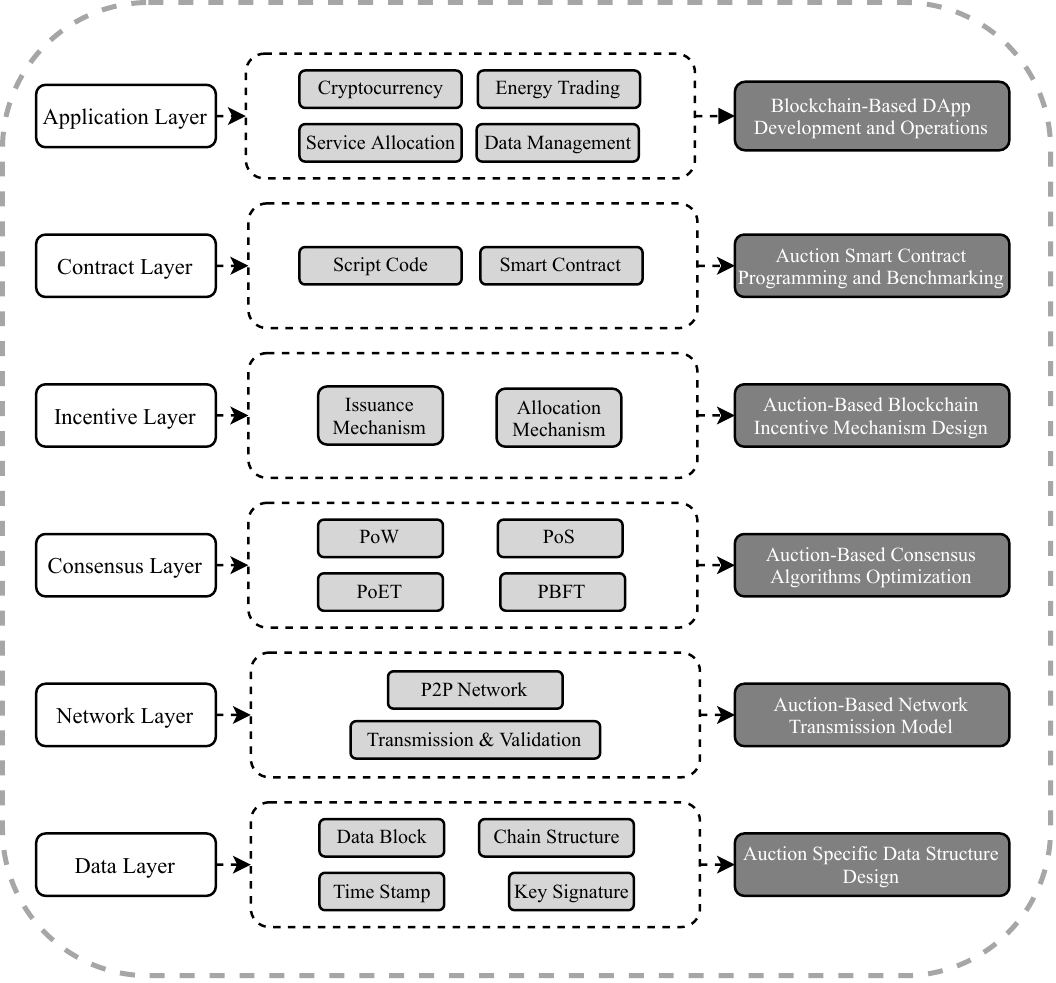}
\caption{Integration opportunities in different blockchain layers.}
\label{architecture}
\end{figure}

On the other hand, auction models can be inserted into any blockchain component to optimize the overall workflow. They have been used to improve blockchain technologies from different aspects, e.g., modeling and optimizing the blockchain transaction fee mechanisms \cite{lavi_redesigning_2019}, selecting miners \cite{devi_using_2020}, and designing new consensus algorithms \cite{ai_abc_2020}. Given the six-layer architecture of a generic blockchain (as introduced in Section \ref{background_blockchain}), detailed opportunities for integrating blockchain with auction models can be discussed at different blockchain layers, as illustrated in Fig. \ref{architecture}. 

\subsubsection{Integration at the Application Layer}
The application layer is where the blockchain encapsulates various application scenarios and use cases \cite{casino2019systematic}. 
To communicate with the blockchain network, applications use either a command-line interface tool provided by the blockchain platform or a specific programming software development kit. In this layer, various auction application scenarios
can be designed and implemented as DApps. The front-end code of an auction DApp can be written in any programming language and make API calls to its back-end (usually blockchain nodes).
Different cryptocurrency DApps are also implemented at this layer to support auction payments.
\subsubsection{Integration at the Contract Layer}
Once deployed on the blockchain, a smart contract cannot be altered. Therefore, an auction smart contract must be carefully analyzed, developed, and tested to ensure that the contract rules meet all requirements before they take effect in a real blockchain. Researchers have designed various auction smart contracts for different auction models (e.g., English auction, Dutch auction, and sealed-bid auction). The operational cost of these contracts, as well as the security and privacy of auction transactions, are the current concerns that need to be addressed urgently. Another consideration for contract layer integration is the development of new programming languages for auction smart contracts \cite{wohrer2020domain}. In addition to general programming languages (e.g., Solidity in Ethereum), domain-specific programming languages have also been proposed to improve the usability of auction smart contracts \cite{sergey2019safer}.

\subsubsection{Integration at the Incentive Layer}
There is great potential to use auction-based incentives in this layer to promote the development of blockchain systems. Most current permissionless blockchains (e.g., Bitcoin and Ethereum) leverage a built-in GFP auction mechanism for the transaction fee market design; users attach a transaction fee when they submit a new transaction to the blockchain, and miners choose the transactions with the highest fee for priority processing. Alternative auction models (e.g., GSP auction) hold the promise of making the blockchain transaction fee market more efficient \cite{basu_towards_2019}.
Another integration direction in this layer is the block reward allocation in mining pools. In reality, auction models are widely used for efficient and fair block reward allocation to motivate more miners to join the mining pool \cite{liu_auction_2019}.

\subsubsection{Integration at the Consensus Layer}
Existing studies have highlighted the need to integrate auction models into blockchain consensus mechanisms.
Auction models can be added to the consensus layer in different manners. 
For example, miners can use auctions to offload the mining tasks to cloud/edge/fog computing servers when traditional PoW consensus requires too many computational resources \cite{xu_hierarchical_2020}. Besides, auction models can be leveraged to model and optimize existing consensus mechanisms. The selection of miner nodes in PoS can be modeled as an auction where miners bid for the new block, and the miner with the highest bid (stake) wins the auction \cite{saad_e-pos_2021}. There are also new consensus mechanisms that are designed using the auction mechanism \cite{ai_abc_2020}. All these directions make the consensus layer a more developed area in terms of integration.

\subsubsection{Integration at the Network Layer}
Researchers are working hard to optimize the blockchain network layer with various techniques to improve security and efficiency \cite{hassan2020differential}. 
However, there are still many concerns about whether an auction model is suitable in this layer.
One promising topic would be the design of data transfer mechanisms with incentives for P2P networks using auction models \cite{ma2004incentive}. 
However, to our knowledge, few studies have specifically incorporated auctions into this blockchain layer. We believe that future research in this area should delve into such mechanisms. 
\subsubsection{Integration at the Data Layer}
It is instinctive that there is little space to optimize the blockchain data structure using the auction model. In contrast, when blockchain is used to support auction applications, all data related to auction activities (including bidding and payment) will be stored in this layer in the form of blockchain transactions. While most researchers use current blockchain data structures to store auction transactions, some others are designing customized data structures for auction scenarios. For example, the authors in \cite{thakur_distributed_2018} added additional fields (e.g., ``Auctioned", ``Expired", ``Price", and ``Consumption") to the Bitcoin transaction data structure to represent energy consumption and auction status. Despite such technical advances, a comprehensive study is still lacking in integrating auction models with the data layer.

In summary, although a blockchain is a multi-layered collaborative system, the research on blockchain-based auction applications mainly focuses on the contract and application layers. By contrast, the research using auction models to enhance blockchain technology mostly targets the incentive and consensus layers. Moreover, the integration in data and network layers is less studied and discussed in the literature. In the following text, we present a detailed and state-of-the-art review of the two integration efforts in Section \ref{blockchain4auction} and Section \ref{auction4blockchain}, respectively.

\subsection{When Does the Integration Make Sense?}
The integration of blockchain and auctions is a two-way connection. Therefore, the impact on the underlying application depends on the context.

\subsubsection{Whether to Use Blockchain for Auctions}
While blockchain-based decentralized auctions are exciting and have the potential to change the way many auctions operate, it doesn't mean that blockchain is the right solution for all auction scenarios. In general, using blockchain only makes sense when multiple mistrustful buyers and sellers want to interact and trade, and are unwilling to use a third-party online auction platform \cite{wust2018you}. In addition, auction organizers and participants need to consider the trade-off between the benefits and costs of centralized and decentralized auctions. For example, decentralized auctions are more difficult to maintain. Once smart contracts that support the auction logic are deployed on the blockchain, they can no longer be removed or manipulated. Therefore, if auction managers have a critical requirement for application updates or bug fixes, they should be careful about using a blockchain for decentralized auctions. In addition, performance in terms of latency and throughput is typically much better in centralized auction systems than in blockchains, due to the additional complexity introduced by the blockchain consensus mechanism. The trade-off between decentralization and throughput should also be considered when deciding whether to use a blockchain-based auction system.

\subsubsection{Whether to Use Auctions for Blockchain}
An auction is a fair and effective price-based incentive mechanism. The incentive module in permissionless blockchain plays a crucial role in the overall economic stability, as it needs to encourage users to continuously join the network to maintain the consensus and secure verification of the blockchain. Incentives are ubiquitous in permissionless blockchain platforms, including but not limited to miner rewards, transaction fee mechanisms, token management, and market forecasting. Therefore, the integration of auction models with blockchain becomes logical and necessary. However, whether auctions are effective and can achieve the desired incentive effect on the blockchain are issues that need to be carefully considered. When better alternative economic incentives exist, auctions will not be needed anymore \cite{ferreira2021dynamic}. It is also important to note that, unlike permissionless blockchains, permissioned blockchains use different consensus algorithms and require that only ``known" nodes can participate in the governance of the ledger. There is no need for built-in cryptocurrencies and complex incentives to maintain the system's stability. As a result, the integration of auction models will not be necessary for permissioned blockchains.

\section{Blockchain-Based Auction Applications}
\label{blockchain4auction}
\begin{figure}[htbp]
  \centering
  \includegraphics[width=\linewidth]{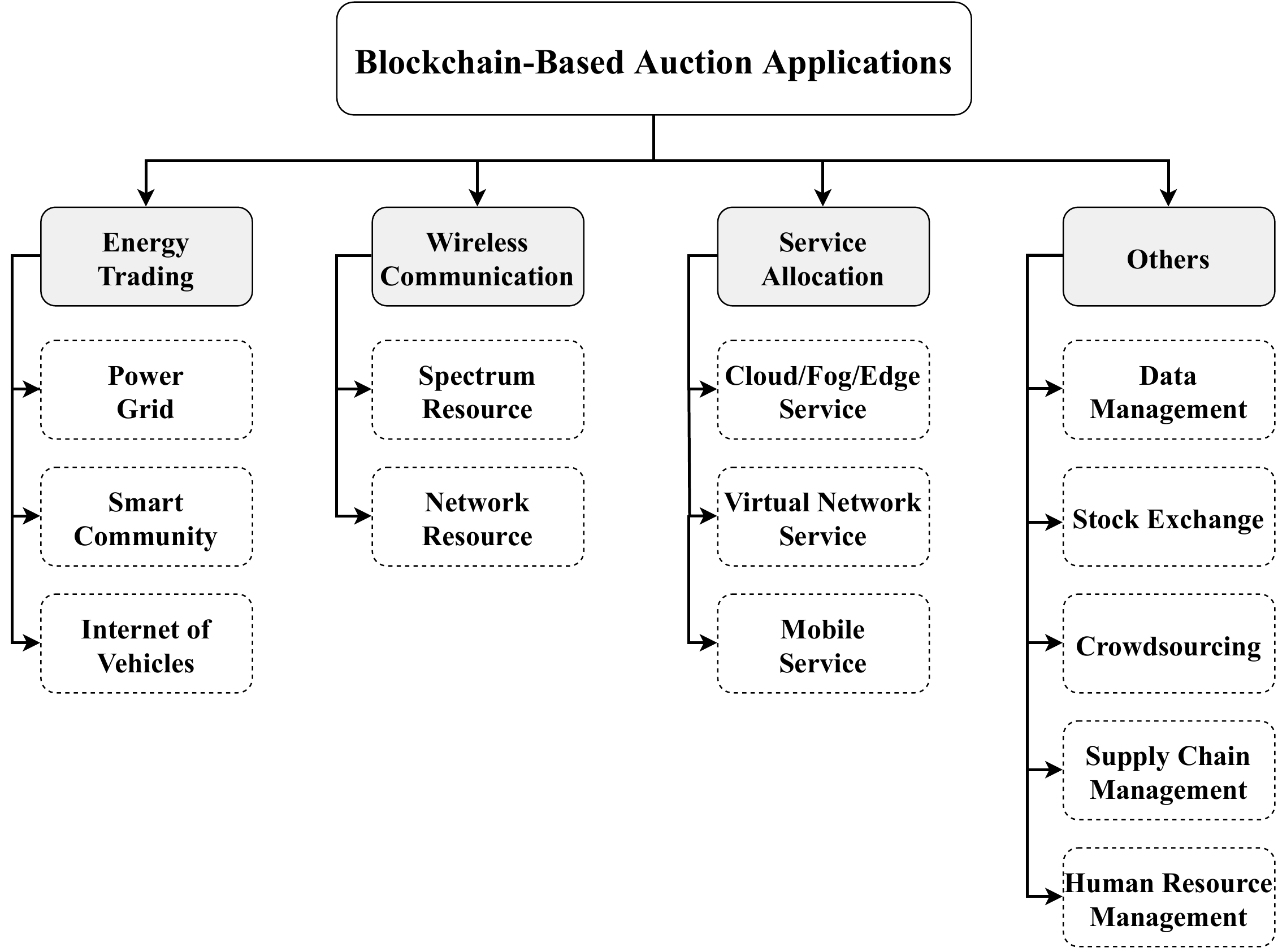}
  \caption{Taxonomy of blockchain-based auction applications.}
  \label{application}
\end{figure}

Existing surveys have indicated the huge potential of blockchain-based auction models in application fields like energy trading~\cite{hassan2021optimizing}.
However, a systematic classification to categorize these applications is still lacking~\cite{casino2019systematic}. In this section, we propose an application-oriented taxonomy for blockchain-based auction applications, which is shown in Fig.~\ref{application}. We identified and reviewed several key application fields, namely energy trading, wireless communication, service allocation, and others. Our classification method is based on a statistical analysis of existing literature and is therefore suitable to analyze current development efforts and illustrate future trends. Table~\ref{tab:application} further summarizes the auction models and blockchain technologies used in different studies.

\subsection{Energy Trading}
Traditional centralized energy transaction models have many shortcomings, including high operating costs, low transparency, and latent risks of transaction data modification \cite{wang_novel_2017}. Integrating blockchain with energy trading is a new paradigm that has recently emerged. As an incentive and pricing mechanism, an auction plays a vital role in ensuring fairness and improving transaction efficiency in energy exchange. However, there are many challenges in integrating traditional energy auctions into blockchain technology. Researchers have proposed different blockchain-based auction models to address those challenges in the energy market \cite{esmat2021novel}. Thematically, the relevant literature can be roughly classified into three categories: power grid, smart community, and Internet of Vehicles (IoV).

\subsubsection{Power Grid}

In traditional centralized power stations (e.g., thermal power, natural gas, and nuclear stations), consumers typically trade indirectly with energy suppliers through retailers in the market. The situation has been improved by a system named microgrid. It is a small-scale power generation and distribution system that comprises distributed power sources, electric loads, distribution facilities, and monitoring devices~\cite{marnay2015microgrid}. By promoting decentralized transactions between distributed generations (DGs) and consumers in a microgrid (instead of letting retailers act as intermediaries), the interests of both parties are increased.
With the development of microgrids, transactive energy paradigms have been proposed to support the development of next-generation energy distribution systems. In this paradigm, customers might also act as suppliers rather than the one-way configuration of suppliers and consumers.
Microgrid systems, in this way, allow customers to store electricity resources, sell them on-demand, and buy them from other customers~\cite{myung_ethereum_2020}.

In this regard, the fusion of blockchain and auction models can provide a transparent and credible trading environment for P2P microgrid energy transactions. The relevant literature has demonstrated that double auctions are more suitable for multi-seller and multi-buyer models in grid transactions. 
In particular, the energy distribution mechanism using double auctions eliminates the need for centralized control, which matches perfectly with the decentralized nature of blockchain.
For example, Wang~\textit{et~al.}~\cite{wang_novel_2017} suggested a model for direct electricity trading between DGs and consumers in microgrids based on blockchain technology and continuous double auctions. 
The model aims to address the potential issues of centralized microgrid trading management, e.g., high operating costs of trading centers, trust issues between trading centers and traders, and huge information security risks.
To allow dynamic adjustment of the auction bids, their model adopts an adaptive aggressiveness bidding strategy. 
Besides, DGs and consumers can exchange digital certificates on the blockchain to settle the auction and guarantee auction security.
Yan~\textit{et~al.}~\cite{yan_novel_2018} used a similar pricing strategy, but they paid more attention to the generation right trade market. They focused on the problem of how to allocate available generation rights to integrate clean energy and reduce thermal power emissions. It should be noted that the energy payments in both of the above-mentioned studies are based on the Bitcoin cryptocurrency protocol. In addition, Thakur~\textit{et~al.}~\cite{thakur_distributed_2018} proposed that the information about energy surplus or deficit can be encoded as blockchain transactions and stored in an optimized Bitcoin data structure to support double auctions.
They argued that blockchain performs a distributed calculation of the winner determination problem, which is more conducive to local energy trading among peers than centralized double auctions. Their simulation experiments showed that distributed double auctions facilitate energy transfer better than centralized double auctions.
Stübs~\textit{et~al.}~\cite{stubs_blockchain-based_2020} 
argued that in a smart grid network, there are multiple data communications between smart devices, edge servers and cloud servers.
So a hierarchical double auction model is proposed for full on-chain implementation of energy transactions. AlAshery~\textit{et~al.}~\cite{alashery_blockchain-enabled_2020} proposed a double auction model with an optimized VCG pricing mechanism for P2P energy trading in power grids on the blockchain.
Zhao~\textit{et~al.}~\cite{zhao_decentralized_2019} proposed a bandit learning-based double auction model that can provide participants with more auction revenues by learning the transaction history. 
Their simulation results showed that the bandit learning approach in a blockchain framework can provide market participants with more revenue than the way energy is traded with centralized entities.

Some traditional single-sided auction models are also presented for microgrid energy trading. Seven~\textit{et~al.}~\cite{seven_peer--peer_2020} proposed a novel P2P energy trading scheme that uses smart contracts for virtual power plants (VPPs). In particular, the authors used an English auction-based workflow to achieve P2P transactions in a VPP. 
The platform is based on a public Ethereum blockchain so that it can be adapted to communications and power distributions on different networks.
Hahn~\textit{et~al.}~\cite{hahn_smart_2017} demonstrated how to implement Vickrey auctions on smart contracts and use them for a trading market, where multiple consumers bid for power resources from photovoltaic arrays. 
Energy consumers may question the fairness, trustworthiness and cyberattack resistance of centralized energy models. Therefore, 
the authors in~\cite{dekhane_greencoin_2019} leveraged both Dutch and Vickrey auction models for user negotiation and power distribution. In addition, a wallet-based cryptocurrency called GreenCoin is created to support energy payments. 

Blockchain-based decentralized systems bring new privacy challenges like the possible leakage of energy usage patterns~\cite{laszka_providing_2017}. So permissioned blockchains with better scalability and identity permission mechanisms are widely discussed in power grids.
In this context, Zhang~\textit{et~al.}~\cite{zhang_privacy_2019} proposed a privacy-preserving scheme for direct power transactions in microgrids, in which a continuous double auction is combined with a permissioned blockchain to reduce costs and improve transaction privacy and efficiency. Hassan~\textit{et~al.}~\cite{hassan_deal_2020}
adopted a permissioned blockchain for the computation of complex on-chain transactions. They argued that the shortcomings of centralized auctioneers in terms of the trust, security, and privacy leakage are more exposed when using VCG auctions. Additionally, they leveraged the differential privacy technology to protect auction privacy.
The authors in \cite{laszka_providing_2017} proposed that transactions and bids can be de-anonymized based on network identifiers (e.g., IP addresses). Therefore, anonymity of the blockchain communication layer is crucial. This can be achieved by anonymous communication techniques such as onion routing.

\subsubsection{Smart Community}
The smart community is another blockchain-based energy auction application field that has attracted much public attention \cite{hassija_blockcom_2019,alcarria_blockchain-based_2018}. In general, a community microgrid is a self-sufficient energy system designed to meet local energy needs (e.g., electricity, heating, and cooling) for communities, villages, towns, and cities. Some households may have extra renewable energy in their community microgrid and can therefore meet the needs of their neighbors. The community can flexibly absorb the peak hours of individual consumers; in this way, the energy demand of the community can be stabilized, and energy resources can be better planned. 
The success of a smart community heavily depends on the function of its auction economic backbone~\cite{hassija_blockcom_2019}. In~\cite{alcarria_blockchain-based_2018}, the authors proposed a model for auctioning energy and water resources between smart communities and smart homes, thus encouraging communities to optimize global consumption. In particular, users can use a Vickrey auction model on the blockchain network during the resource negotiation stage. Guo~\textit{et~al.}~\cite{guo_combined_2020} considered the issue of energy trading in combined cooling, heating, and power (CCHP) systems
and developed a non-cooperative Stackelberg game between power grid agents and the system to model energy transactions.
Their system consists of an Internet of Energy (IoE) subsystem and a blockchain subsystem, where P2P communication and energy transactions between power agents and CCHP systems can be performed efficiently and securely.

Other studies focus on improving the scalability of the blockchain to improve the performance of community energy auctions. Saxena~\textit{et~al.}~\cite{saxena_design_2019} presented a permissioned blockchain implementation of a P2P energy trading system for residential communities. In this system, a single house owner can place his/her energy bid in the district within discrete time intervals on the blockchain. A more scalable local grid system for smart communities is enerDAG~\cite{gros_enerdag_2020}, in which a blockchain with tangled data structures is leveraged to overcome issues such as expensive transaction fees and limited throughput. 
Their decentralized local energy trading platform achieves higher reliability; only a massive disruption of the communication network would cause a system collapse.
However, there are still many debates regarding this blockchain since it deviates from the traditional blockchain's ``chained block" data structure.

Quartierstrom~\cite{brenzikofer_privacy-preserving_2019} is a blockchain-based project for community energy trading. It is designed to manage the exchange and payment of electricity resources between consumers, producers, and local grid suppliers without any intermediaries. In Quartierstrom, a real-world prototype system has been implemented and tested in the town of Wallenstadt in Switzerland (a community with 37 families involved). 
The pricing mechanism of the Quartierstrom market is a double auction with discriminative pricing, while Tendermint serves as the underlying blockchain~\cite{ableitner_quartierstrom_2019}.
Tendermint is highly flexible and customizable to accommodate specific application requirements. It offers reduced communication, empty block creation, and customized time delays between blocks.

\subsubsection{Internet of Vehicles}
Vehicle-to-vehicle (V2V) describes a trading model in which plug-in electric vehicles (EVs) communicate with each other to exchange electricity energy. 
It can enhance the cooperation between vehicles, extend the driving endurance, and avoid the grid overload problem~\cite{xia_bayesian_2020,wang_electric_2020}. 
However, conducting non-transparent energy transactions in IoV without trust is risky.
Most existing IoV energy trading platforms and facilities are centralized, and they rely on TTPs to manage power dispatch, transaction payments, and security issues; nevertheless, these third parties are costly and can be corrupted~\cite{sun_blockchain-enhanced_2020}. In a blockchain-enabled decentralized IoV network, Xia~\textit{et~al.}~\cite{xia_bayesian_2020} argued that Bayesian games with incomplete information have significant advantages over complete information games in terms of communication overhead. Therefore, they presented a V2V electricity trading strategy using Bayesian game-based bidding and pricing. Sun~\textit{et~al.}~\cite{sun_blockchain-enhanced_2020} further considered transaction privacy and efficiency issues. They proposed that centralized IoV energy trading platforms suffer from a single point of failure and lack privacy protection. In addition, power centers are inefficient in controlling large-scale and geographically distributed EVs, especially in social hotspots far from charging stations. They adopted a permissioned blockchain in the designed V2V energy trading architecture. Additionally, a novel DPoS consensus mechanism is utilized to boost trade efficiency. 
In \cite{ali_efficient_2020} and~\cite{guo_double_2020}, the authors argued that the high computational cost required in the classic permissionless blockchain is not suitable for IoV. Therefore, they adopted a blockchain with a DAG data structure for charging scheduling among EVs.
Furthermore, Choubey~\textit{et~al.}~\cite{choubey_energytradingrank_2019} introduced a new cryptocurrency called ETcoin to facilitate energy transactions among EVs on the permissioned blockchain. 

\begin{figure}[!t]
  \centering
  \includegraphics[width=\linewidth]{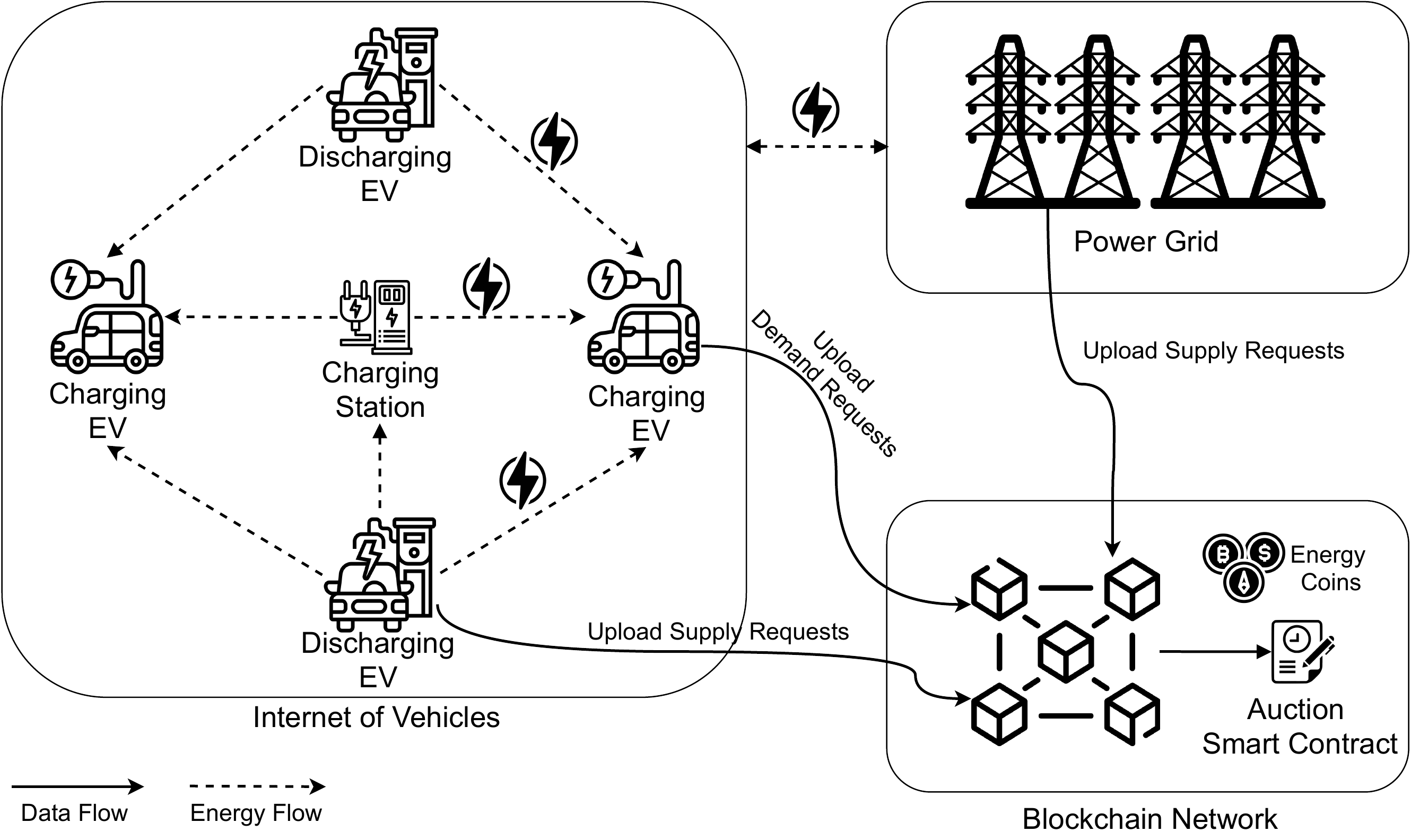}
  \caption{An illustration of a blockchain-based energy trading model for IoV. Charging/discharging EVs and power grids upload the demand and supply requests as well as the bids/offers to the blockchain. After transactions are confirmed on the blockchain, the energy resources are traded between different entities and paid in cryptocurrencies.}
  \label{IoVenergytrading}
\end{figure}

Another related topic is vehicle-to-grid (V2G), which describes a system in which plug-in EVs communicate with the grid by returning electricity or limiting their charging rate to sell demand response services.
Hassija~\textit{et~al.}~\cite{hassija_blockchain-based_2020} proposed a scheme utilizing the IOTA blockchain for data sharing and energy trading in V2G networks. The scheme implements an auction-based game-theoretic approach for the price competition between EVs and grid users.
Similarly, Liu~\textit{et~al.}~\cite{liu_electric_2019} developed a reverse auction-based dynamic pricing model for V2G networks in order to improve social welfare and transaction efficiency. In their model, unfilled charging EVs are powered by the smart grid, while charging and discharging transactions are executed on the smart contract. Pustišek~\textit{et~al.}~\cite{pustisek_blockchain_2016} presented a model that allows independent selection/dispatch of the most convenient charging stations for EVs in V2G networks via blockchain. 
Compared to traditional centralized approaches, such a solution does not require any central entity and can be fully automated, including the payment of energy.
The model is implemented using the Ethereum blockchain and an FPSB auction model. To summarize, a general blockchain-based energy trading model for IoV is illustrated in Fig. \ref{IoVenergytrading}.

\subsection{Wireless Communication}
As wireless systems develop with new mobile communication technologies, they become increasingly complex in terms of architecture and management. Auctions have been proposed as practical mechanisms for assigning a wide range of wireless resources (e.g., spectra, subchannels, time slots, and transmit power levels). By designing and employing various auction procedures, wireless resources can be efficiently allocated between consumers and resource providers~\cite{zhang2012auction}.

\subsubsection{Spectrum Resource}
\begin{figure}[!t]
\centering
    \subfigure[A Typical Centralized Model]{
    \begin{minipage}{\linewidth} %
    \centering\includegraphics[width=\linewidth]{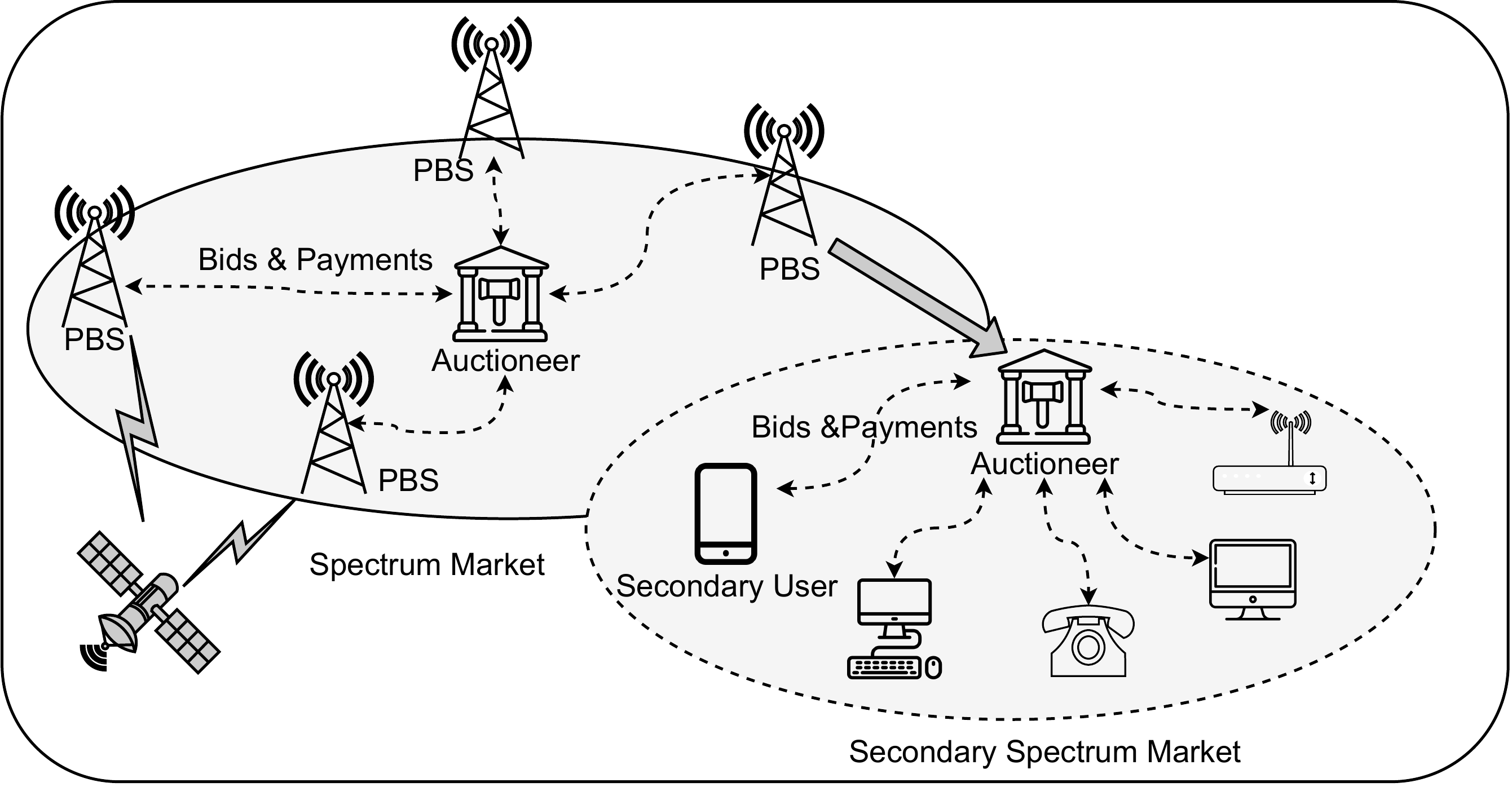} \\
    \end{minipage}
    \label{centralized}
    }
    \subfigure[A Typical Decentralized Model]{
    \begin{minipage}{\linewidth}%
    \centering\includegraphics[width=.99\linewidth]{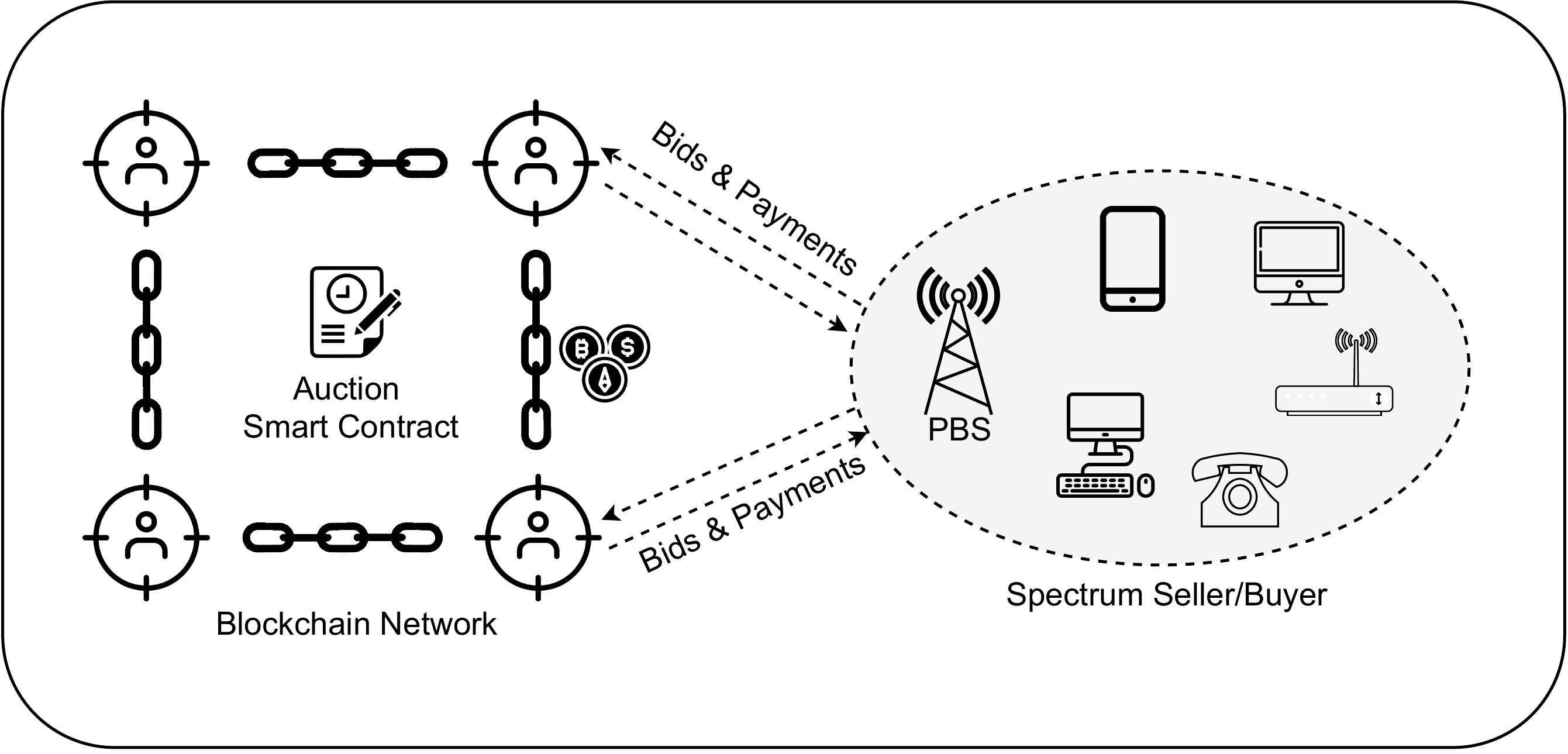} \\
    \end{minipage}
    \label{decentralized}
    }
\caption{A comparison of centralized and decentralized spectrum auction models. A primary base station (PBS) obtains or transfers the spectrum ownership through a centralized auction managed by an auctioneer.
While in a decentralized auction, spectrum users can conduct P2P spectrum transactions on the blockchain without the need for a third-party auctioneer.} 
\label{spectrumauction}
\end{figure}

With the rapid development of communication technology, users' demand for spectrum resources continues to increase, making spectrum a scarce resource in the trading market. However, the traditional government-led static spectrum allocation approach has failed to fully utilize the limited spectrum resources. According to the report from FCC, the utilization of the licensed spectrum can only be maintained between 15\% to 85\% with static spectrum allocation solutions~\cite{federal2002spectrum}. As a result, market-driven spectrum auctions have emerged as promising solutions for spectrum allocation~\cite{wang_secure_2020}. A spectrum auction can be centralized or decentralized, and Fig. \ref{spectrumauction} shows a comparison of the two approaches. 

In this context, Fan and Huo~\cite{fan_blockchain_2020} suggested a blockchain-based framework for license-free spectrum resource management in cyber-physical-social systems (CPSS). In particular, two ways of obtaining a spectrum access license (i.e., mining and auction) are designed. A new virtual currency, called Xcoin, is also introduced in this process to enhance spectrum trading. Yu~\textit{et~al.}~\cite{yu_smart_2019} focused on the space communication field and presented a spectrum auction model for heterogeneous spacecraft networks based on blockchains.
They argued that the communication between different organizations in a heterogeneous spacecraft network is multi-hop compared to traditional space communication networks, which makes coordination difficult.
Recent studies have further highlighted the security and privacy challenges~\cite{zheng_smart_2020}. For example,
Tu~\textit{et~al.}~\cite{tu_blockchain-based_2020} designed a privacy-preserving double auction mechanism for blockchain-enabled spectrum sharing using the differential privacy technology.
Wang~\textit{et~al.}~\cite{wang_secure_2020} designed a secure spectrum auction protocol that utilizes Intel Software Guard Extensions (SGX) technology and the Paillier cryptosystem.
In their system, each bidder can use remote authentication to establish a secure communication channel with the SGX enclave thereby enabling the transmission and computation of sensitive data.

It should be noted that spectrum auctions are different from traditional auctions due to the reusable nature of spectrum resources. 
In most traditional auctions, the same items (e.g., artworks, antiques, and estates) can only be auctioned to a specific buyer.
Spectrum auctions, by contrast, can allow the sharing of an auctioned channel as long as the buyers do not interfere with each other. 
In this context, dynamic spectrum management in cognitive radio (CR) networks can address the lack and underutilization of spectrum resources. 
CRs can be dynamically programmed and configured to use the best wireless channel nearby to avoid users interference and congestion. Based on the cognition and reconfiguration of CRs, the primary users can share their licensed spectrum with secondary users to improve spectrum utilization~\cite{zheng_smart_2020}. The authors in~\cite{kotobi_blockchain-enabled_2017,kotobi_secure_2018} argued that the current centralized spectrum allocation is wasteful since license holders do not consistently utilize their allocated spectrum resources. They therefore introduced the idea of using blockchain as a decentralized database to verify spectrum sharing and auctions in CR networks. 
For secondary spectrum auctions in a CR network, an automatic pricing strategy based on a blockchain token called ``spectrum dollars" is introduced in~\cite{khan_blockchain_2020}.

\subsubsection{Network Resource}
In addition to spectrum resources, researchers have paid attention to other network resources in wireless networks. SAFE~\cite{chen_safe_2020} is a framework designed for users to customize auction formats and allocate general wireless network resources, e.g., spectrum channels, femtocell access permissions, and resource blocks of device-to-device connections. 
Numerous experimental results have shown that the communication cost of SAFE is quite low, so it is practical in real-life network environments.
Afraz~\textit{et~al.}~\cite{afraz_distributed_2019} proposed a distributed resource market mechanism for future telecommunications networks, in which a double auction model and a permissioned blockchain are combined to enhance the scenarios of bilateral trading markets that exist in the telecommunications industry such as resource allocation in network functions virtualization (NFV), mobile crowd sensing, and femtocell access.
Besides, cooperative relaying can be an effective way to improve the capacity, reliability, and security of wireless networks.
It either helps establish communications between the source and destination or improves the established communications by adding diversity. In~\cite{khan_blockchain-based_2019}, relay operators are designed to be responsible for the relay/jammer selection and resource allocation. A double auction mechanism is used to simulate the interaction between transmitters and relay operators.
Furthermore, User congestion in wireless networks is a severe problem to be solved. A Vickrey auction-based user offloading mechanism between macrocell base stations and small cell access points has been proposed in~\cite{chen_blockchain_2020} to improve the capacity of heterogeneous wireless networks.
Their blockchain-enabled decentralized auction solution avoids multiple malicious behaviors caused by auctioneers (third-party agents), sellers (macrocell base stations), and buyers (small cell access points).

An unmanned aerial vehicle (UAV), also known as a drone, is a newly emerging flying antenna system with a critical requirement for network resource allocation. Accordingly, a drone-mounted base station is primarily responsible for the communication between the UAV backhaul and access networks. In this field, Hassija~\textit{et~al.}~\cite{hassija_drone_2020} introduced the idea of using dynamic auctions to allocate the bandwidth of drone-mounted base stations to different users to improve availability and reduce costs.
They argued that communications between drone-mounted and regular base stations are vulnerable to wiretapping or man-in-the-middle attacks, so using blockchain to record the data exchange of wireless communication in a tamper-proof ledger would be a good choice.
Khan~\textit{et~al.}~\cite{khan_trusted_2020} proposed a multi-UAV network framework, which can: 1) outsource network coverage in specific areas based on the required service requirements; 2) enable each network entity to use the blockchain intelligently; and 3) provide an auction mechanism to make autonomous decisions. To model the interaction between UAV operators and business agents, a reputation-based truthful auction method is also presented. 

\subsection{Service Allocation}
Recent developments in service computing allow the use of blockchain to allocate heterogeneous services, where blockchain can be used as decentralized auditing devices, and cryptocurrencies can secure money payments. However, most of the existing models do not provide incentives for matching service customers and providers; they often rely on manual and inefficient solutions~\cite{sonnino_asterisk_2019}. Therefore, different auction models are proposed together with blockchain to provide secure, credible, and economical service allocation platforms.
\subsubsection{Cloud/Fog/Edge Service}
\label{edge}
With the rapid growth of the cloud computing industry, more and more application operators are now using the cloud for service hosting, computing offloading, and data storage. Some large cloud service providers (e.g., AWS, Azure, and Google Cloud) have already supported spot instance pricing, allowing users to bid on unused capacity in cloud data centers. In this way, some users can even save up to 90\% of the cost compared with the traditional on-demand instance pricing \cite{spotinstance}.
However, since cloud service providers usually sell services in a centralized and opaque manner, the fairness of the auction is challenging to guarantee in reality.
A trustworthy transaction and payment mechanism is urgently needed to motivate service providers/customers and improve service utilization. AStERISK~\cite{sonnino_asterisk_2019} is a framework designed to fill this gap; it automatically determines the best price for cloud services and assigns customers to the most appropriate providers by implementing sealed-bid auctions on the blockchain. 
Similarly, Chen~\textit{et~al.}~\cite{chen_fair_2020} introduced a blockchain-based auction and trading model for cloud virtual machine allocation. Their model can achieve fairness in auction transactions by implementing commitment-based state mechanisms, smart contracts, and cryptocurrency technologies. In~\cite{gu_cloud_2018,gu_decentralized_2018}, the authors paid attention to the cloud storage problem and proposed VCG auction-based resource trading models for distributed cloud storage. This is based on the context that traditional storage resource trading systems typically operate in a centralized model, leading to high costs, vendor lock-in and single point of failure risks.

The paradigm of connecting things to the cloud to receive a centralized service is not always the best option, which leads to the context in which edge and fog computing are widely discussed.
Basically, they both intend to distribute the computing capacity and assist the cloud server with additional resources located near the end users~\cite{Zigurat_2019}. In this respect, DeCloud~\cite{zavodovski_decloud_2019} is a secure and decentralized auction system specifically built for open edge computing infrastructures. It integrates a truthful double auction and a bidding language to match highly heterogeneous edge resources with different service requests. 
Compared to recently proposed decentralized cloud/fog solutions such as iExec, Sonm and Golem, DeCloud is more focused on designing an effective marketplace for decentralized open infrastructures.
Debe~\textit{et~al.}~\cite{debe_blockchain-based_2020} demonstrated a blockchain-based reverse auction solution for public fog service allocation. Yu~\textit{et~al.}~\cite{yu_building_2019} also leveraged the reverse auction model and presented a blockchain-based edge crowdsourcing service system. Specifically, a changeable auction algorithm is designed so that each request from the user will find a winner that can provide the appropriate edge service.

\subsubsection{Virtual Network Service}
Network function virtualization (NFV) has come into view for its ability to provide multiple network functions at a low cost \cite{han2016virtual}.
The traditional NFV marketplace relies on third-party companies for the provisioning, distribution, and execution of NFV resources.
BRAIN~\cite{figueredo_brain_2019} is a blockchain-based reverse auction solution with a focus on NFV scenarios. It is introduced to address the challenge of discovering and selecting infrastructures that can efficiently host NFV services based on specific user needs. Virtual network embedding (VNE) is one of the most important problems in network virtualization and is responsible for mapping virtual networks to underlying physical networks. Many auction methods have been presented in the literature to achieve efficient resource allocation in VNE. 
Rizk~\textit{et~al.}~\cite{rizk_brokerless_2018} argued that although a centralized VNE approach demonstrates high efficiency in slice allocation, it suffers from scalability issues since everything depends on one virtual network provider. Therefore, they designed a decentralized VNE system that uses smart contracts and a Vickrey auction model for trustworthy virtual network partitioning and allocation.

\subsubsection{Mobile Service}
The number of mobile devices and compute-intensive mobile applications has exploded in recent decades. The focus of these mobile applications is to improve the quality of service (QoS) for end users; however, by improving the QoS, these applications generate a large amount of mobile traffic, thus posing a huge challenge to mobile network providers. 
One of the most promising ways to deal with this issue is mobile data offloading. For example, Hassija~\textit{et~al.}~\cite{hassija_mobile_nodate} created a mobile data offloading model in which mobile devices and users can securely perform computation offloading services on the blockchain. 
The simulation results showed that their model achieves low communication costs and optimized scheduling performance compared to other offloading schemes.
FlopCoin~\cite{chatzopoulos_flopcoin_2018} is a virtual currency specially designed for compensating mobile devices when they execute device-to-device offloading services.

On the other hand, the widespread dissemination of programmable sensor-employed smartphones has facilitated mobile crowdsensing applications such as environmental monitoring, crowd journalism, and public safety. These applications require effective incentives to compensate and reward mobile users for their resource contributions. Chatzopoulos~\textit{et~al.}~\cite{chatzopoulos_privacy_2018} suggested the use of blockchain and smart contracts to manage spatial crowdsensing interactions between mobile service providers and customers. A truthful and cost-optimal auction model is also designed on the blockchain to reduce payments from crowdsensing providers to mobile users.
Their experimental results showed that the time overhead of using blockchain in short-term crowdsourcing tasks is negligible compared to centralized server solutions.

\subsection{Others}
\subsubsection{Data Management}
The uncertainty of data value makes it difficult to make accurate estimates of the appropriate price for data. 
An auction is a powerful approach to protect the interests of both data sellers and buyers while maintaining the fundamental principles of the marketplace. To eliminate systemic risks caused by collusion in large-scale data auctions, the authors in \cite{xiong_anti-collusion_2020} introduced a decentralized data auction system that uses an anti-collusion auction algorithm executed on the smart contract. The system ensures that buyers and sellers can engage in data auctions without relying on TTPs. An~\textit{et~al.}~\cite{an_truthful_2019} implemented a crowdsourcing data trading system using blockchain and reverse auctions. They used carefully designed smart contracts to replace third-party data brokers, thus providing a trustworthy environment for data sellers and consumers.
Besides, a permissioned blockchain-based model is used in~\cite{chen_secure_2019} to enable secure and efficient IoV data transactions. An iterative double auction model is also presented to optimize data pricing and improve data transaction volume.

\subsubsection{Stock Exchange}
A stock exchange is a marketplace where traders can buy and sell securities, e.g., stocks, bonds, options. 
Traditional stock markets are performed in a centralized manner. This structure ensures the authenticity and security of transactions, but is vulnerable to attacks and lack of transparency in the trading process.
To address the single point of failure in centralized stock exchange platforms, Al-Shaibani~\textit{et~al.}~\cite{al-shaibani_consortium_2020} introduced a permissioned blockchain-based decentralized stock exchange platform. Similarly, Pop~\textit{et~al.}~\cite{pop_decentralizing_2018} suggested addressing the shortcomings of centralized stock trading to reduce transaction costs caused by brokers and central institutions. An Ethereum-based decentralized Bucharest stock exchange model is further
proposed and validated. 
Their experimental results indicated that for partially filled order books, the blockchain-based solution has a significant price advantage compared to the centralized solution.
Recently, dark pool trading, as an anonymous and decentralized stock trading approach, has become an increasingly important component of traditional stock exchanges. The decentralized and secure transaction properties of blockchain are well suited to provide support for anonymous dark pool transactions.
AuditChain~\cite{vishnia_auditchain_2020} is an auditing and record-keeping platform for financial markets using blockchain. 
In particular, a periodic double auction-based dark pool use case is used to demonstrate the platform's feasibility for stock trading.
When a private corporation wants to raise capital by issuing new stocks, it can issue shares to the public by conducting an initial public offering (IPO). 
Purchasers usually acquire multiple shares from the seller at the same price in an IPO, which is a typical example of a uniform price auction. 
In \cite{halevi_initial_2019}, the authors introduced a uniform price auction model for IPOs on the permissioned blockchain. 
They designed an additional communication chaincode to provide applications with limited access to P2P APIs in the built-in communication layer.
The model further leverages secure multi-party computation technology to protect the privacy of IPO transactions.

\subsubsection{Crowdsourcing}
Crowdsourcing is a specific business model for acquiring resources in which an individual or organization can leverage a large number of users to obtain desired services. Traditional centralized crowdsourcing platforms face many challenges, including motivating workers to share their truthful costs and guaranteeing trusted interactions among users and the platform. To cope with those challenges, ABCrowd~\cite{kadadha_abcrowd_2020} is a fully decentralized crowdsourcing framework that implements a repeated single-minded VCG auction mechanism on the blockchain. 
BitFund~\cite{hassija_bitfund_2020} is a platform designed to connect developers and investors in the global crowdfunding environment, where a novel ascending-price progressive auction algorithm is implemented for cost-effective task allocation.

\subsubsection{Supply Chain Management}
In a supply chain, decentralized auctions can be widely used to coordinate transactions between suppliers and consumers. BitCom~\cite{gupta_bitcom_2020} is a decentralized supply chain model built on the blockchain to provide a clean and efficient trading environment. Martins~\textit{et~al.}~\cite{martins_fostering_2020} proposed a customer-driven supply chain marketplace on the blockchain, where customers post their proposals and suppliers strive to outbid each other in a reverse auction model. 
Similarly, Koirala~\textit{et~al.}~\cite{koirala_supply_2019} introduced a solution to improve transparency and traceability in the carrier procurement process. Their solution considers multiple attributes of carriers in the supply chain during the reverse auction bidding process.
The traditional English auction model has also been found in the literature. In~\cite{shwetha2021auction}, an online English auction system is implemented to sell and buy food products using the Ethereum blockchain.

\subsubsection{Human Resource Management}
Employment and labor industries become more and more important since the value of human resources is directly related to a company's profitability. 
However, employee background check remains a controversial field in HR operations, particularly in the cases of employment, education, and skills verification~\cite{liu_blockchain-based_2020}. E\textsuperscript{2}C-Chain \cite{liu_blockchain-based_2020}~\cite{liyuan_e2_2019} is a two-stage blockchain designed to assist the improvement of human resource management. In the first stage, the employees' background records can be stored in the blockchain in an immutable manner. After that, a VCG auction mechanism is leveraged to encourage verifiers to join in the skill verification of employees. Another application field is employee recognition program, where employers reward employees for their achievements, milestones, and anniversaries \cite{Employee_2019}. In such a context, Ward~\textit{et~al.}~\cite{ward_establishing_nodate} argued that employees could liquidate their unwanted gifts to others through auction mechanisms. Blockchain and smart contract technologies can be used in this process of matching individuals for exchanging gifts.

We also identified individual applications in blockchain-based auction models, e.g., federated learning (FL), IoT collaboration, and code ownership management. For instance, a centralized aggregator is usually needed to maintain and update the global state in a traditional FL model. BAFFLE \cite{ramanan_baffle_2020} is a decentralized framework for non-aggregator FL. It uses smart contracts to coordinate FL tasks and a user scoring and bidding mechanism to reach the FL goal.
For FL in edge computing, Fan~\textit{et~al.}~\cite{fan_hybrid_2020} proposed a resource trading system using a hybrid blockchain.
Their main idea is to establish a transparent, decentralized, and high-performance trading platform that can encourage more edge nodes to join in the FL model training. Another interesting topic is collaborative IoT. As IoT projects become more and more complex, IoT managers, experts, and non-technical staff are expected to collaborate in the IoT development cycle \cite{iotcollaboration_2020}. In \cite{cheng_auction-based_2020}, a novel blockchain-based reverse auction model is proposed to prompt active cooperation among IoT participants. Besides, the current centralized code ownership management scheme is cumbersome and opaque. Therefore, a blockchain-based approach for managing code ownership is proposed in \cite{seike_blockchain-based_2018}, where auctions are used for ubiquitous code allocation. 

\subsection{Key Observations} The key observations we obtained in this section are summarized as follows:

\begin{itemize}
\item Blockchain-based decentralized auctions offer great potential to optimize the traditional centralized auction model, which is particularly reflected by different application scenarios. Different researchers have used different auction models and blockchain technologies to handle auctions for specific application scenarios. These applications exist mainly in energy trading, wireless communication, and service allocation.

\item The centralized auction model has long been the dominant trading model in energy trading. However, the development of traditional centralized energy markets has gradually encountered bottlenecks. For example, the performance of energy trading is highly dependent on the servers and networks of centralized third-party platforms in the traditional model. It is therefore vulnerable to single points of failure. In addition, centralized auction management leads to high operational costs, low transparency, and the potential risk of tampering with energy transaction data. Finally, centralized long-distance energy transmission makes the power supply vulnerable to disruptions \cite{wang2019energy}. In contrast, decentralized P2P energy trading is a more desirable solution in modern power systems to improve efficiency and stability. 

\item Efficient allocation of scarce network resources has always been a hot research topic in wireless communications. Although resource sharing architectures using both centralized and decentralized auctions can improve resource utilization, the security issue of conducting transactions between untrusted entities is severe in a centralized model. In addition, most traditional solutions can only maintain a single specific auction format and lack a common framework that can accommodate a variety of auction formats. Automation of business processes is becoming increasingly critical as it facilitates dynamic utilization of network resources. 

\item In terms of service allocation, the traditional centralized approach to service trading suffers from several weaknesses. For example, most existing cloud auction solutions have vendor lock-in issues, where the vendor acts as an auctioneer. In such cases, auction fairness is difficult to guarantee because large cloud providers can abuse their dominant market position, forcing users to trust their services and adapt to the rules and prices. In addition, some service providers and customers may collude with third-party auctioneers to learn about users' bids and use that knowledge to gain more profit or exit the market in time.

\item All of the above issues are driving the application of blockchain-based decentralized auctions as a future trend. Overall, blockchain as an enabling technology in the transition from centralized to decentralized auctions provides the following advantages: 1) Decentralized trust management. Blockchain provides a decentralized, transparent, and trustworthy auction trading environment. Such a design does not require a centralized auctioneer and optimizes the design and operation of the trading platform; 2) Secure, private, and cost-effective transaction. Compared to traditional centralized auctions, blockchain-based auctions can achieve the trust requirements of auction participants at a much lower cost; 3) Tokenized auction payment. Blockchain has a cryptocurrency market with a broad user base to support auctions, and some application-specific tokens are designed to be used for specific auctions; 4) Customizable auction format. With the powerful programmability provided by smart contracts, almost any auction format can be programmed to meet specific application and business requirements; and 5) Automated auction execution. Smart contracts can help automate the auction process so that all participants can immediately get results according to established rules without any intermediary involvement or loss of time.
\end{itemize}

\begin{figure}[!t]
\centering
    \subfigure[Distribution of Blockchain Platforms]{
    \begin{minipage}{.5\textwidth} %
    \centering
    \includegraphics[width=\textwidth]{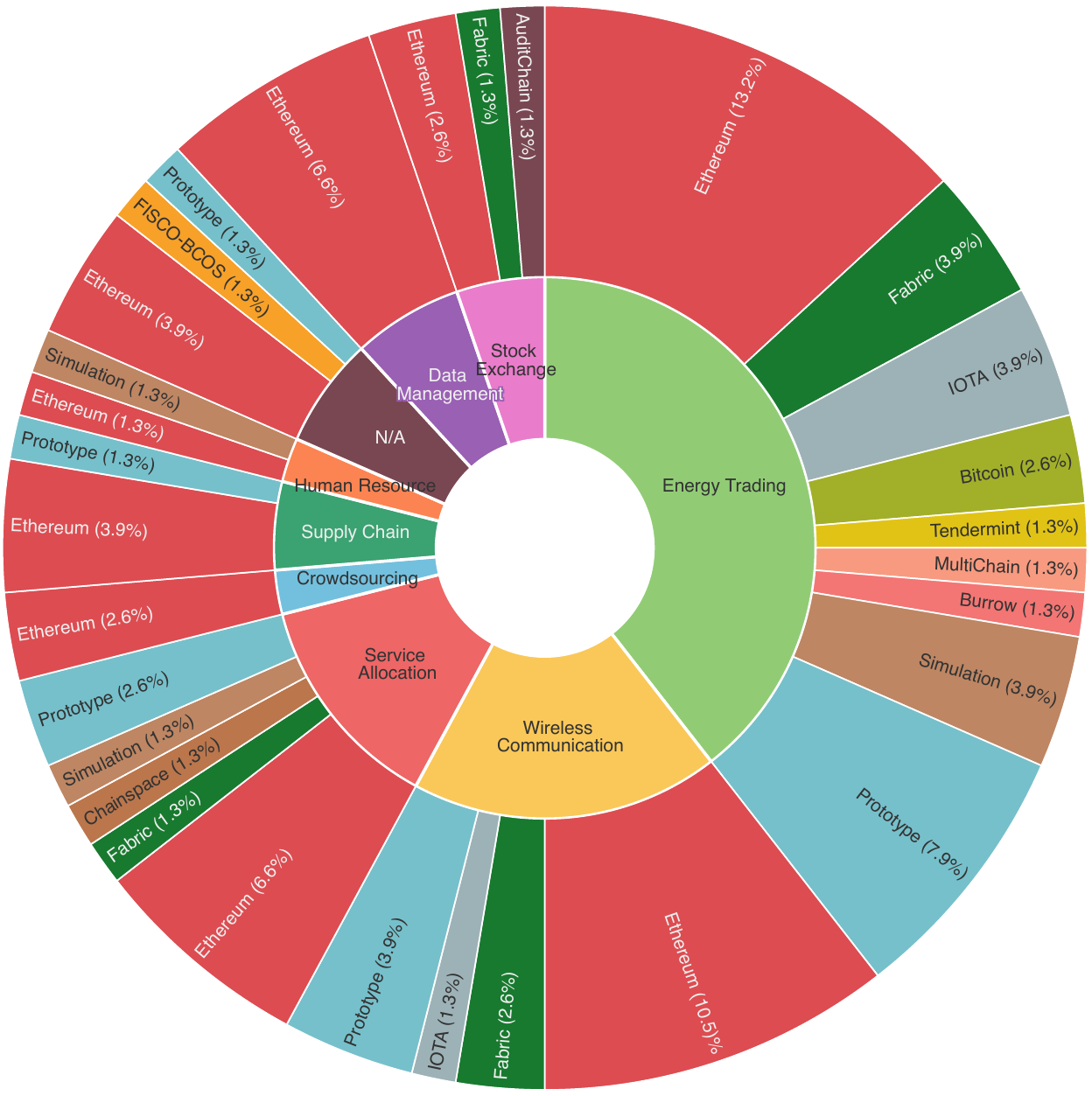} \\
    \end{minipage}
    \label{blockcaindis}
    }
    \subfigure[Distribution of Auction Models]{
    \begin{minipage}{.5\textwidth}%
    \centering
    \includegraphics[width=\textwidth]{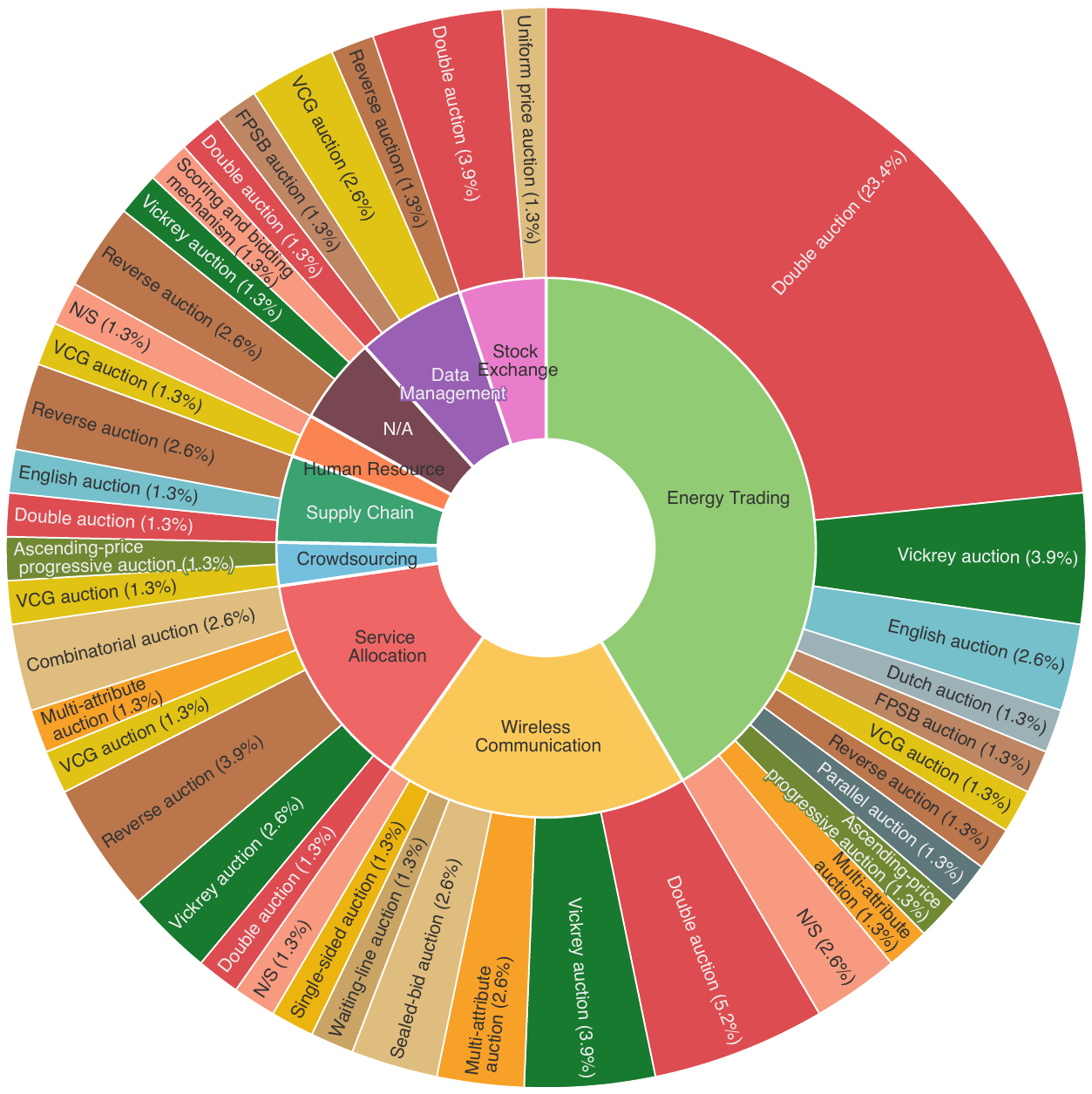} \\
    \end{minipage}
    \label{auctiondis}
    }
\caption{The distribution of blockchain technologies and auction models in existing studies regarding different application fields. The results are obtained by quantitative statistics based on their number of appearances in the literature. Details of the auction model and blockchain technology used in each paper are listed in the Table \ref{tab:application}.}
\label{distribution}
\end{figure}

To get an overview of how blockchain and auction models are integrated,
we summarized the auction models and blockchain technologies used in different studies, as shown in Table \ref{tab:application}.
In general, although different blockchain technologies have their trade-offs, researchers tend to have specific selection requirements and preferences when actually performing the model construction. We find that the largest share of studies (34.7\%) adopt permissioned blockchain technologies. It is widely believed that the access control mechanism in the permissioned blockchain can protect business secrets better. In addition, the high throughput of permissioned blockchains can accommodate large-scale transactions, making them more suitable for real industrial applications.
Slightly fewer studies (28.0\%) use permissionless blockchains. In these studies, researchers argue that the fully decentralized nature of permissionless blockchains can make the auction platform more trustworthy, and the built-in cryptocurrency can directly support transactions within the blockchain platform. We find that only three studies choose the hybrid blockchain. Although cross-chain solutions have been proposed for several years, they have rarely been studied in auction applications. Nevertheless, we believe this could be a promising direction for future research. Finally, in one-third of the studies, no specific blockchain technology was determined. Those authors leave the choice of implementing blockchain technologies to users.

Fig. \ref{blockcaindis} further illustrates the distribution of blockchain platforms used in different auction application fields. Our finding is that more than half of the studies use Ethereum as the underlying blockchain infrastructure. Apart from energy trading, Ethereum is also the most popular blockchain platform in all application fields. 
Some researchers argue that microtransactions in P2P energy trading require high system throughput, so it is more favorable to implement a permissioned blockchain (e.g., Hyperledger Fabric) or DAG blockchain (e.g., IOTA) platform.
In addition, we notice that a small number of authors do not choose established commercial blockchains; instead, they use simulation tools (e.g., Python or Matlab) to validate their models or frameworks. Other studies only present the conceptual proof of their blockchain-based auction models without on-chain implementations.

As shown in Fig. \ref{auctiondis}, the most commonly used auction models are double auction (36.4\%), reverse auction (11.7\%), Vickrey auction (11.7\%), and VCG auction (7.8\%).
We notice that double auctions are most frequently used in energy trading and stock exchange. This is mainly because the energy trading and stock exchange markets with multiple sellers and multiple buyers are well suited to integrate with double auctions. Among other application fields, most researchers prefer traditional single-sided auctions (e.g., reverse, Vickrey, and VCG auctions).
Another interesting finding is that reverse auctions are popular in service allocation and supply chain management.
This is mainly because the reverse auction can bring substantial cost savings to buyers in those two application fields. A reverse auction also helps streamline the auction process; auction time is saved because buyers do not need to send requests to different sellers one by one.

\input{table/Table_Application4}

\section{Auction-Based Solutions for Blockchain Enhancement}
\label{auction4blockchain}
\begin{figure}[ht]
  \centering
  \includegraphics[width=\linewidth]{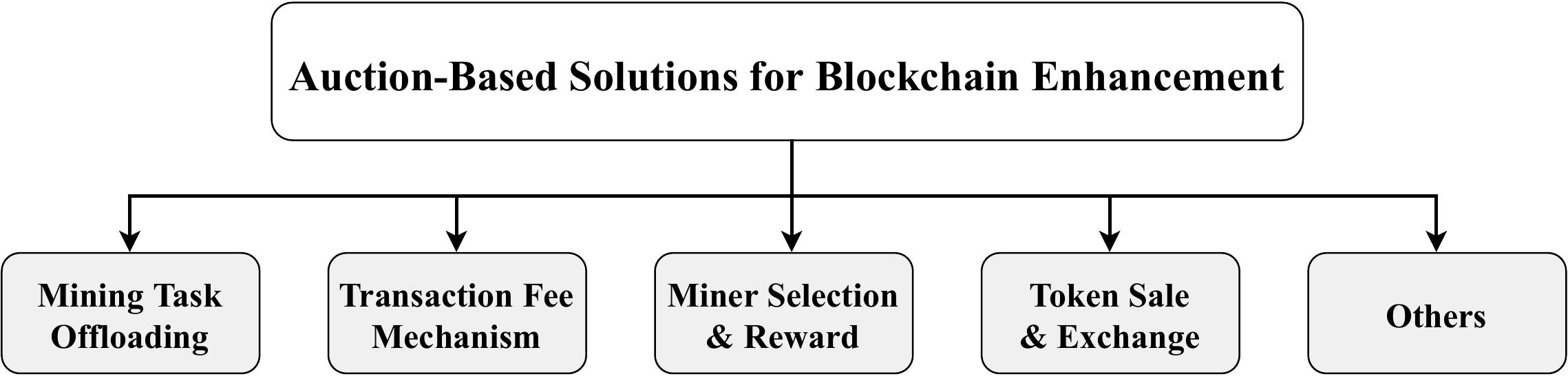}
  \caption{Taxonomy of auction-based solutions for blockchain enhancement.}
  \label{application4blockchain}
\end{figure}

Recent studies have shown that the auction model has great potential for optimizing blockchain technology \cite{xia_etra_2018, dimitri_transaction_2019, liu_auction_2019}.
However, a systematic review of existing enhancement solutions in different scenarios is still missing. 
In this section, we present a taxonomy of the current auction-based solutions for blockchain enhancement, as shown in Fig. \ref{application4blockchain}. Based on the purpose of using auction models in the blockchain, five application domains are identified: mining task offloading, transaction fee mechanism design, miner selection \& reward distribution, token sale \& exchange, and others. In addition, a summary of the auction models and main contributions in different studies is shown in Table \ref{tab:auction4blockchain}.

\subsection{Mining Task Offloading}
\begin{figure}[t]
  \centering
  \includegraphics[width=.95\linewidth]{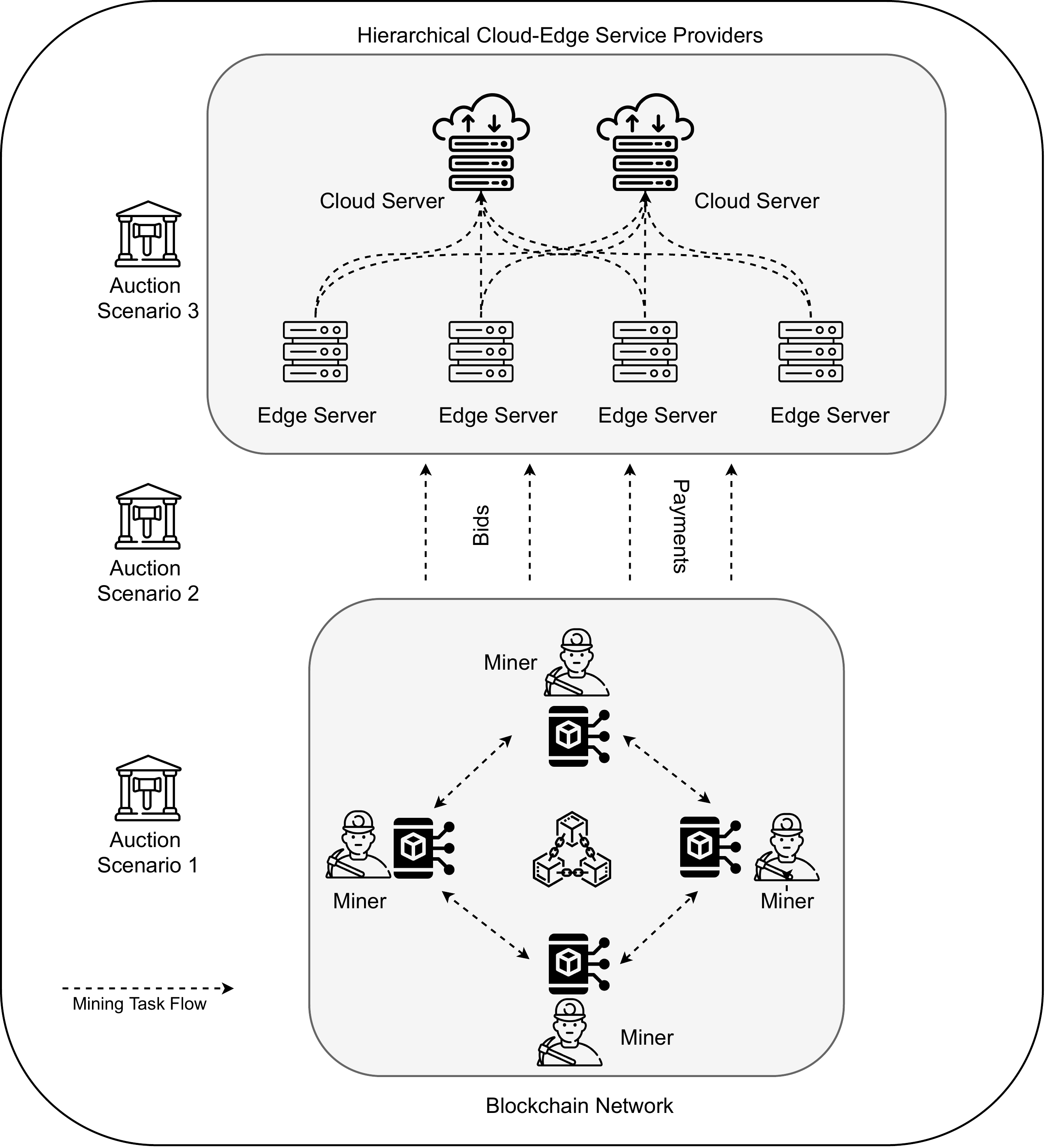}
  \caption{There are three auction scenarios for offloading mining tasks: 1) overloaded miners can offload mining tasks to other unused miners; 2) miners can offload mining tasks to edge providers; and 3) mining tasks can be further offloaded to cloud providers when edge resources are not enough.}
  \label{miningoffloading}
\end{figure}

The use of blockchain in IoT and mobile computing scenarios can increase security and trust assurance and prevent malicious attackers from entering the network \cite{lao2020survey}.
Solving the PoW puzzle needs continuous computation. 
Unfortunately, lightweight IoT and mobile devices have difficulty participating in the PoW consensus process due to a lack of computing capacity~\cite{guo_differential_2021}.
As a result, rational miners (i.e., consensus nodes) will naturally offload PoW computational tasks to cloud/edge/fog computing servers. An illustration of the auction-based mining task offloading is shown in Fig. \ref{miningoffloading}.
Generally, this offloading model has the following two assumptions: 1) the blockchain network is permissionless
and adopts the classical PoW consensus protocol; and 2) 
miners cannot use their own devices, such as lightweight or mobile devices, to complete all mining tasks~\cite{jiao_auction_2019}.
In \cite{jiao_social_2017}, the authors used an auction model to study resource management and pricing mechanisms for mobile blockchains. They demonstrated that their VCG-based auction model could maximize social welfare and satisfy several important auction properties. Similarly, Xia~\textit{et~al.}~\cite{xia_etra_2018} proposed a VCG-based auction mechanism for mobile blockchain resource allocation. Their auction includes three stages: 1) matching potential winners; 2) matching cloudlets for access points; and 3) allocating the resource. 
Taking into account the diverse resource demands, bids, and usage patterns of mobile users,
the authors in~\cite{gao_dynamic_2019} used an optimized Vickrey auction to acquire dynamic resource allocation strategies in mobile blockchain networks. 

Our investigation also shows that many researchers focused on allocating mining resources between multiple blockchain users/miners and service providers through double auction models. 
Based on the combinatorial double auction, a two-level allocation mechanism is designed for cloud and edge computing resources in~\cite{li_double_2021}. Specifically, mobile users compete to allocate edge-level resources first, and then cloud-level resources can be used as supplements. 
This model satisfies the requirements of multiple users and optimizes resource utilization compared to single-level computing offloading.
Similarly, Xu~\textit{et~al.}~\cite{xu_hierarchical_2020} and~Li~\textit{et~al.}~\cite{li_resource_2019} argued that 
mobile edge computing servers with limited resources could request resources from cloud computing servers.
They considered two auction scenarios, namely single-seller multiple buyers and multi-seller multi-buyer. Correspondingly, a hierarchical auction model (including a single-sided combinatorial auction and a double combinatorial auction) was presented.
Guo~\textit{et~al.}~\cite{guo_blockchain_2020} proposed that
the non-mining devices and idle resources on the edge cloud can be selected to create a so-called collaborative mining network (CMN) to perform mining tasks.
Thus, a double auction can be used to manage resource allocation between mining and sharing devices in a CMN. 
In addition, the interactions between edge cloud operators and CMNs are modeled with a Stackelberg game, and both uniform and differentiated pricing strategies are analyzed.
Zhang~\textit{et~al.}~\cite{zhang2022optimal} investigated both normal task and mining task offloading problems for blockchain-enabled beyond 5G networks. In particular, they used a double auction to study normal task offloading among mobile devices and further proposed a mining task offloading scheme based on the Stackelberg game.

The following studies also investigate the different demands of miners, privacy considerations, bid rigging issues, and algorithms for winner determination problems.
Jiao~\textit{et~al.}~\cite{jiao_auction_2019} and Zhang~\textit{et~al.}~\cite{zhang2022truthful} both studied two types of miners with different demands on computing resources: 1) miners with constant demand; and 2) miners with multiple demands. Correspondingly, two auction algorithms are proposed and tested. 
Ahmadi~\textit{et~al.}~\cite{ahmadi2022locality} considered the locality and priority of auction-based resource allocation in blockchain networks. They designed priority and locality algorithms for mining task offloading, with both fixed and multi-demand miners considered. Their approach provides priority and locality features to the traditional auction method. The priority algorithm can serve VIP customers, while the locality algorithm can improve performance by around 78\%.
In \cite{guo_differential_2021}, the authors argued that the competition between IoT devices and edge servers should be truthful and privacy-preserving in terms of personal data.
Therefore, they designed a truthful multi-item double auction.
The proposed auction model is also extended with the differential privacy technology to protect sensitive bidding information. 
Qiu and Li~\cite{qiu2022auction} pointed out that it is difficult to identify and avoid bid rigging behaviors by users in existing mining task offloading auctions, which can lead to revenue loss for edge providers. Therefore, they introduced an auction method to address this issue.
Liu~\textit{et~al.}~\cite{liu_efficient_2019} argued that the solution of combinatorial double auctions can be modeled as NP-hard winner determination problems. Thus two different greedy algorithms are designed and implemented to solve these problems. 

In addition to the above research, mining offloading is also discussed in blockchain-based vehicle-to-everything (V2X) networks \cite{jameel_efficient_2020}. Besides, machine learning-based optimal auction models have been discussed. For example, the authors in~\cite{luong_optimal_2018,luong_machine-learning-based_2020} proposed a deep learning-based optimal auction model for allocating edge resources in mobile blockchain networks.
Their model contains a multi-layer neural network, and the training data is composed of bidder valuation profiles of the miners.
Qiu~\textit{et~al.}~\cite{qiu2019online} argued that traditional mining task scheduling methods (e.g., based on auctions or game theory) cannot adapt to changing circumstances or achieve long-term performance. Therefore, a deep reinforcement learning-based mining offloading method was proposed and tested.

\subsection{Transaction Fee Mechanism Design}
\begin{figure}[t]
  \centering
  \includegraphics[width=.95\linewidth]{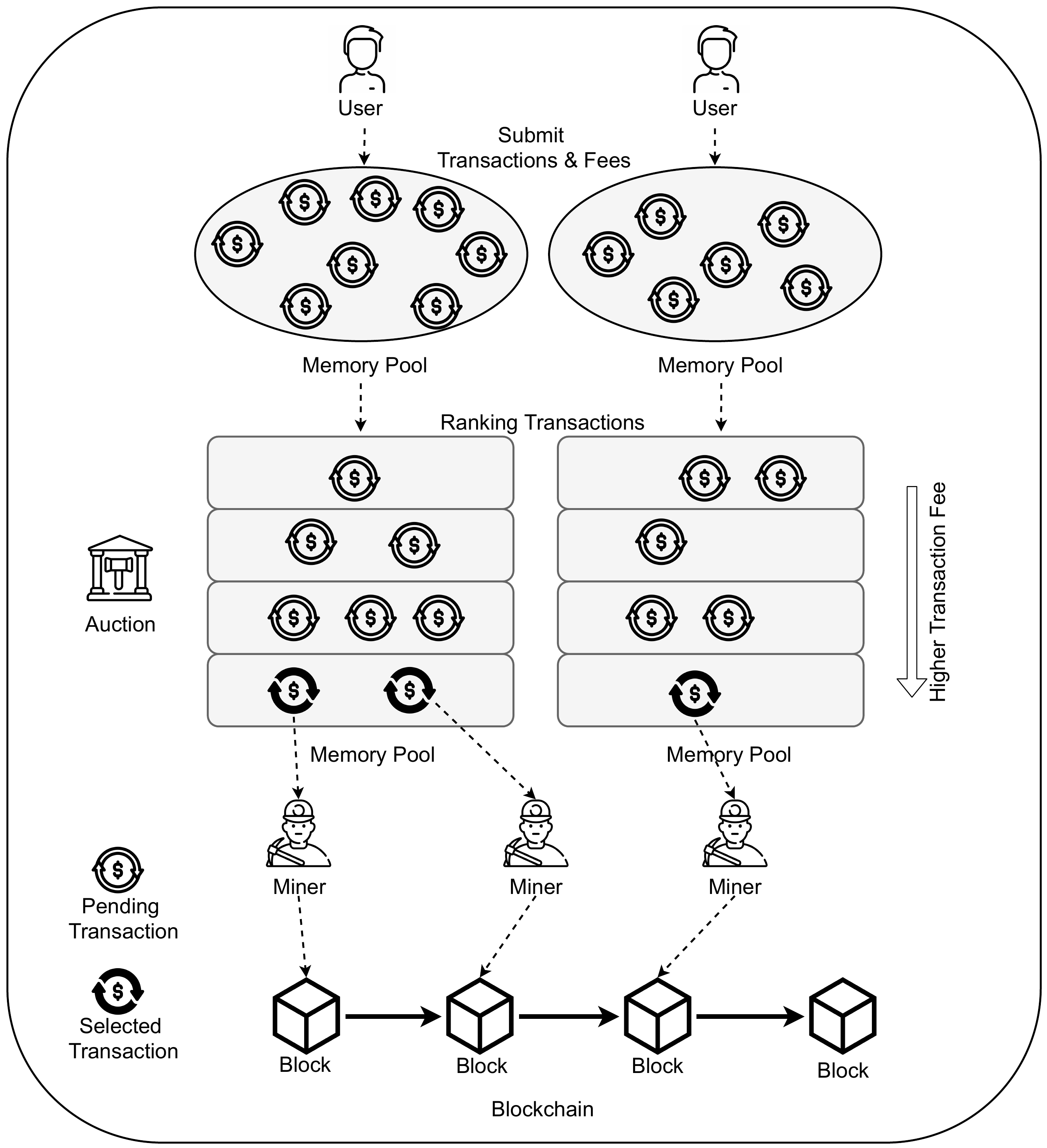}
  \caption{The basic process of a blockchain transaction confirmation auction. Users submit transactions and fees (bids) to the memory pool. Then, miners select and validate the transactions with higher transaction fees for priority processing.}
  \label{transactionfeedesign}
\end{figure}

Most current permissionless blockchains utilize the same transaction fee mechanism for transaction prioritization. Each transaction is charged a fee from the user, and miners choose the transactions with the highest fees to include in the block (as illustrated in Fig. \ref{transactionfeedesign}). 
Such a mechanism is critical in cryptocurrencies since it subsidizes miners to keep building the blockchain and ensures efficient use of network resources.
As the cryptocurrency becomes more popular and the baseline subsidy (block reward) to miners gradually decreases, the revenue from transaction fees will play a more prominent role in ensuring network stability~\cite{basu_towards_2019}. In this context, auction models are widely used to model and optimize the blockchain transaction fee mechanism. Huberman~\textit{et~al.}~\cite{huberman_monopoly_2017} found that the Bitcoin protocol, despite the absence of an auctioneer, implicitly includes a priority auction. Besides, users' bids have the characteristic of a VCG mechanism, i.e., each user offers a bid equal to his/her externalities (the transaction delays he/she caused to others). They simulated this auction activity and demonstrated that the Bitcoin payment system could serve as a prototype for protecting customers from monopolies.
Dimitri~\cite{dimitri_transaction_2019} 
modeled the blockchain transaction fee as the Nash equilibrium result of a complete information auction game. Successful miners function as auctioneers in the game, selling block space to users who bid for shares to confirm their transactions.
Their analysis shows that the optimal block size limit for successful miners is determined by the transaction confirmation fee that users are willing to pay. 
Similar to this study, Kruminis~\textit{et~al.}~\cite{kruminis2022game} modeled the inclusion of transactions into blocks as an auction game and studied the impact of potential block size changes on the system. In particular, they consider a dynamic environment in which the fees offered are based on bids from other users.
Daian~\textit{et~al.}~\cite{daian_flash_2020} focused on the decentralized exchange (DEX) field and observed that bots in DEXes participate in so-called priority gas auctions, where bots compete to raise their transaction fees in order to confirm transactions faster. 
Their analysis of the priority gas auction demonstrates that protocol details (e.g., miner selection criteria and P2P network composition) could directly affect smart contracts' application-layer security and fairness properties.
FairTraDEX \cite{mcmenamin2022fairtradex} is a DEX protocol that provides formal game-theoretic guarantees against extractable values. The protocol extends traditional frequent batch auctions and proposes a width-sensitive frequent batch auction for supporting exchanges between clients and market makers.

It is commonly believed that the transaction confirmation process in Bitcoin and Ethereum is equivalent to a GFP auction~\cite{roughgarden_transaction_2020}. This auction is first proposed in online advertising auctions where advertisers bid for more prominent advertising positions. 
However, the market practice of Bitcoin has demonstrated that the GFP mechanism may have some defects \cite{li_novel_2019}. For example, the GFP mechanism causes Bitcoin users to pay unnecessary transaction fees.
When the Bitcoin memory pool is stuffed with a huge number of transactions, it can lead to severe congestion where users have to wait for days to confirm transactions and pay additional transaction fees. In addition, the empirical analysis also shows that the transaction fees charged vary significantly from block to block and from day to day, leading to large fluctuations in miners' income. 
All of these facts prove that the GFP mechanism is not a perfect auction mechanism for the Bitcoin transaction fee market.
Therefore, researchers have proposed different alternative auction models to optimize the current blockchain transaction fee market. Although related studies have focused only on Bitcoin and Ethereum, the findings can be generalized to any permissionless blockchain that uses PoW mining mechanism to validate new transactions.

In~\cite{basu_towards_2019}, the authors argued that it is the lack of a dominant strategy equilibrium in the existing transaction fee market that caused instability and low efficiency. Therefore, a GSP auction-based transaction fee mechanism can be used to replace the GFP one. Implementing such a mechanism remains challenging as miners can include transactions using any criteria and manipulate the auction results after seeing the proposed transaction fees. Nevertheless, they demonstrated that the suggested approach is immune to manipulation as the user base grows.
Similarly, two cases of GSP auctions in Bitcoin, namely complete information under synchronous submission and incomplete information under asynchronous submission, have been analyzed in~\cite{li_novel_2019,li_novel_2020}. Their results suggest that this new GSP mechanism can help users save transaction fees compared to the currently used GFP approach.
Furthermore, Yan~\textit{et~al.}~\cite{yan_dynamic_2020} took time series into consideration and studied the time-dependent dynamic game model of the GSP transaction fee mechanism. 
Their analysis result shows that there exists a perfect Bayesian game equilibrium so that the whole system can remain stable.

Some other auction models have been identified as possible alternative mechanisms for blockchain transaction inclusion.
In a monopolistic auction, given a series of bids, miners selectively include transactions in the block. All selected transactions are subject to the same charge, which is the lowest bid in the current block. 
The authors in~\cite{lavi_redesigning_2019} observed that the monopolistic auction is immune to malicious auctioneers in the Bitcoin network. Besides, such an auction is easy to implement for transaction issuers, and miners' revenue does not decrease as the maximum block size increases. Yao~\cite{yao_incentive_2018} further proved that the monopolistic auction mechanism is almost truthful for any i.i.d. distribution when the number of users becomes large, making it a good candidate for the blockchain transaction fee market.
Basu~\textit{et~al.}~\cite{basu2019stablefees} proposed StableFees, a transaction fee mechanism based on the uniform price auction. They showed that as the number of users and miners increases, their protocol is immune to manipulation as the gains from manipulation are negligible in practice.
Gibaja~\textit{et~al.}~\cite{gibaja2022auction} are more concerned with the execution of smart contracts. They modeled the competition between miners for a single contract as a time auction where the winner is the one who sets the minimum time to pick up the contract. They concluded that the cost is negatively related to execution time in the equilibrium state.

\subsection{Miner Selection \& Reward Distribution}
In a PoW-based permissionless blockchain, anyone with the necessary computing capacity and network connection can become a miner. Miners compete with each other to provide transaction processing services to the blockchain and receive corresponding rewards~\cite{huberman_monopoly_2017}. 
For some blockchain platforms with new consensus algorithms, however, miners need to be selected and authorized to verify new blocks.
PoS is one of the popular solutions to replace energy-intensive PoW. 
It specifies that users can mine or validate new block transactions based on the stakes they hold, and therefore the network may suffer from centralization and unfairness issues.
Endurthi and Khare~\cite{endurthi2022two} proposed a two-layer consensus mechanism that combines PoW and PoS. In the PoS layer, each node has to bid based on the lowest unique integer bid strategy. In this way, 10\% of the nodes are selected as the next layer of PoW miners, thus reducing the energy consumption by 90\%.
Saad~\textit{et~al.}~\cite{saad_e-pos_2021} presented an enhanced version of PoS called e-PoS. In particular, a blind block auction is integrated into e-PoS to offload mining opportunities to more users, thus improving fairness and resisting centralization.
Another interesting consensus mechanism called ABC is proposed in~\cite{ai_abc_2020}, in which a continuous double auction is leveraged to determine and select miners to write new blocks. Through extensive experiments, the authors concluded that ABC has better performance than PBFT in blockchain networks with a large number of nodes.
Devi~\textit{et~al.}~\cite{devi_using_2020} developed a novel mechanism for miner selection in order to encourage miners to participate in block validation and optimize the blockchain performance. Especially, a multi-attribute two-stage auction model is designed and implemented to select miners; only nodes with high credibility and data quality can be selected as miners for block verification. 
Amin~\textit{et~al.}~\cite{amin_secured_2020} argued that users can leverage a discretionary mining strategy in an IOTA blockchain network. 
This means that a user with low computational power can outsource his/her verification tasks to a mining pool. The nomination of a specific number of miners can be conducted using an FPSB auction.

Other studies focus on the design of distribution mechanisms for miners' rewards.
Typically, PoW-based blockchain systems distribute rewards based on the blocks discovered by miners~\cite{liu_auction_2019}.
Nadendla and Varshney~\cite{nadendla_difficulty_2020} suggested that blockchain mining can as characterized as an all-pay auction, where the computational efforts of miners are interpreted as bids. In this way, the reward distribution function is defined as the chance of solving the cryptographic puzzle in a single try with unit computational power. Based on such assumptions, they constructed a mining auction mechanism that generates a logarithmic equilibrium among miners. 
The analysis shows that no allocation function in equilibrium prevents miners from bidding higher costs. 
As a result, it is crucial to penalize miners who choose a larger computational cost to maintain a trustworthy system.
In real-life blockchains, miners can also join a number of mining pools to share the rewards they earn in time. In~\cite{liu_auction_2019}, the authors investigated the reward distribution mechanism of mining pools.
They compared several block reward allocation strategies in a long-term scenario and showed that no existing technique could guarantee continuous mining for miners in a pool. To address this issue, they proposed an auction-based approach that increases miners' enthusiasm and the mining pool's stability.
Xue~\textit{et~al.}~\cite{xue2021incentive} provided a public cost model and a private cost model for Bitcoin mining pool systems with rational miners. They used a Stackelberg game to represent the mining process in the public cost model. For the private cost model, they developed a budget-feasible reverse auction to handle the reward optimization problem.

\subsection{Token Sale \& Exchange}
\begin{figure}[!t]
  \centering
  \includegraphics[width=.95\linewidth]{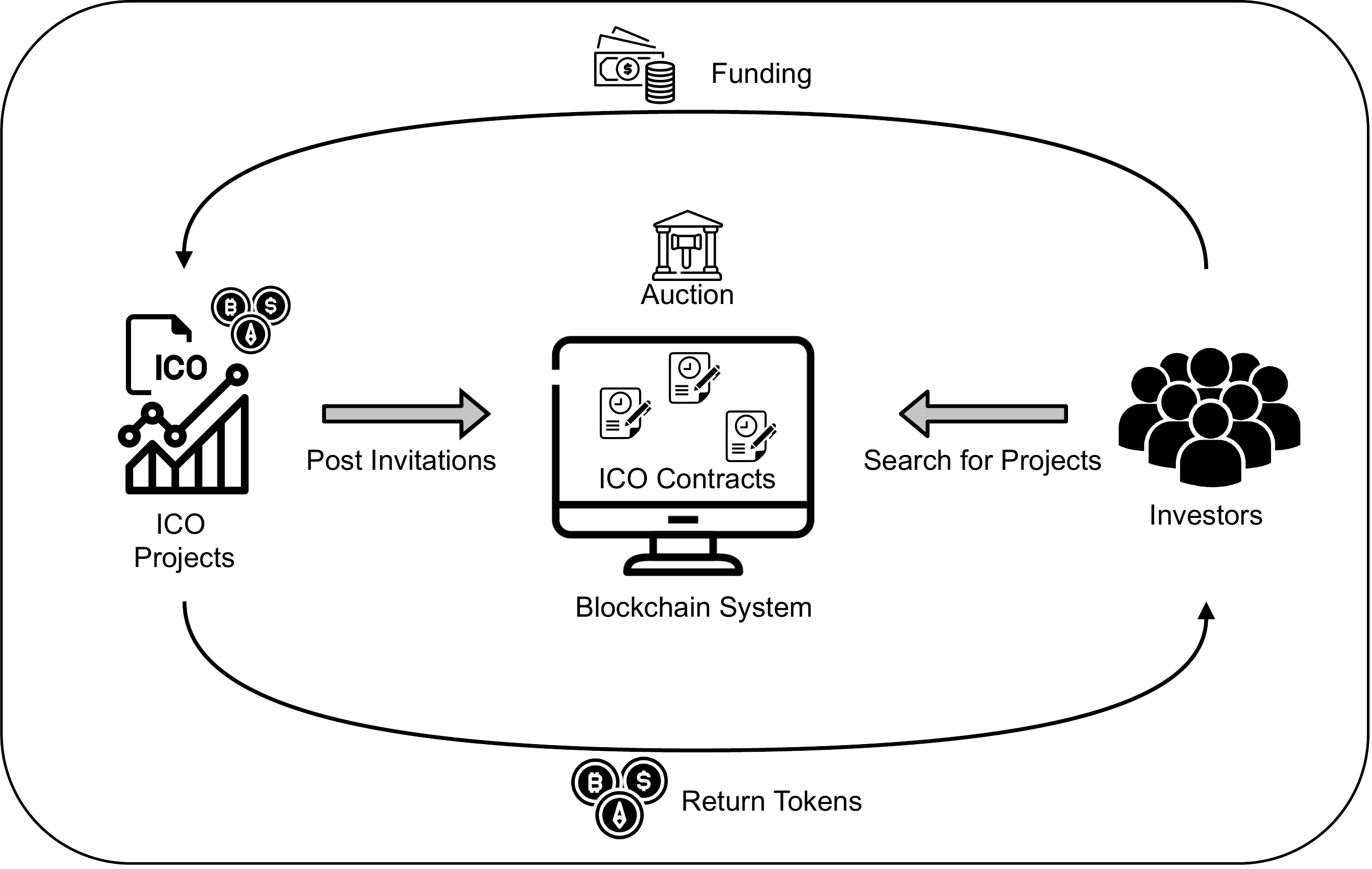}
  \caption{The basic process of an ICO. The developer of an ICO project publishes cryptocurrency information and invites bids on the blockchain and smart contracts to raise funds. Investors pay their fiat money or other cryptocurrencies to get pre-mined tokens from the ICO project.}
  \label{icofigure}
\end{figure}

Auction models are playing important roles in token sale and exchange programs. An initial coin offering (ICO) (also known as a token sale) is the process of raising funds from the public for the development of a new cryptocurrency project. It is a particular application of IPO in the cryptocurrency industry. With the integration of blockchain and smart contract technologies, it is possible to raise external funds for cryptocurrencies without any intermediaries \cite{momtaz2019token}. The basic process of an ICO is shown in Fig. \ref{icofigure}. When a new cryptocurrency is created, pre-mined tokens are sold to the public through an ICO in exchange for other cryptocurrencies or fiat money \cite{somin2018network}. The issuing agency can arbitrarily set a fixed price or determine the sale price through an auction \cite{kranz2019blockchain}. 
According to a survey, the most commonly used auctions include the Dutch auction and its variants, such as the Vickrey-Dutch auction and the reverse Dutch auction \cite{fridgen2018don}. For example, an optimal ICO mechanism based on the multi-unit Vickrey-Dutch auction is proposed in \cite{cerezo2017optimal} to guarantee truthful bidding. Some third-party organizations, e.g., CoinList and Gnosis, are also actively working on providing auction services for token sales \cite{tokensale}. However, a survey revealed that only a small percentage of ICO projects use auction mechanisms \cite{howell2020initial}. While researchers have noted that various types of auctions can be used in theory, there are few real examples of attempts in ICOs except for Dutch auctions so far \cite{teutsch2019interactive}.
A non-fungible token is a particular type of cryptocurrency associated with a specific digital or physical asset. In recent years, NFT auctions have become popular among the cryptocurrency community as people have become aware of their business potential. In \cite{fazli2021under}, the authors analyzed 65,000 NFT auction records from the Foundation platform to explore potential auction activity (e.g., possible speculation in the sale of used art). They discovered that only 3.97\% of NFTs were resold at the second auction despite 36.10\% of NFTs being sold at the first auction. This indicates that speculation is not frequently seen in NFT auctions.

An atomic swap (also called atomic cross-chain trading) is an automated, self-enforcing cryptocurrency exchange contract that allows P2P transactions of cryptocurrencies without the need for TTPs \cite{Atomic_2021}. In this respect, 
Zhang~\cite{zhang_optimum_2020} demonstrated an exchange mechanism that uses uniform price auctions for atomic swaps. He proved that a uniform price auction could save data cost, but the optimal transaction collection of the atomic swap auction is NP-hard. Liu~\textit{et~al.}~\cite{liu2021aucswap} presented an efficient protocol for cross-chain asset transfers using Vickrey auction and atomic transfer techniques.
Furthermore, Black~\textit{et~al.}~\cite{black_atomic_2019} proposed the notion of atomic loans, which make use of atomic swap technology to enable market participants to build cryptocurrency debt instruments. They also introduced a competitive bidding process for the fair distribution of collateral when defaults occur. 
Zhang~\textit{et~al.}~\cite{zhang2021enabling} proposed a reverse Vickrey auction-based routing scheme to optimize the selection of connectors in cross-blockchain exchange protocols. Their solution can effectively enhance the atomic swap of different types of cryptocurrencies in healthcare payments.

There are some commercial efforts dedicated to building an auction-based atomic swap trading platform. For example, Freimarkets~\cite{friedenbach_freimarkets_nodate_2013} is a protocol that adds primitives required for implementing non-currency financial transactions on Bitcoin. Specifically, three auction models (i.e., English auction, Dutch auction, and double auction) are demonstrated on how to conduct atomic cross-chain trade. By contrast,
Gnosis~\cite{walther_multi-token_2018} is a trading platform based on the Ethereum Plasma framework in which a uniform price multi-batch auction model is implemented among different ERC-20 (Ethereum Request for Comments 20) tokens.
Each batch, in particular, allows requests to purchase any ERC-20 tokens in exchange for other ERC-20 tokens. All orders are gathered at predetermined time intervals and then settled using a uniform clearing price computed across all token pairs.

\subsection{Others}
There are some individual studies that could not be assigned to the above classification, and we thereby listed them in this section. 
An important research direction is the optimization of blockchain networks through auction-based incentives.
For example, most existing studies assume that the network state of the blockchain is perfect; however, this is not the real case. When miners have limited network bandwidth, their utility, the block broadcasting, and the throughput of the entire blockchain are heavily affected \cite{jiao_auction_2019}. In this context, 
Wang~\textit{et~al.} \cite{wang_incentivizing_2021} argued that peers in a blockchain network should be paid for forwarding a large number of transactions, just as miners put computational resources into block mining and get rewards. Therefore, a new incentive can be designed for blockchain network participants with limited bandwidth. In their study, an FPSB auction is used to build this relay payment scheme.
ABIDE \cite{mondal2007abide} is a bid-based incentive model that encourages non-cooperative peers to provide services in mobile P2P networks. In ABIDE, each peer node can provide data to other peer nodes, and there is a price attached to such a service. ABIDE is not explicitly designed for blockchain, but such a model is applicable to the underlying P2P network of the mobile blockchain.
Another direction is blockchain domain name trading.
The Ethereum Name Service (ENS) is a distributed, open and scalable naming system based on the Ethereum blockchain. It maps human-readable names to machine-readable identifiers like wallet addresses, content hashes, and metadata via a Vickrey auction-based name registration approach defined in Ethereum Improvement Proposal (EIP)-162~\cite{johnson_ens_nodate}.

\subsection{Key Observations} The key observations we obtained in this section are summarized as follows:

\begin{itemize}
    \item Auctions have been proven to be promising solutions for blockchain enhancement; the efficient distribution and fair trade features of auctions can facilitate different blockchain workflows. There are four main application areas and directions identified in the literature: mining task offloading, transaction fee mechanism design, miner selection \& reward distribution, and token sale \& exchange. These four areas are mainly focused on optimizing the blockchain's incentive and consensus layers because those two layers are more likely to see interactions between different stakeholders (e.g., miners, users, resource providers, and token holders).
    \item Integrating auction models into the blockchain incentive layer is very beneficial. A blockchain is, by its nature, a new technology for maintaining a public ledger among a large group of participants. Such a property determines that the success of blockchain will depend heavily on how humans interact with it. Therefore, it is essential to design optimal incentives to encourage decentralized peers to actively participate in the security verification of the blockchain. 
    Using auctions to design and optimize the blockchain transaction fee mechanism is a promising direction. Traditional permissionless blockchains (e.g., Bitcoin and Ethereum) use a GFP auction model to build their transaction fee market. Despite the potential for overbidding, the success of the first-price auction-based transaction fee mechanism has been witnessed in the Bitcoin and Ethereum booms. Such a mechanism allows for the rapid disposal of large volumes of transactions, reducing transaction congestion and improving blockchain performance. GSP, monopolistic, and uniform price auctions are often recommended in the literature as alternative mechanisms with more benefits. However, the usability and reliability of these new mechanisms need to be further tested in real-world blockchains. Similarly, by designing a reasonable incentive through auctions, miners can get a fair share of the mining rewards in a mining pool, thus ensuring their continuous mining in the blockchain network. Another direction that integrates auctions into the blockchain incentive layer is the issuance, sale, and exchange of tokens. On the one hand, new tokens can be sold to the public by auctions through ICOs. On the other hand, the quick exchange among different tokens can be performed through atomic swap technology using various auction models. Auction-based token trading eliminates long negotiation periods and guarantees fairness: buyers/sellers know they are competing fairly and on the same terms as all other buyers/sellers.
    \item Integrating auction models into the blockchain consensus layer also shows promise, mainly demonstrated in optimizing traditional PoW consensus algorithms (e.g., mining task offloading) and enhancing other alternative consensus algorithms (e.g., miner selection). The huge energy consumption caused by the PoW consensus has always been a big challenge for blockchain. Some mobile blockchain or IoT-based blockchain miners are not capable of using their own resources to complete the mining task. By using auction models, PoW mining tasks can be assigned to different devices to reduce the computational burden. In this case, edge and cloud servers become ideal targets for mining task offloading. Offloading mining resources can be performed using a variety of mechanisms, including various forms of auctions or direct negotiation. An auction produces a level playing field for miners and service providers through competitive bidding. In addition, it eliminates lengthy negotiation periods and streamlines the offloading process. Besides, auctions can be used to optimize the workflow of some specific blockchain consensus algorithms (e.g., selecting the right miners for bookkeeping in PoS). The miner selection process is dominated by high-level stakeholders in PoS, but this is considered very dangerous in some cases. For instance, if malicious miners gain enough stakes, they would disrupt fair block validation and cause the system to collapse. An efficient auction model can improve blockchain performance and security by designing a safer and fairer mechanism to select the most suitable miners.
    \item By contrast, existing studies using the auction model to enhance blockchain technology are less involved in other blockchain layers (i.e., data layer, network layer, contract layer, and application layer). This is mainly determined by technical characteristics; these layers usually set up data structures, network protocols, and contract specifications without the active participation of blockchain stakeholders. As a result, the auction model, as an economic incentive that drives human behavior, can only play a minimal role in these layers.
\end{itemize}

\include{table/Table_Auction4Blockchain2}

\section{Challenges and Future Directions}
\label{challenge}
\begin{figure*}[htb]
  \centering
  \includegraphics[width=\linewidth]{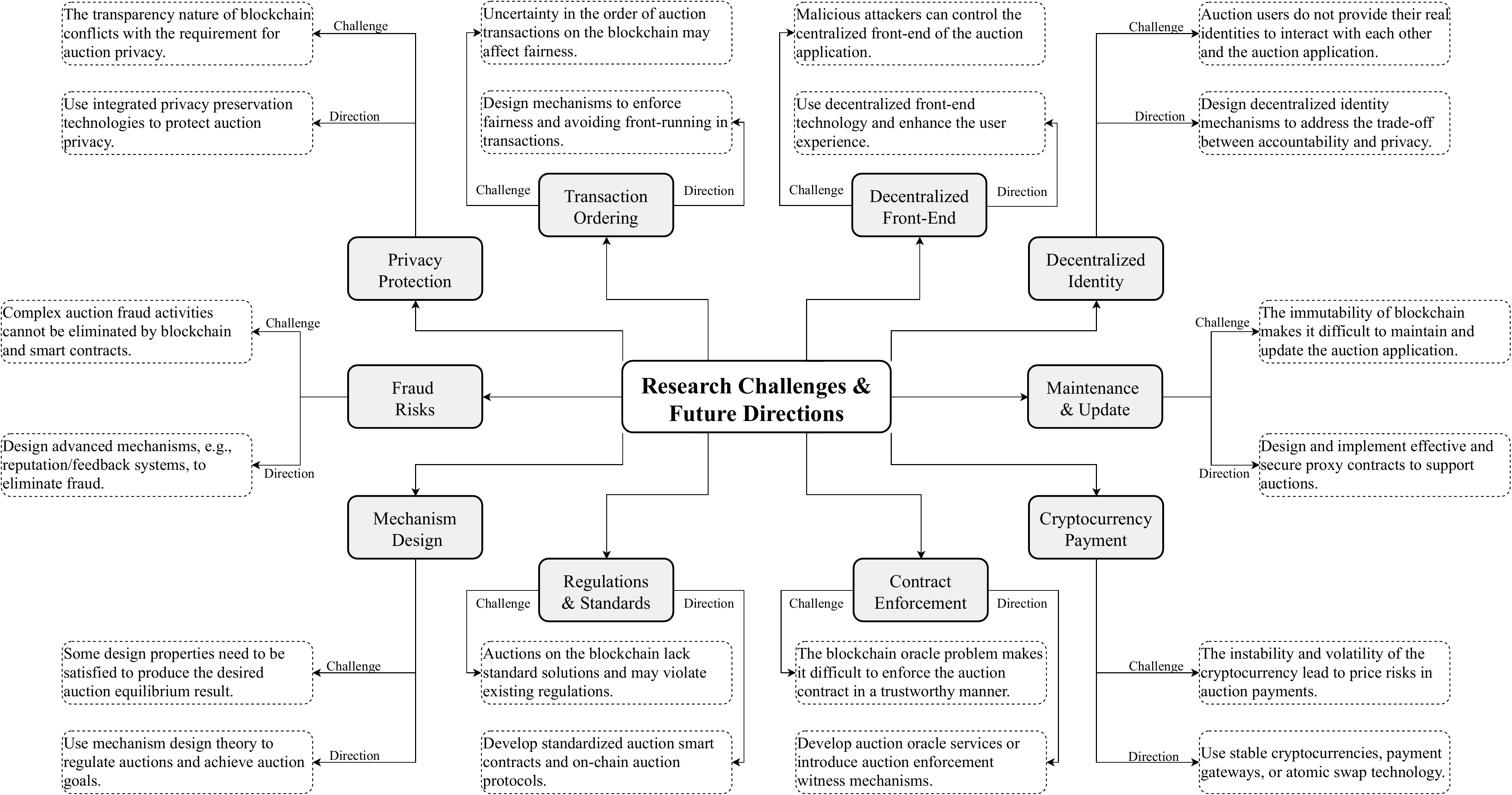}
  \caption{Taxonomy of challenges and solutions for integrated blockchain-auction models.}
  \label{fig:challenges}
\end{figure*}

Despite the great potential of integrating blockchain with auction models, there are several research challenges that need to be addressed. In this section, we highlight and summarize ten open challenges identified in the literature, as shown in Fig. \ref{fig:challenges}. Specifically, the first eight challenges, i.e.,
auction privacy protection, transaction ordering \& fairness, decentralization of auction front-end, decentralized identity management, auction maintenance \& update, auction payment with cryptocurrency, auction contract enforcement, and auction regulations \& standards
correspond to the topic of blockchain-based auction applications (Section \ref{blockchain4auction}); while the last two challenges, i.e., auction mechanism design and auction fraud risks, are both involved in blockchain-based auctions (Section \ref{blockchain4auction}) and auction-based solutions for blockchain enhancement (Section \ref{auction4blockchain}).

\subsection{Auction Privacy Protection}
\label{security}

All data stored on the blockchain must be public to all blockchain nodes in order to ensure traceability, verifiability, and immutability. This conflicts with the privacy requirements of most auction applications, especially for those with important trade secrets. Normal users will be discouraged from using the blockchain for auctions if privacy can not be fully guaranteed.

\begin{figure}[!b]
  \centering
  \includegraphics[width=\linewidth]{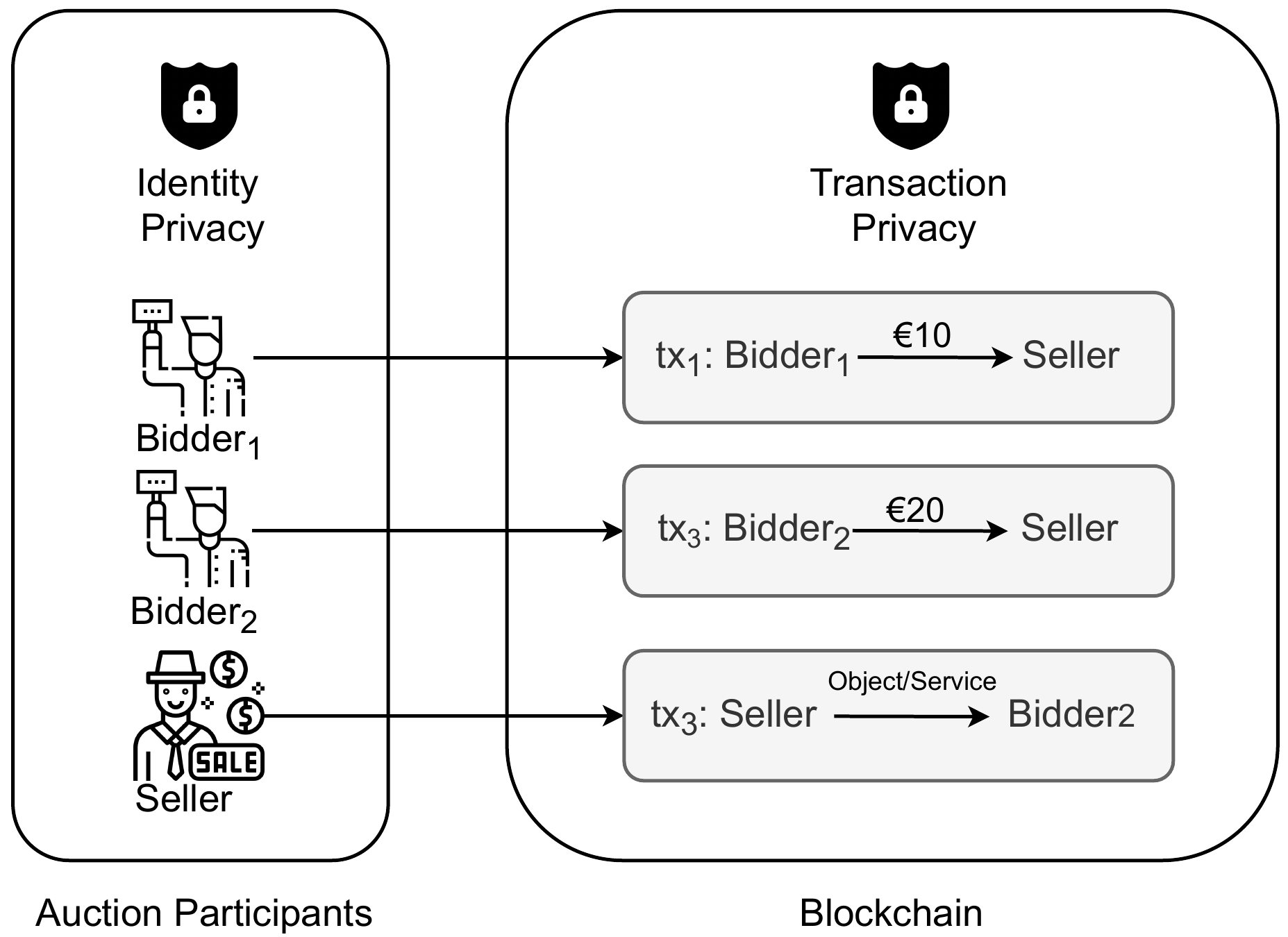}
  \caption{There are two privacy concerns identified for blockchain-based auctions: identity privacy and transaction privacy. The former concerns the privacy of auction participants, while the latter concerns the privacy of various auction transactions (e.g., bids, payments, and contract details).}
  \label{privacyissue}
\end{figure}

As illustrated
in Fig. \ref{privacyissue}, there are generally two types of privacy concerns for blockchain-based auctions \cite{NBERw27634}. 
The first one is identity privacy, which considers participants' privacy and prevents transactions from being associated with specific auction users and their blockchain addresses. The second one is transaction privacy, which covers the privacy of auction information about bids, auction contracts, payments, and other transaction details.
We find that most researchers target both types of privacy concerns in their models through a combination of various techniques. 
Based on the relevant literature, the existing privacy protection solutions and their challenges are summarized in the following text. A more detailed comparison of the techniques used in different studies can be found in Table \ref{tab:my-table2}.

\subsubsection{Cryptographic Primitives}
Cryptographic techniques can effectively protect privacy in
blockchain-based auction models. The most common cryptographic primitives used in the literature can be summarized as follows:

\begin{itemize}
    \item \textit{Multi-Party Computation (MPC)}: Multiple parties securely compute an objective function without TTPs, while each party does not have access to any input information from other parties except for the computation result.
    \item \textit{Zero-Knowledge Proof (ZKP)}: One person (the prover) demonstrates to another person (the verifier) that he/she knows a value without providing any information other than the fact that he/she knows the value.
    \item \textit{Commitment Scheme (CS)}: One person can commit to a chosen value (or statement) while hiding it from others and being able to reveal the promised value later.
    \item \textit{Asymmetric Encryption (AE)}: A key pair is required for encryption and decryption. So it is also known as public key encryption. Ciphertexts encrypted with a public key can only be decrypted with the associated private key and vice versa. 
    \item \textit{Homomorphic Encryption (HE)}: An encryption method that enables users to retrieve, compare and calculate encrypted data without decrypting it in advance.  
    \item \textit{Digital Signature (DS)}: Two operations are designed to verify the authenticity of digital messages or documents: signature and verification. One can use the private key to encrypt (generate a signature), and others can use the public key to decrypt (verify the signature).
\end{itemize}

Since these cryptographic techniques differ in terms of effects and application scenarios, some studies choose to integrate multiple algorithms on the blockchain to build a secure and privacy-preserving auction system. For example, ZKP, MPC, AE, and CS and their variant algorithms have been widely combined in recently proposed frameworks in \cite{galal_verifiable_2018,blass_borealis_2020,damle_practical_2019,sanchez_raziel_2017}. Such an integrated model can reduce the potential risk of using one single encryption algorithm, thus presenting an overall good privacy-preserving effect.
On the other hand, while cryptographic primitives can protect auction privacy, they suffer from high computational complexity and high transaction costs when implemented on the blockchain.
It is reported that a non-interactive ZKP
verification roughly takes more than 3 million gas on the Ethereum blockchain \cite{galal_succinctly_2018}. The huge transaction fees make it impractical for auction users to use these algorithms and join the auction. To effectively protect privacy, the performance of the cryptographic algorithms used in the blockchain needs to be significantly improved \cite{sarfaraz_tree_2021}. Recent studies have focused on designing lightweight cryptographic protocols that are weaker than traditional ones (e.g., MPC/ZKP) \cite{blass_strain_2018,damle_practical_2019,liu2020blockchain,ma_fully_2019}. These protocols can perform specific auction tasks and achieve optimized on-chain performance.

\subsubsection{Mixing/Tumbler Service}

Blockchain addresses do not guarantee the anonymity of auction participants. The pseudonymous addresses used for transactions can be publicly verified, 
So it is possible for anyone to analyze an auction user's address and link transaction records to his/her real identity \cite{zhang2019security}. In this context, tumbler or mixing, as a service that prevents users' addresses from being linked to their real identities, has been leveraged to protect bidders' privacy. AStERISK \cite{sonnino_asterisk_2019} is a blockchain-based auction framework that uses the mixing function of the Coconut contract \cite{sonnino2018coconut} to break the link between bidders and their bids and tokens, thus protecting the bidders' privacy. It should be noted that centralized mixing services may pose a significant risk to users, as all operations are handled centrally. 
Attackers may also utilize big data analytics and machine learning techniques to compromise mixing services and auction privacy.

\subsubsection{Differential Privacy}
Although the blockchain addresses of auction participants can be hidden using cryptography and mixing services, hackers can infer the true identity of users through side-channel information (i.e., side-channel attacks). This is why differential privacy, as a new privacy-preserving technology, has recently been investigated to protect bidder's privacy by adding noise to the auction transactions. DEAL \cite{hassan_deal_2020} is a decentralized auction model for microgrid energy trading. It can protect the privacy of auction participants through the combination of Laplacian and exponential noise on a permissioned blockchain. 
However, the introduction of noise can reduce the data utility, which becomes more severe for small data sets.
Recent studies focused on how to add noise to a data set to maintain confidentiality while maximizing the data utility \cite{hassan2020differential}.

\subsubsection{Trusted Execution Environment (TEE)}

TEEs can be used to create secure auction computation environments and protect the sensitive data involved in blockchain-based auctions.
Intel SGX, as one of the most popular hardware-based commercial TEE solutions, has been widely used in building blockchain-based auction systems \cite{yuan_shadoweth_2018, wang_secure_2020,galal_trustee_2019}. 
To effectively safeguard auction smart contracts with TEEs, however, challenges such as rollback attacks, state continuity, and TEE protocol integration must be properly addressed \cite{strackx2014ice}.
In \cite{brandenburger_blockchain_2018} \cite{brandenburger_trusted_2019}, the authors introduced a framework for executing Hyperledger Fabric chaincodes in Intel SGX to deal with rollback attacks. A sealed-bid auction use case is also evaluated to demonstrate the framework's feasibility.

\subsubsection{Permissioned Blockchain}

Permissioned blockchains usually have an extra authentication mechanism for permission management and, therefore, can provide privacy protection against non-member users. 
For example, Hyperledger Fabric implements the ``channel" technology, which is essentially a private ledger between specific network members; nodes within the same ``channel" can share data, but nodes outside the channel cannot access it.
In \cite{desai_hybrid_2019}, a hybrid blockchain-based auction architecture is proposed, in which a permissioned blockchain is used to publish sensitive bids and a permissionless blockchain is used to make the auction accountable.
It should be noted that this solution offers privacy protection against non-member peers, but still suffers from possible data leakage from malicious nodes within the same network \cite{benhamouda_supporting_2019}. Therefore, it is often used in combination with other encryption techniques.

\textbf{Research Gap.} 
Integrating advanced privacy protection protocols and solutions has become an essential practice for protecting the privacy of blockchain-based auctions. The current research challenge focuses on how to preserve auction privacy in an efficient and economical manner. Implementing cryptographic protocols on the blockchain introduces additional execution overhead. 
In addition to the idea of reducing the algorithm complexity, researchers have proposed that cryptographic protocols can be implemented off-chain as separate modules to reduce on-chain execution costs. However, this increases the risk of data corruption during the transmission and communication phases from on-chain to off-chain. To address this issue, Benhamouda~\textit{et~al.}~\cite{benhamouda_supporting_2019} proposed an approach to integrate MPC protocols into the Hyperledger Fabric blockchain architecture to support secure auctions. Nonetheless, few studies have discussed the problems posed by this new paradigm and the effect on auction privacy protection. This could be an important research gap that should be addressed by the research community.

\input{table/Table_Privacy3}

\subsection{Transaction Ordering \& Fairness}
One of the significant challenges of decentralized auctions is the time synchronization. In a decentralized system, each node user has its own clock. There is no such an absolute clock, so it is difficult to enforce a precise time window to manage particular auction applications. Permissionless blockchains typically use PoW-based consensus algorithms to determine the order of transactions on the blockchain and avoid the double-spending problem. However, this approach introduces uncertainty in the order of auction transactions, i.e., auction users do not know whether their transactions will be prioritized or deferred. For example, different bidders may submit concurrent operations regarding a competing auction smart contract; one decides that the auction has ended, and the other is still trying to bid. Sometimes transactions may experience delays due to network congestion, and some transactions may even be canceled due to the process of the consensus mechanism. 
This uncertain waiting process can lead to the exposure of trading intentions, making front-running easy to occur in blockchain-based auctions. In traditional finance, front-running is a type of cheating, where information that will affect the price of an asset is known in advance from non-public information. In blockchain-based auctions, it means that while an auction transaction is waiting to be packaged, other users can profit by setting a higher blockchain fee to preempt the transaction \cite{orda2021enforcing}. Front-running is unfair and undermines the trading strategies of normal auction participants, harming their trading interests. 
In summary, the successful implementation of a blockchain-based decentralized auction application must deal with front-running issues to ensure the transaction's fairness. This is especially critical for auction models that are sensitive to the order of bids.

\textbf{Research Gap.} 
Researchers have identified that the order of transactions has a significant impact on the fairness of auction results. A commonly used remediation is to remove the benefit of front-running in auction applications, i.e., weaken the importance of bid priority \cite{frontrunning}. For example, bidders can limit the visibility of their bids on the blockchain through a commitment scheme. However, there is no practical solution to this problem regarding time-sensitive auctions in the existing studies. Therefore, we determine that enforcing fairness and avoiding front-running in transactions is a significant research gap that needs further investigation.

\subsection{Decentralization of Auction Front-End}
From a software development perspective, auction applications require excellent front-end components to assist users and improve the user experience. Decentralized smart contracts allow any compliant auction transaction to be executed securely and continuously as long as the blockchain exists. However, while smart contracts and the underlying blockchain are fully decentralized, the front-end of most on-chain auction applications is still implemented using traditional centralized Internet architecture. This allows attackers to influence the user experience by taking control of the front-end web page. An example is Whisky Auctioneer's claim that in 2020 that they had to shut down their auction website and stopped the online auction of thousands of bottles of rare whisky due to a constant Distributed Denial of Service (DDoS) attack \cite{whiskyauction}. Imagine if the auction application was truly decentralized, then attackers would not be able to prevent most users from accessing the front-end pages through DDoS attacks. An attacker could also do other malicious acts, such as making users connect to unaudited malicious contract code on the blockchain by controlling the front-end, even though the original auction smart contract is both audited and secure. To have a truly decentralized experience, users need to be able to control their front-end. This is important because it protects users from malicious attackers and achieves the goal of a fully decentralized auction application. 

\textbf{Research Gap.} 
Although researchers have proposed and implemented different decentralized auction models, few have considered the impact of a centralized front-end operations on the auctions. Solutions combining decentralized domains (e.g., ENS) and decentralized storage (e.g., IPFS) can be used to design and implement decentralized web front-ends \cite{decentralizedfrontend}. However, whether these solutions can provide a user experience comparable to a centralized auction front-end is still a gap that needs to be thoroughly investigated and studied.

\subsection{Decentralized Identity Management}
User identity data is controlled by an authority in traditional centralized auction applications; therefore, the verification, authorization, and accountability of users are also implemented and guaranteed by a centralized authority. Different auction platforms usually have their own identity systems and account management databases, which are not interoperable. While blockchain systems offer the advantages of flexible distributed auction collaboration, their identity management also faces significant challenges. In a blockchain network, users can obtain an address without presenting their real identity and apply it to any auction identity authentication. Such a design protects the privacy of auction users. Still, it increases the potential for spoofing since users do not provide their real identities to interact with the auction application and other users \cite{liu2020blockchainidentity}. For example, bidders can increase their utility by submitting multiple false-name bids without being held accountable for their real identity. These behaviors have already been identified in various online auctions and have led to fraud and unfairness, which may be further exacerbated in blockchain-based auctions. In some auction applications, a real-name user authentication mechanism is required to achieve participant access control and ensure that transactions comply with regulatory requirements.

\textbf{Research Gap.} 
To address this issue, the ERC-725 identity standard is proposed to fill the gap in identity authentication in the Ethereum blockchain. It is an identity identification mechanism that can provide digital contracts with more secure identity authentication \cite{ERC725}. Permissioned blockchains also propose corresponding supervisory anonymous authentication mechanisms in identity management. For example, Hyperledger Fabric adopts a digital certificate-based approach to manage users' digital identity in real name by deploying Certificate Authority (CA) services. However, the trade-off between accountability and user privacy is not well studied in existing solutions. Considering the needs of auction participants, decentralized identity management with accountability and privacy protection is a research gap and an important future research direction.

\subsection{Auction Maintenance \& Update}
In a traditional auction platform, developers can perform regular software maintenance and updates efficiently since the software code is stored on a centralized server. In contrast, in a blockchain-based decentralized auction, all back-end code and data are immutable and maintained by the decentralized nodes. While bugs may arise or the auction business logic may need to be changed, it is impossible to make changes to the original smart contract \cite{wu2022evolution}. As a result, there are many complexities and challenges in maintaining and updating decentralized auction applications. In fact, the development of decentralized auction applications is more like hardware development than software development. When bugs occur in an auction application, redeployment is costly and can seriously damage the application's reputation, leading to a crisis of trust. To address this challenge, a practical approach is introducing a proxy contract mechanism to deploy the underlying auction contracts \cite{chen2021maintenance}. The deployer of an auction application can set up a proxy contract architecture and then deploy a new contract to upgrade the auction logic. In the proxy model, all message calls are made through the proxy contract, and the proxy contract redirects the call request to the latest deployed contract. OpenZeppelin \cite{proxycontract} provides a range of standard libraries to handle the complex proxy contracts described above. 

\textbf{Research Gap.} 
Proxy contracts are widely used to help with the maintenance and update of smart contracts. However, this approach incurs additional transaction costs. There is also a security risk, as an attacker could try to attack the proxy contract and change the target contract to a modified malicious one. Apart from proxy contracts, few studies have focused on developing more economical and secure methods to support the maintenance and updates of decentralized auction applications. This is another crucial research gap that needs to be addressed in the future.

\subsection{Auction Payment with Cryptocurrency}
\label{payment}
Following the end of an auction, the exchange of goods and money between buyers and sellers is expected to happen. A cryptocurrency is often leveraged to complete the auction payment due to its easy and secure transaction properties. Besides, auction payments can be enforced automatically through the token in a smart contract. With such a design, payments can be processed within the blockchain, and transactions containing the corresponding values can be processed between different wallets \cite{figueredo_brain_2019}. On the other hand, the price volatility of cryptocurrencies is a big challenge. Due to the speculative nature of cryptocurrencies, their market values are constantly fluctuating. This makes it difficult for auction sellers to accept cryptocurrencies as the payment method without considering the price risk. Buyers who expect the cryptocurrency value to increase will also be hesitant to use their own tokens as auction payments \cite{Cryptocurrency_2018}. 
In this regard, artificial intelligence (AI) techniques can assist users in predicting cryptocurrency prices and hedging the risk of auction payments. For example, the authors in \cite{awotunde2021machine} proposed a cryptocurrency price prediction model using long short-term memory neural networks. They tested it on three different cryptocurrencies and showed that the proposed model provides good predictive performance.
Market liquidity is another concern regarding using cryptocurrencies for auction payments. There are already cryptocurrencies that are designed to support application-specific auctions, e.g., GreenCoin \cite{dekhane_greencoin_2019} for energy trading and Xcoin \cite{fan_blockchain_2020} for spectrum trading. However, the trading market of these emerging cryptocurrencies is quite small and therefore lacks liquidity. This means that in some cases, cryptocurrencies may not be considered equivalent to fiat money. Another issue is that different blockchain platforms support different cryptocurrencies, which makes it difficult for cross-chain payments.

\textbf{Research Gap.} 
The introduction of cryptocurrency payment gateways \cite{gateway} and atomic swap technologies \cite{black_atomic_2019} can partially address these challenges. However, current research lacks studies on the acceptance of technologies that solve the problem of cryptocurrency payments. There is also a lack of research on the factors that significantly influence the acceptance of cryptocurrencies for auctions. These research gaps need to be addressed in the near future.

\subsection{Auction Contract Enforcement}
\label{oracle}
Blockchain and smart contracts cannot confirm the veracity of external data, which is known as the blockchain oracle problem. This is a big challenge that prevents the widespread adoption of smart contracts for auction applications on the blockchain. 
It should be noted that many of the (non-digital) auctioned items and services cannot be managed by the blockchain directly. For instance, in an art auction, while the ownership of artworks can be recorded by the blockchain, the blockchain cannot directly enforce the transfer of off-chain artworks. 
Basically, a blockchain oracle is a secure middleware that facilitates communication between the blockchain and any off-chain system \cite{Chainlink_2020}. 
Using oracles in an auction fills this gap and ensures that the real-world data fed into the blockchain (e.g., whether the auction item/service is delivered as agreed) is accurate and the auction contract is triggered properly \cite{omar2021implementing}.
This is why some smart contract-based auction platforms have a built-in oracle component \cite{lafourcade_auctionity_2018}. 
Current blockchain oracle services are often provided by third-party companies. Some successful solutions include Chainlink, Provable, and Witnet \cite{Oracles_2021}. These oracle services usually require additional commission fees, and a single oracle may suffer from a single point of failure. In \cite{omar2021implementing}, a decentralized oracle network is integrated into an auction system. The oracles act as external timers to trigger the start/end of the auction in a trustworthy way. 
Another similar solution to the oracle problem is to introduce a decentralized witness mechanism to monitor the delivery of auctioned goods/services. In this case, game theory can be used to design incentive mechanisms to motivate normal blockchain users to join the network and work as witnesses \cite{zhou2019blockchain}. In \cite{matsushima_mechanism_2020}, a self-enforcing contract witness mechanism is proposed. The basic idea is that the smart contract can be enforced through the mutual judgment of auction participants. 
In addition, machine learning technologies demonstrate great potential to enable smarter oracle services \cite{smarteroralces}. For example, by analyzing auction market datasets, an oracle can make short-term predictions about auctions and warn users of upcoming trading peaks. However, AI-based oracle solutions are not currently widely adopted due to the low throughput of blockchain; this may change as blockchain technology improves. It is foreseeable that AI-based solutions will improve traditional oracle services based on incentives and manual verification.

\textbf{Research Gap.} An efficient and economical decentralized oracle/witness mechanism will significantly facilitate the enforcement of blockchain-based auction applications. A general research gap lies in how to construct incentives for decentralized oracles/witnesses to maintain incentive compatibility so that they are incentivized to submit the correct auction report.

\subsection{Auction Regulations \& Standards}
\label{regulation}
There is no authority in a decentralized blockchain network to avoid possible transaction disputes. In an auction application, decentralized users may generate transaction data in different formats. It would be a huge challenge to ensure that the information uploaded by auction users complies with the relevant laws and regulations. For instance, a key part of the EU General Data Protection Regulation (GDPR) lies in the citizen's right to data erasure, i.e., the GDPR claims that individuals have the right to delete the data associated with them \cite{berberich2016blockchain}. However, due to the immutable nature of the blockchain, it is difficult to remove on-chain sensitive information once uploaded to the blockchain. 
Currently, different countries and regions are actively developing new blockchain industry regulations to promote blockchain applications.
The compliance with current laws and regulations needs to be carefully considered when designing blockchain-based auction applications. 
Another pressing challenge is standardization. Currently, different blockchain platforms have different architectures and design patterns, and there are hundreds of auction models to support different application scenarios. There is an urgent need for a standardized solution to set, maintain and merge standards across blockchain platforms to enable seamless integration. Standardized solutions for auction applications have great potential to address challenges such as interoperability, user experience, social acceptance, scale, governance, cost consumption, digital identity, privacy protection, and developer shortcomings \cite{Zeeve_2019}.

\textbf{Research Gap.} As one of the largest blockchain communities, Ethereum has developed several standards (e.g., ERC-20 for token development) to help maintain project interoperability across different implementations \cite{Standards_2021}. However, current practice lacks high-quality standards, libraries, and reference codes for designing blockchain-based decentralized auctions. We believe that the development and operations of standardized auction smart models will be an active research direction in the near future.

\subsection{Auction Mechanism Design}
\label{DesignProperties}
Mechanism design is a branch of economics and game theory that takes an objective-oriented approach to design economic mechanisms or incentives to achieve desired outcomes. As a result, mechanism design is also commonly referred to as reverse game theory \cite{han_niyato_saad_2011}. Auction mechanism design allows a designer to organize specific auction rules to produce the desired equilibrium outcome (e.g., maximize the auction social welfare). Generally, the main properties of designing an auction model can be summarized as follows:

\input{table/Table_Economic}

\begin{itemize}
    \item \textit{Individual Rationality}: An auction is individually rational if no person loses from joining the auction. This is a basic assumption in economic theory when modeling auctions with game theory.
    \item \textit{Incentive Compatibility} (also known as truthfulness or strategy-proofness): An auction is incentive-compatible (or truthful) if every participant can achieve the best outcome for themselves just by acting according to their true preferences.
    \item \textit{Balanced Budget}: An auction is budget-balanced if all money transfers are conducted only between buyers and sellers; the auctioneer should not gain or lose money.
    \item \textit{Economic Efficiency}: An auction is economically efficient if the total social welfare of the auction is maximized. Social welfare can be defined as the sum of individual utilities of all auction participants \cite{hassan2021optimizing}.
    \item \textit{Computational Efficiency}: An auction is computationally efficient if the auction result, including the winning buyer/seller, the price charged to the buyer, and the payment to the seller, can be obtained in polynomial time \cite{guo_double_2020}.
    \item \textit{Allocative Efficiency} (also known as system efficiency):  An auction is allocatively efficient if the overall value of the items awarded to bidders is maximized.
    \item \textit{Cost-Optimal}: An auction is cost-optimal if it minimizes the cost incurred by sellers \cite{chatzopoulos_privacy_2018}. This property is usually associated with user satisfaction, which implies revenue maximization or cost minimization for one side of the auction user (seller or buyer).
\end{itemize}

An auction mechanism may expect several design goals to meet different market requirements. Some classical auction models have intrinsic properties. For instance, the FPSB auction is by default a non-incentive-compatible auction, while the Vickrey auction is an incentive-compatible one. The VCG mechanism satisfies three basic properties, namely individual rationality, economic efficiency, and incentive compatibility. The ability to realize economic efficiency while ensuring truthful biding makes it a unique mechanism that has attracted much discussion \cite{hassan_deal_2020,kadadha_abcrowd_2020,lee_trustful_2020}. Some studies focus on the optimization of existing auction mechanisms. In this context, a novel pricing rule to remedy balanced budget property in the VCG auction is proposed in \cite{alashery_blockchain-enabled_2020}. Among the above economic design properties, incentive compatibility is always considered a top priority in auction design because malicious bidders have instinctive incentives to manipulate the market and harm honest bidders in a non-incentive-compatible auction. An incentive-compatible auction can simplify the decision-making of auction bidders since truth-telling is their dominant strategy \cite{chen_safe_2020}. Therefore, most of the current studies focus on designing an incentive-compatible auction while ensuring other properties. A detailed summary of auction design properties in existing blockchain-auction integrated models is shown in Table \ref{tab:mechanism}.

\textbf{Research Gap.} According to the Myerson-Satterwhite theorem, four basic properties (i.e., individual rationality, incentive compatibility, balanced budget, and economic efficiency) cannot be satisfied in a single auction market mechanism \cite{myerson1983efficient}. Although there is not a ``perfect" auction, designing an auction mechanism to satisfy as many economic objectives as possible is a future research direction. Another promising direction is automated mechanism design (AMD), which aims to shift the design burden from humans to machines. AI techniques can be widely used to assist in AMD. For example, the authors in \cite{shen2018automated} proposed the first model that uses neural networks to discover optimal auction mechanisms. However, using AI techniques to solve the AMD problem in an integrated blockchain-auction model is still a research gap that needs more attention.

\subsection{Auction Fraud Risks}
\label{fraud}
Auction fraud is a complex research topic that has received much attention in traditional auction theory studies. Typically, an auction involves three parties: the bidder, the seller, and the auctioneer. Each party can collude with anyone on the opposite side or on its own side in a variety of manners \cite{Jinglan_2018}. 
The most common auction fraud activities include collusive bidding and shill bidding. In \cite{chua2004fighting}, 11 types of auction fraud are identified and summarized, including failure to ship, failure to pay, misrepresentation, loss or damage claims, and three-party fraud. Blockchain and smart contracts offer a new perspective to partially solve these problems. By providing an open, transparent and trustworthy environment, blockchain eliminates the information asymmetry that exists in traditional auctions. 
In addition, many advanced security mechanisms, e.g., access control, insurance/guarantee mechanisms, reputation/feedback systems, and certification authorities, can be integrated with blockchain-based auction models to alleviate auction fraud \cite{shu_blockchain_2017}. However, there is no one-size-fits-all solution to this problem. When faced with more sophisticated fraud variations, blockchain and smart contacts could be powerless.
Machine learning techniques can also be used for auction fraud detection and prediction. For example, the authors in \cite{abidi2021real} proposed a shill bidding detection model in online auctions using machine learning algorithms (i.e., support vector machines and artificial neural networks). However, despite the high detection accuracy, these machine learning methods can only assist in fraud detection and post-auction remediation, and cannot prevent fraud at the root.

\textbf{Research Gap.} 
Since blockchain-based auctions are a new paradigm, there is still a lack of understanding of whether these frauds are enhanced or weakened by the introduction of blockchain. Therefore, comparative studies are urgently needed to investigate the difference between various fraud behaviors in blockchain-based decentralized auctions and traditional centralized auctions. We expect more studies to fill this gap in the future.

\section{Conclusion}
\label{conslusion}
In this paper, we review existing auction models and blockchain technologies, and provide a conceptual schema to analyze research and innovation opportunities from their integration. Specifically, we provide an overview of main application areas for blockchain-based auction models, e.g., energy trading, wireless communication, and service allocation. 
Moreover, existing auction-based solutions for blockchain enhancement are classified into several categories through extensive investigations, e.g., mining task offloading, transaction fee mechanism design, miner selection \& reward distribution, and token sale \& exchange. There are many open research challenges identified for integrated blockchain-auction models, e.g., 
auction privacy protection, transaction ordering \& fairness, decentralization of auction front-end, decentralized identity management, auction maintenance \& update, auction payment with cryptocurrency, auction contract enforcement, auction regulations \& standards, auction mechanism design, and auction fraud risks,
should be further investigated in the near future. 

In summary, recent research on the integration of blockchain and auction models is quite extensive. Scientific communities have recognized the great potential of integrating the two to solve problems in various application scenarios. While there are still many challenges, such an integration trend will be beneficial to both industry and academia. This paper attempts to explore how blockchain technology and auction models work and when they should be fused together to tackle corresponding challenges. We believe that the main findings of this survey will offer theoretical support and practical guidance for researchers and auction practitioners.

\section*{Acknowledgment}
This research is funded by the European Union’s Horizon 2020 research and innovation program under grant agreements 825134 (ARTICONF project), 862409 (BlueCloud project), and 824068 (ENVRI-FAIR project). The research is also supported by the Chinese Scholarship Council, and EU LifeWatch ERIC.

\clearpage

\ifCLASSOPTIONcaptionsoff
  \newpage
\fi

\bibliographystyle{IEEEtran}
\bibliography{IEEEabrv,ref}

\clearpage

\begin{IEEEbiography}[{\includegraphics[width=1in,height=1.25in,clip,keepaspectratio]{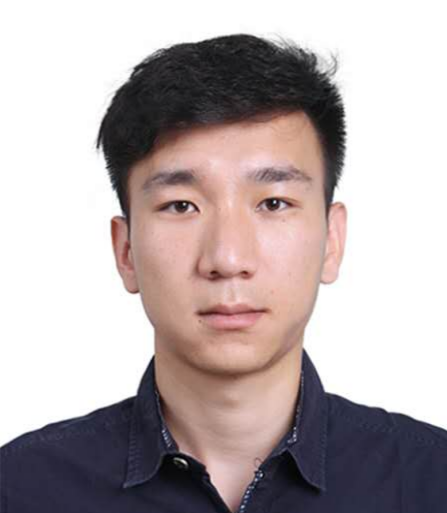}}] {Zeshun Shi} is currently pursuing his Ph.D. degree in the MultiScale Networked Systems (MNS) group, University of Amsterdam (UvA), Netherlands. 
He received his Master's degree from Beijing Normal University (BNU) in 2018.
His research interests include blockchain, auction theory, cloud computing, and DevOps.
His Ph.D. research focuses on the use of blockchain, smart contracts, game theory, and auction models to create a decentralized cloud marketplace, and to orchestrate trustworthy transactions for cloud services.
\end{IEEEbiography}

\begin{IEEEbiography}[{\includegraphics[width=1in,height=1.11in,clip]{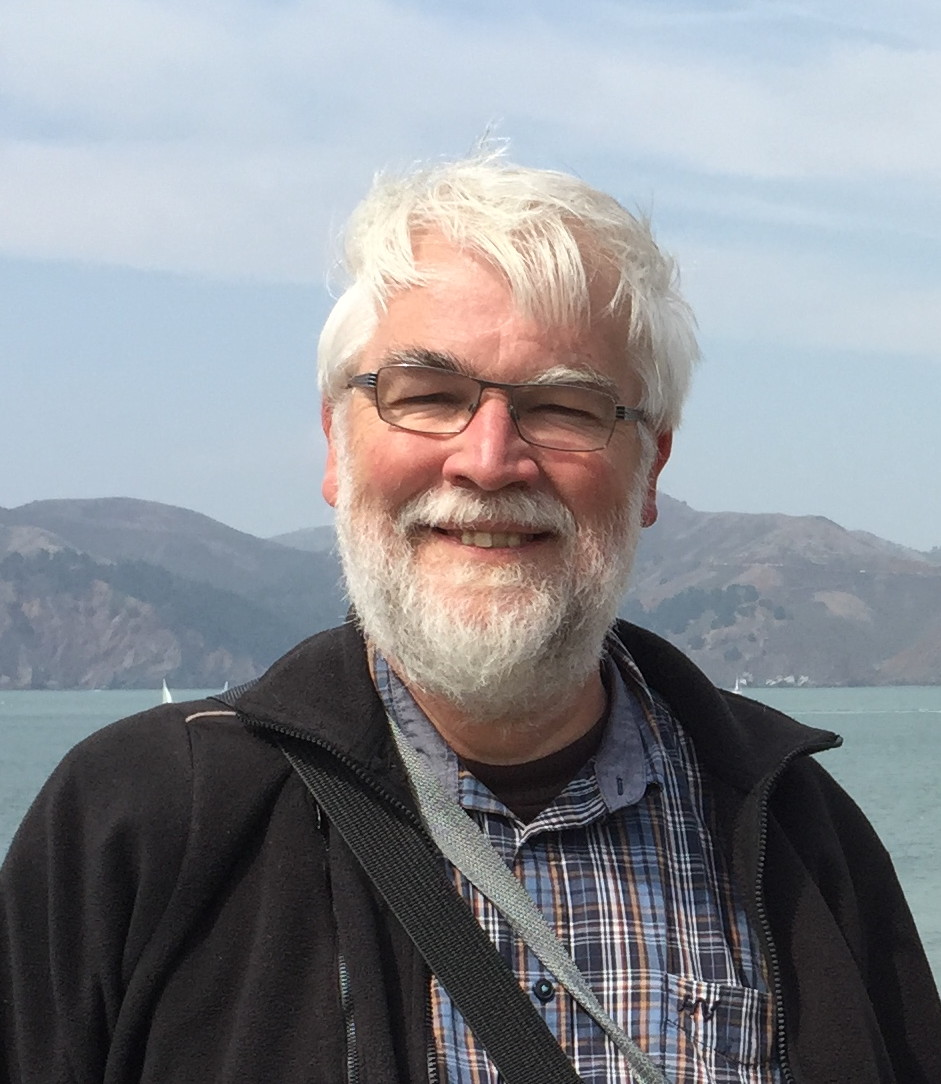}}]{Cees de Laat} is the chair of the System and Network Engineering laboratory at the University of Amsterdam. The SNE lab conducts research on leading-edge computer systems of all scales, ranging from global-scale systems and networks to embedded devices. His own work focuses on Secure Trusted Distributed Data Processing Systems. Prof. de Laat served on the Lawrence Berkeley Laboratory Policy Board for ESnet, the scientific advisory board of SURF, was (co-)founder of the Global Lambda Integrated Facility (GLIF), GRIDforum.nl, and CineGrid.org. He is a member of the Advisory Board Internet Society Netherlands. See: http://delaat.net/.
\end{IEEEbiography}

\begin{IEEEbiography}[{\includegraphics[width=1in,height=1.25in,clip,keepaspectratio]{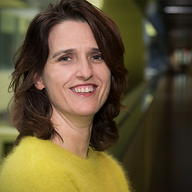}}] {Paola Grosso} Dr. Paola Grosso is an associate professor at the Institute for Informatics at the University of Amsterdam. She leads the Multiscale Networked Systems (MNS) group, which researches the emerging architectures that can support the operations of multiscale systems across the Future Internet. She has an extensive record of contribution to international projects and she is currently involved with her group in numerous EU-funded projects, among them GN4-3, FEd4FIRE+.
\end{IEEEbiography}

\begin{IEEEbiography}[{\includegraphics[width=1in,height=1.25in,clip,keepaspectratio]{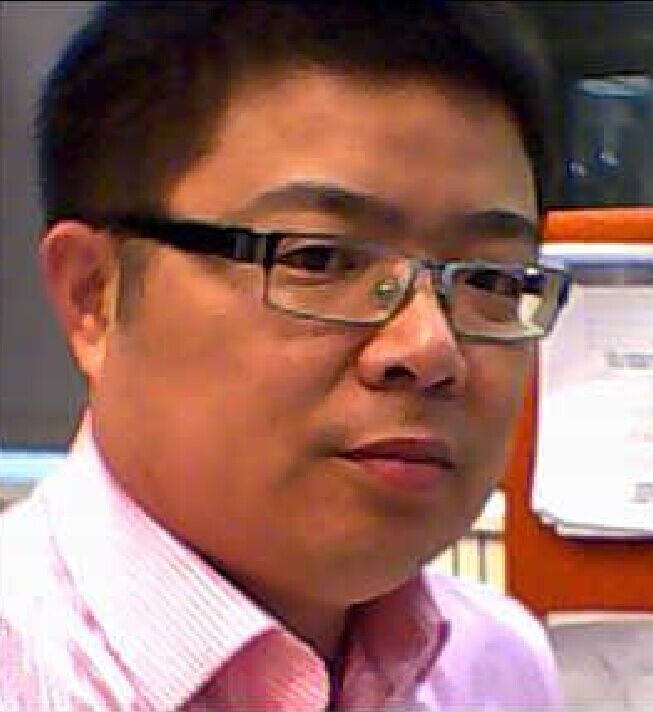}}]{Zhiming Zhao} received his Ph.D. in computer science in 2004 from the University of Amsterdam (UvA). He is currently an assistant professor in the MultiScale Networked Systems (MNS) group at UvA. He coordinates research efforts on quality critical systems on programmable infrastructures in the context of ARTICONF, SWITCH, ENVRI-FAIR and several other EU H2020 projects. His research interests include blockchain, SDN, workflow management systems, multi-agent systems, and big data research infrastructures.
\end{IEEEbiography}

\end{document}

%% file: pics/Related.tex
\scalebox{0.51}{
\begin{tikzpicture}
[thick, node distance=.15cm,start chain=going below]
    \node[punktchain, fill=black!15] (sc) {Smart Contract \cite{Hu_2021}};
    \node[punktchain, fill=black!15] (health) {Healthcare \cite{de2020survey}};
      \begin{scope}[start branch=venstre,]
            \node [punktchain, fill=black!15, on chain=going left] (city) {Smart City \cite{xie2019survey}};
        \end{scope}
    \node[punktchain, fill=black!15] (sec) {Security \& Privacy \cite{hassan2020differential, peng2020privacy}};
        \begin{scope}[start branch=venstre,]
            \node [punktchain, fill=black!15, on chain=going left] (sec) {Security \& Privacy \cite{zhang2019security, lee2019systematic}};
        \end{scope}
    \node[punktchain, fill=black!15] (model) {Models \& Tools \cite{huang2021survey}};
        \begin{scope}[start branch=venstre,]
            \node [punktchain, on chain=going left] (market) {Market Design \cite{doi:10.1146/annurev-economics-080218-025818}};
        \end{scope}
    \node[punktchain, fill=black!15] (iot) {IoT \cite{lao2020survey}};
        \begin{scope}[start branch=venstre,]
            \node [punktchain, fill=black!15, on chain=going left] (iot) {IoT \cite{ferrag2018blockchain}};
        \end{scope}
    \node[punktchain, fill=black!35] (energy) {Energy Trading \cite{oprea2020local}};
        \begin{scope}[start branch=venstre,]
            \node[punktchain, on chain=going left, fill=black!35] (energy) {Energy Trading \cite{wang2019energy}};
        \end{scope}    
        \begin{scope}[start branch=venstre,
            every join/.style={->, thick, shorten <=1pt}, opacity= 0]
            \node[punktchain, on chain=going left=by {<-}] (positionnode1) {};
            \node[punktchain, on chain=going left=by {<-}] (positionnode2) {};
            \node[punktchain, on chain=going left=by {<-}] (positionnode3) {};
        \end{scope}
        \begin{scope}[start branch=venstre,
            every join/.style={->, thick, shorten <=1pt}] 
            \node[punktchain,left of= positionnode3] (wireless) {Crowdsensing \cite{zhang2015incentives}};
        \end{scope}
    \node[punktchain, fill=black!15] (data) {Big Data \cite{deepa2020survey}};
        \begin{scope}[start branch=venstre,
            every join/.style={->, thick, shorten <=1pt}, ]
            \node[punktchain, on chain=going left=by {<-}, fill=black!15] (edge) {Edge Computing \cite{yang2019integrated}};
            \node[punktchain, on chain=going left=by {<-}] (wireless) {Wireless Systems \cite{habiba2018auction}};
            \node[punktchain, on chain=going left=by {<-}] (wireless1) {Wireless Systems \cite{zhang2012auction}};
        \end{scope}
    \node[punktchain, fill=black!15] (cloud) {Cloud Computing \cite{gai2020blockchain,10.1145/3403954}};
        \begin{scope}[start branch=hoejre,]
            \node (energy) [punktchain, fill=black!35, on chain=going right] {Energy Trading \cite{hassan2021optimizing}};
        \end{scope}
        \begin{scope}[start branch=venstre,
            every join/.style={->, thick, shorten <=1pt}, ]
            \node[punktchain, on chain=going left=by {<-}, fill=black!15] (consensus) {Consensus Mechanisms \cite{wang2019survey}};
            \node[punktchain, on chain=going left=by {<-}, fill=black!15] (sec) {Security \& Privacy \cite{conti2018survey}};
            \node[punktchain, on chain=going left=by {<-}] (eng) {English Auction \cite{wahaballa2015taxonomy}};
        \end{scope}
    \node (overview) [punktchain, fill=black!15]  {Blockchain Overview \cite{butijn2020blockchains,kolb2020core}};
        \begin{scope}[start branch=venstre,
            every join/.style={->, thick, shorten <=1pt}, ]
            \node[punktchain, on chain=going left=by {<-}, fill=black!15] (app) {Blockchain Applications \cite{casino2019systematic}};
            \node[punktchain, on chain=going left=by {<-}, fill=black!15] (overview) {Blockchain Overview \cite{zheng2018blockchain}};
            \node[punktchain, on chain=going left=by {<-}] (theory) {Auction Theory \cite{klemperer2004auctions, jain2006classification}};
            \end{scope}
      \begin{scope}[start branch=hoejre,]
            \node (finans) [punktchain, fill=black!15, on chain=going right] {Cryptocurrency Mining \cite{hacioglu2021crafting}};
      \end{scope}
  
\matrix [draw,fill=black!1,below left] at (-7.5,0) {
  \node [auctionrelated,label=right:Auction-Related] {}; \\
  \node [blockchain,label=right:Blockchain-Related] {}; \\
  \node [both,label=right:Both-Related] {}; \\
};

   \draw[thick,->] (-11.7,-10.7) -- (5.4,-10.7) node[anchor=north west] {};
   \draw[thick,->] (-11.7,-10.7) -- (-11.7,1) node[anchor=south east] {};
   
    \begin{scope}[on background layer]
      \draw [thick, dotted, ->] (-8.28,-10.7) -- (-8.28, 1);
    \end{scope}
   \draw [thick, dotted, ->] (-4.97,-10.7) -- (-4.97, 1);
   \draw [thick, dotted, ->] (-1.65,-10.7) -- (-1.65, 1);
   \draw [thick, dotted, ->] (1.65,-10.7) -- (1.65, 1);
   \draw [thick, dotted, ->] (4.96,-10.7) -- (4.96, 1);
   
  \node at (-9.7,-11) {\textbf{Before 2017}};
  \node at (-6.5,-11) {\textbf{2018}};
  \node at (-3.3,-11) {\textbf{2019}};
  \node at (-0.1,-11) {\textbf{2020}};
  \node at (3.4,-11) {\textbf{2021 (Until March)}};
  \node at (-9.5,1.2) {\textbf{Topic \& Number of Surveys}};

  \node at (-9.7,-5.3) {\textbf{N = 5}};
  \node at (-6.5,-6.6) {\textbf{N = 3}};
  \node at (-3.3,-0.4) {\textbf{N = 9}};
  \node at (-0.1,0.8) {\textbf{N = 12}};
  \node at (3.4,-7.9) {\textbf{N = 2}};

\end{tikzpicture}
}

%% file: table/Table_Abbreviation.tex
\begin{table}[!ht]
\setlength{\extrarowheight}{0.23pt}
\centering
\caption{Summary of Abbreviations}
\label{tab:ACRONYMS}
\begin{tabularx}{\columnwidth}{|m{.25\columnwidth}|m{.65\columnwidth}|}
\hline
AE    & Asymmetric Encryption                      \\ \hline
AI    & Artificial Intelligence                    \\ \hline
AMD    & Automated Mechanism Design                 \\ \hline
API   & Application Programming Interface          \\ \hline
BB    & Balanced Budget                            \\ \hline
BFT    & Byzantine Fault Tolerance                 \\ \hline
CA    & Certificate Authority                      \\ \hline
CCHP  & Combined Cooling, Heating, and Power       \\ \hline
CE    & Computational Efficiency                   \\ \hline
CMN   & Collaborative Mining Network               \\ \hline
CPSS  & Cyber-Physical-Social Systems             \\ \hline
CR    & Cognitive Radio                            \\ \hline
CS    & Commitment Scheme                          \\ \hline
DAG   & Directed Acyclic Graph                     \\ \hline
DApp  & Decentralized Application                  \\ \hline
DDoS  & Distributed Denial of Service              \\ \hline
DEX   & Decentralized Exchange                     \\ \hline
DG    & Distributed Generation                     \\ \hline
DHT   & Distributed Hash Table                     \\ \hline
DP    & Differential Privacy                       \\ \hline
DPoS  & Delegated Proof of Stake                   \\ \hline
DS    & Digital Signature                          \\ \hline
ECC   & Elliptic Curve Cryptography                \\ \hline
ECDSA & Elliptic Curve Digital Signature Algorithm \\ \hline
EE    & Economic Efficiency                        \\ \hline
EIP   & Ethereum Improvement Proposal              \\ \hline
ENS   & Ethereum Name Service                      \\ \hline
ERC   & Ethereum Request for Comments              \\ \hline
EV    & Electric Vehicle                           \\ \hline
FCC   & Federal Communications Commission          \\ \hline
FL    & Federated Learning                         \\ \hline
FPSB  & First-Price Sealed-Bid                     \\ \hline
GDPR  & General Data Protection Regulation         \\ \hline
GFP   & Generalized First-Price                    \\ \hline
GSP   & Generalized Second-Price                   \\ \hline
HE    & Homomorphic Encryption                     \\ \hline
IC    & Incentive Compatibility                    \\ \hline
ICO   & Initial Coin Offering                     \\ \hline
IoE   & Internet of Energy                         \\ \hline
IoT   & Internet of Things                         \\ \hline
IoV   & Internet of Vehicles                       \\ \hline
IPFS   & InterPlanetary File System              \\ \hline
IPO   & Initial Public Offering                    \\ \hline
IR    & Individual Rationality                     \\ \hline
MPC   & Multi-Party Computation                    \\ \hline
NFV   & Network Function Virtualization            \\ \hline
P2P   & Peer-to-Peer                               \\ \hline
PB    & Permissioned Blockchain                    \\ \hline
PBFT  & Practical Byzantine Fault Tolerance        \\ \hline
PBS   & Primary Base Station                       \\ \hline
PoA   & Proof of Authority                         \\ \hline
PoET  & Proof of Elapsed Time                      \\ \hline
PoS   & Proof of Stake                             \\ \hline
PoW   & Proof of Work                              \\ \hline
QoS   & Quality of Service                         \\ \hline
SGX   & Software Guard Extensions                  \\ \hline
TEE   & Trusted Execution Environment              \\ \hline
TPS   & Transactions Per Second                    \\ \hline
TTP   & Trusted Third Party                        \\ \hline
UAV   & Unmanned Aerial Vehicle                    \\ \hline
V2G   & Vehicle-to-Grid                            \\ \hline
V2V   & Vehicle-to-Vehicle                         \\ \hline
V2X   & Vehicle-to-Everything                      \\ \hline
VCG   & Vickrey–Clarke–Groves                      \\ \hline
VNE   & Virtual Network Embedding                  \\ \hline
VPP   & Virtual Power Plant                        \\ \hline
ZKP   & Zero-Knowledge Proof                       \\ \hline
\end{tabularx}%
\end{table}

%% file: table/Table_Auction3.tex
\begin{table*}[!t]
\setlength{\extrarowheight}{5pt}
\setlength{\tabcolsep}{1pt}

\centering
\caption{Summary of Representative Auction Types}
\label{tab:auctions}
\begin{tabularx}{\linewidth}{|m{2cm}|m{2.7cm}|m{6.9cm}|m{6.2cm}|}

\hline
  \multicolumn{1}{|c|}{\textbf{Auction Type}} &
  \multicolumn{1}{c|}{\textbf{Alternative Name}} &
  \multicolumn{1}{c|}{\textbf{Auction Mechanism}} &
  \multicolumn{1}{c|}{\textbf{Properties/Suitable Scenarios}} \\ \hline
\textbf{English auction} &
  Open-outcry ascending-price auction &
  \begin{minipage}{2.8in} 
  \textbullet\- The price starts low and increases as buyers bid. \\ \textbullet\- The auction continues until no higher bids are received.\end{minipage} &
  \textbullet\- Support a dynamic price discovery process and maximize sellers' profits. \\ \hline
\textbf{Dutch auction} &
  Clock auction; Open-outcry descending-price auction &
  \begin{minipage}{2.7in}\textbullet\- The auctioneer starts the auction with a high asking price.\\ \textbullet\- The price is gradually reduced until one bidder accepts it. \end{minipage} &
  \textbullet\- Suitable for perishable auction items or auctions that need to be completed quickly. \\ \hline
\textbf{FPSB auction} &
  Blind auction &
  \begin{minipage}{2.7in}\textbullet\- All bidders simultaneously submit a sealed bid.\\ \textbullet\- The highest bidder wins and pays his or her bid. \end{minipage} &
  \textbullet\- Prior to making their own offers, bidders can collect details about their competitors' bids. \\ \hline
\textbf{Vickrey auction} &
  Second-price sealed-bid auction &
  \begin{minipage}{2.7in}\textbullet\- All bidders simultaneously submit a sealed bid.\\ \textbullet\- The highest bidder still wins but only pays the second-highest bid.\end{minipage} &
  \textbullet\- Well studied in theory due to the truthful bidding property, but uncommon in practice. \\ \hline
\textbf{Double auction} &
  Double-sided auction &
  \begin{minipage}{2.7in}\textbullet\- Multiple sellers and buyers submit their bids/offers.\\ \textbullet\- The auctioneer chooses a price that clears the market.\end{minipage} &
  \textbullet\- Real word marketplaces with multiple sellers and buyers, e.g., stock exchanges. \\ \hline
\textbf{Combinatorial auction} &
  Multi-lot auction &
  \begin{minipage}{2.7in}\textbullet\- Several heterogeneous items are sold.\\ \textbullet\- Bidders can place bids on combinations of items.\end{minipage} &
  \textbullet\- Suitable when bidders have non-additive valuations on bundles of items, e.g, spectrum allocation. \\ \hline
\textbf{Uniform price auction} &
  Clearing price auction &
  \begin{minipage}{2.7in}\textbullet\- Multiple homogeneous items are sold.\\ \textbullet\- Winners pay the same price regardless of their actual bid.\end{minipage} &
  \textbullet\- Bidders tend to shade their bids when they demand multiple units.\\ \hline
\textbf{Pay-as-bid auction} &
  Discriminatory price auction &
  \begin{minipage}{2.7in}\textbullet\- Multiple homogeneous items are sold.\\ \textbullet\- Winners pay their bids based on the items they won.\end{minipage} &
  \begin{minipage}{2.4in}\textbullet\- A common way to allocate assets and commodities.\\ \textbullet\- Bidders face no uncertainty about the price they will receive if they win.\end{minipage} \\ \hline
\textbf{All-pay auction} &
  - &
  \begin{minipage}{2.7in}\textbullet\- Every bidder must pay regardless of whether they win.\\ \textbullet\- The auction is still awarded to the highest bidder.\end{minipage} &
  \begin{minipage}{2.4in}\textbullet\- Very popular among governments and central banks.\\ \textbullet\- Overbidding is a common behavior.\end{minipage} \\ \hline
\textbf{Multi-attribute auction} &
  - &
  \begin{minipage}{2.7in}\textbullet\- The bids may have multiple attributes. \\ \textbullet\- A scoring mechanism calculates the attributes' value.\end{minipage}&
  \begin{minipage}{2.4in}\textbullet\- Suitable when multiple attributes (e.g., service time, quality) need to be considered in an auction. \end{minipage} \\ \hline
\textbf{Reverse auction} &
  Buyer-determined auction; Procurement auction &
  \begin{minipage}{2.7in}\textbullet\- The buyer makes a request for the required goods. \\ \textbullet\- Sellers place bids for the goods they are willing to buy.\end{minipage} &
  \textbullet\- Suitable for procurement by governments and companies, as it causes sellers' competition. \\ \hline
\textbf{GFP auction} &
  - &
  \begin{minipage}{2.7in}\textbullet\- $n$ bidders compete for $k$ slots/positions. \\ \textbullet\- The highest bidder gets the first slot (with his bid), the second-highest gets the second, and so on. \end{minipage}&
  \begin{minipage}{2.4in}\textbullet\- The auction structure is naturally unstable. \\ \textbullet\- The first mechanism introduced in sponsored search auctions.\end{minipage} \\ \hline
\textbf{GSP auction} &
  - &
  \begin{minipage}{2.7in}\textbullet\- $n$ bidders compete for $k$ slots/positions. \\ \textbullet\- The highest bidder gets the first slot and pays the second highest bid, and so on.\end{minipage} &
  \begin{minipage}{2.4in}\textbullet\- An extension of Vickrey auction for multiple units.\\ \textbullet\- The most commonly used mechanism for sponsored search auctions.\end{minipage} \\ \hline
\textbf{VCG auction} &
  - &
  \begin{minipage}{2.7in}\textbullet\- Bidders submit bids that report their true value. \\ \textbullet\- Each bidder pays for the losses he or she causes to others. \\ \textbullet\- Items are assigned in a socially optimal way. \end{minipage} &
  \begin{minipage}{2.4in} \textbullet\- An extension of Vickrey auction for multiple units.\\ \textbullet\- More complex to interpret and implement than the GSP auction in sponsored search auctions.\end{minipage} \\ \hline

\end{tabularx}%
\end{table*}

%% file: table/Table_Blockchain.tex
\begin{table}[!ht]
\setlength{\tabcolsep}{0.75pt}
\centering
\caption{Characteristics of the Three Types of Blockchain}
\label{blockchaintypes}
{\renewcommand{\arraystretch}{2.5}
\begin{tabularx}{\linewidth}{|m{1.6cm}|m{2.1cm}|m{2.8cm}|m{2.1cm}|}
\hline
  \null  & 
  \centering{\textbf{Permissionless Blockchain}} &
  \centering{\textbf{Permissioned Blockchain}} &
  {\centering\textbf{Hybrid \\ \quad \, Blockchain}} \\ \hline
\textbf{Participants} &
  \begin{minipage}{1in}\textbullet\- Public\\ \textbullet\- Anonymous\end{minipage} &
  \begin{minipage}{1.5in}\textbullet\- Private/Consortium\\ \textbullet\- Known identities\end{minipage} &
  \textbullet\- Public + Private \\ \hline
\textbf{Access Mechanism} &
  \begin{minipage}{1in}\textbullet\- Anyone\\ \textbullet\- Decentralized\end{minipage} &
  \begin{minipage}{1.5in}\textbullet\- Selected users\\ \textbullet\- Partially decentralized\end{minipage} &
  \begin{minipage}{1in}\textbullet\- Customized\end{minipage} \\ \hline
\textbf{Consensus} &
  \begin{minipage}{1in}\textbullet\- PoW, PoS\\ \textbullet\- Energy-intensive\end{minipage} &
  \begin{minipage}{1.5in}\textbullet\- PBFT, Raft, and PoET \\ \textbullet\- Energy-efficient \end{minipage} &
  \textbullet\- Integrated  \\ \hline
\textbf{Performance} &
  \textbullet\- Low &
  \textbullet\- High &
  \textbullet\- Medium \\ \hline
\textbf{Examples} &
 \begin{minipage}{1in}\textbullet\- Bitcoin\\ \textbullet\- Ethereum \end{minipage}&
  \begin{minipage}{1in}\textbullet\- Hyperledger Fabric\end{minipage} &
  \textbullet\- Aergo\\ \hline
  
\end{tabularx}%
}
\end{table}

%% file: table/Table_Application4.tex
\onecolumn
{
\footnotesize
\renewcommand{\arraystretch}{1.65}
\setlength{\tabcolsep}{4pt}
\setlength\LTright{-1in plus 1 fill}

\setlength\LTleft{0pt}
\setlength\LTright{0pt}
\begin{longtable}[c]{|m{1cm}|m{1.1cm}|m{0.5cm}|m{0.5cm}|m{4.7cm}|m{3.3cm}|m{2.0cm}|m{2.5cm}|}
\caption{Summary of Blockchain-Based Auction Applications}\label{tab:application} \\
\hline
\multicolumn{2}{|l|}{\textbf{Application Field}} &
  \textbf{Ref.} &
  \textbf{Year} &
  \textbf{Addressed Issue} &
  \textbf{Auction Model} &
  \textbf{Blockchain Type} &
  \textbf{Blockchain Platform} \\ \hline
\endfirsthead

 \hline
\multicolumn{8}{|l|}%
{Continued from previous page} \\
\hline
\multicolumn{2}{|l|}{\textbf{Application Field}} &
  \textbf{Ref.} &
  \textbf{Year} &
  \textbf{Addressed Issue} &
  \textbf{Auction Model} &
  \textbf{Blockchain Type} &
  \textbf{Blockchain Platform} \\ \hline
\endhead

\hline \multicolumn{8}{|r|}{{Continued on next page}} \\ \hline
\endfoot

\hline
\endlastfoot

\multirow{40}{*}{\begin{minipage}{0.5in}\textbf{Energy Trading}\end{minipage}} &
  \multirow{18}{*}{\begin{minipage}{0.5in}\textbf{Power Grid}\end{minipage}} &
  \cite{wang_novel_2017} &
  2017 &
  Microgrids energy trading &
  Continuous double auction &
  Permissionless &
  Bitcoin \\ \cline{3-8} 
 &
   &
  \cite{yan_novel_2018} &
  2018 &
  Generation right trading &
  Continuous double auction &
  Permissioned &
  MultiChain \\ \cline{3-8} 
 &
   &
  \cite{thakur_distributed_2018} &
  2018 &
  Microgrids energy trading &
  Double auction &
  Permissionless &
  Bitcoin \\ \cline{3-8} 
 &
   &
  \cite{stubs_blockchain-based_2020} &
  2020 &
  Smart grids energy trading &
  Hierarchical double auction &
  Permissionless &
  Ethereum \\ \cline{3-8} 
 &
   &
  \cite{alashery_blockchain-enabled_2020} &
  2020 &
  Smart grids energy trading &
  Double auction &
  Permissioned &
  Simulation \\ \cline{3-8} 
 &
   &
  \cite{zhao_decentralized_2019} &
  2019 &
  Transactive energy trading &
  Double auction with bandit learning &
  N/S &
  Ethereum \\ \cline{3-8} 
 &
   &
  \cite{seven_peer--peer_2020} &
  2020 &
  Energy trading in virtual power plants &
  English auction &
  Permissionless &
  Ethereum \\ \cline{3-8} 
 &
   &
  \cite{hahn_smart_2017} &
  2017 &
  Transactive energy trading &
  Vickrey auction &
  N/S &
  Ethereum \\ \cline{3-8} 
 &
   &
  \cite{dekhane_greencoin_2019} &
  2019 &
  Smart power distribution &
  Dutch auction \& Vickrey auction &
  N/S &
  Ethereum \\ \cline{3-8} 
 &
   &
  \cite{laszka_providing_2017} &
  2017 &
  Transactive energy trading &
  N/S &
  Permissionless &
  Prototype \\ \cline{3-8} 
 &
   &
  \cite{zhang_privacy_2019} &
  2019 &
  Microgrids energy trading &
  Continuous double auction &
  Permissioned &
  Simulation \\ \cline{3-8} 
 &
   &
  \cite{hassan_deal_2020} &
  2020 &
  Microgrids energy trading &
  Modified VCG auction &
  Permissioned &
  Prototype \\ \cline{3-8} 
 &
 &
  \cite{esmat2021novel} &
  2017 &
  Decentralized energy trading market &
  Short-term parallel auction &
  Permissioned &
  Hyperledger Burrow \\ \cline{3-8} 
 &
   &
  \cite{myung_ethereum_2020} &
  2020 &
  Microgrids energy trading &
  English auction \& Continuous double auction &
  N/S &
  Ethereum \\ \cline{3-8} 
 &
   &
  \cite{foti_blockchain_2019} &
  2019 &
  Decentralized energy trading market &
  Uniform-Price double auction &
  Permissioned &
  Ethereum \\ \cline{2-8} 
 &
  \multirow{7}{*}{\begin{minipage}{0.5in}\textbf{Smart Community}\end{minipage}} 
   &
  \cite{hassija_blockcom_2019} &
  2019 &
  Smart communities energy trading &
  Double auction &
  Hybrid &
  Prototype \\ \cline{3-8} 
 &
 &
  \cite{alcarria_blockchain-based_2018} &
  2018 &
  Smart communities energy trading &
  Vickrey auction &
  Permissioned &
  Ethereum \\ \cline{3-8} 
 &

   &
  \cite{guo_combined_2020} &
  2020 &
  Energy trading in CCHP systems &
  N/S &
  Permissionless &
  Prototype \\ \cline{3-8} 
 &
   &
  \cite{saxena_design_2019} &
  2019 &
  Residential communities energy trading &
  Periodic double auction &
  Permissioned &
  Hyperledger Fabric \\ \cline{3-8} 
 &
   &
  \cite{gros_enerdag_2020} &
  2020 &
  Local energy trading market &
  Double auction &
  Permissionless &
  IOTA \\ \cline{3-8} 
 &
   & \cite{brenzikofer_privacy-preserving_2019} \cite{ableitner_quartierstrom_2019} &
  2019 &
  Local energy trading market &
  Double auction &
  Permissioned &
  Tendermint \\ \cline{2-8} 
 &
  \multirow{10}{*}{\begin{minipage}{0.5in}\textbf{Internet of Vehicles}\end{minipage}} &
  \cite{xia_bayesian_2020} &
  2020 &
  V2V energy trading &
  Double auction &
  Permissioned &
  Hyperledger Fabric \\ \cline{3-8} 
 &
   &
  \cite{wang_electric_2020} &
  2020 &
  EV group energy trading &
  Double auction &
  N/S &
  Prototype \\ \cline{3-8} 
 &
   &
  \cite{sun_blockchain-enhanced_2020} &
  2020 &
  V2V energy trading  &
  Iterative double auction &
  Permissioned &
  Simulation \\ \cline{3-8} 
 &
   &
  \cite{ali_efficient_2020} &
  2020 &
  Energy trading in IoV  &
  Multi-attribute auction &
  Permissionless &
  IOTA \\ \cline{3-8} 
 &
   &
  \cite{guo_double_2020} &
  2020 &
  EV charging scheduling &
  Constrained double auction &
  Permissionless &
  Prototype \\ \cline{3-8} 
 &
   &
  \cite{choubey_energytradingrank_2019} &
  2019 &
  V2V energy trading &
  Double auction &
  Permissioned &
  Hyperledger Fabric \\ \cline{3-8} 
 &
   &
  \cite{hassija_blockchain-based_2020} &
  2020 &
  V2G data sharing and energy trading &
  Ascending-price progressive auction &
  Permissionless &
  IOTA \\ \cline{3-8} 
 &
   &
  \cite{liu_electric_2019} &
  2019 &
  V2G energy trading &
  Reverse sealed-bid auction &
  Permissioned &
  Ethereum \\ \cline{3-8} 
 &
   &
  \cite{pustisek_blockchain_2016} &
  2016 &
  V2G charging scheduling &
  FPSB auction &
  N/S &
  Ethereum \\ \hline
\multirow{14}{*}{\begin{minipage}{0.5in}\textbf{Wireless Communication}\end{minipage}} &
  \multirow{10}{*}{\begin{minipage}{0.5in}\textbf{Radio Spectrum}\end{minipage}} &
  \cite{wang_secure_2020} &
  2020 &
  Spectrum resource allocation &
  Single-sided auction &
  N/S &
  Ethereum \\ \cline{3-8} 
 &
   &
  \cite{fan_blockchain_2020} &
  2020 &
  Spectrum resource management in CPSS &
  N/S &
  Permissioned &
  Prototype \\ \cline{3-8} 
 &
   &
  \cite{zheng_smart_2020} &
  2020 &
  Multiple-operators spectrum sharing &
  Double auction &
  Permissioned &
  Ethereum \\ \cline{3-8} 
 &
   &
  \cite{tu_blockchain-based_2020} &
  2020 &
  Dynamic spectrum sharing &
  Double auction &
  Permissioned &
  Ethereum \\ \cline{3-8} 
 &
   &
  \cite{kotobi_blockchain-enabled_2017} \cite{kotobi_secure_2018} &
  2017 2018 &
  Spectrum sharing in CR networks &
  Waiting-line auction &
  Permissionless &
  Prototype \\ \cline{3-8} 
 &
   &
  \cite{khan_blockchain_2020} &
  2020 &
  Secondary spectrum trading market &
  Periodic sealed-bid auction &
  N/S &
  Prototype \\ \cline{3-8} 
 &
   &
  \cite{yu_smart_2019} &
  2019 &
  Spectrum allocation in spacecraft networks &
  Generalized Vickrey auction &
  Permissionless &
  Ethereum \\ \cline{2-8} 
 &
  \multirow{9}{*}{\begin{minipage}{0.5in} \textbf{Network Resource}\end{minipage}} 
   &
  \cite{chen_safe_2020} &
  2020 &
  Wireless network resource allocation &
  General sealed-bid auction &
  Permissionless&
  Ethereum \\ \cline{3-8} 
 &
   &
  \cite{afraz_distributed_2019} &
  2019 &
  Trade market for telecommunication networks &
  Double auction &
  Permissioned &
  Hyperledger Fabric \\ \cline{3-8} 
 &
 &
  \cite{khan_blockchain-based_2019} &
  2019 &
  Ccooperative relaying resource allocation &
  Double auction &
  Permissionless &
  Ethereum \\ \cline{3-8} 
 &
  &
  \cite{chen_blockchain_2020} &
  2020 &
  User offloading in wireless networks &
  Vickrey auction &
  N/S &
  Ethereum \\ \cline{3-8} 
 &
   &
  \cite{hassija_drone_2020} &
  2020 &
  Bandwidth allocation for UAV base stations &
  Multi-attribute auction &
  Permissioned &
  Ethereum \\ \cline{3-8} 
 &
   &
  \cite{khan_trusted_2020} &
  2020 &
  UAV network resource allocation &
  Vickrey auction &
  Permissioned &
  Hyperledger Fabric \\ \cline{3-8} 
 &
   &
  \cite{hassija_framework_2020} &
  2020 &
  Bandwidth allocation between EVs and roadside units &
  Multi-attribute auction &
  Permissionless &
  IOTA \\ \hline
\multirow{12}{*}{\begin{minipage}{0.4in}\textbf{Service Allocation}\end{minipage}} &
  \multirow{8}{*}{\begin{minipage}{0.5in}\textbf{Cloud/ \\ Fog/ \\ Edge \\ Service}\end{minipage}} &
  \cite{sonnino_asterisk_2019} &
  2019 &
  Shared economy service allocation &
  Vickrey auction &
  Permissionless &
  Chainspace \\ \cline{3-8} 
 &
   &
  \cite{chen_fair_2020} &
  2020 &
  Cloud VM allocation &
  Combinatorial auction &
  N/S &
  Ethereum \\ \cline{3-8} 
 &
&
  \cite{gu_cloud_2018} &
  2018 &
  Cloud data storage resource trading &
  VCG auction &
  N/S &
  Ethereum \\ \cline{3-8} 
 &
   &
  \cite{gu_decentralized_2018} &
  2018 &
  Distributed data storage &
  Reverse VCG auction &
  N/S &
  Ethereum \\ \cline{3-8} 
 &
   &
  \cite{zavodovski_decloud_2019} &
  2019 &
  Edge/Cloud service trading &
  Double auction &
  N/S &
  Prototype \\ \cline{3-8} 
 &
   &
  \cite{debe_blockchain-based_2020} &
  2020 &
  Fog service trading &
  Reverse auction &
  Permissionless &
  Ethereum \\ \cline{3-8} 
 &
   &
  \cite{yu_building_2019} &
  2019 &
  Edge service crowdsensing &
  Reverse auction &
  N/S &
  Prototype \\ \cline{3-8} 
 &
   &
  \cite{lee_trustful_2020} &
  2020 &
  Service allocation in fog-enabled IoV  &
  VCG auction &
  Permissioned &
  Hyperledger Fabric \\ \cline{2-8} 
 &
  \multirow{2}{*}{\begin{minipage}{0.5in}\textbf{Network Service}\end{minipage}} &
  \cite{figueredo_brain_2019} &
  2019 &
  Virtual network services in NFV markets &
  Reverse FPSB auction &
  Permissionless &
  Ethereum \\ \cline{3-8} 
 &
   &
  \cite{rizk_brokerless_2018} &
  2018 &
  Brokerless virtual network embedding &
  Vickrey auction &
  Permissioned &
  Ethereum \\ \cline{2-8} 
 &
  \multirow{2}{*}{\begin{minipage}{0.5in}\textbf{Mobile Service}\end{minipage}} 
   &
  \cite{hassija_mobile_nodate} &
  2020 &
  Mobile data offloading &
  Multi-attribute auction &
  Permissionless &
  Simulation \\ \cline{3-8} 
   &
  &
  \cite{chatzopoulos_privacy_2018} &
  2018 &
  Mobile service crowdsensing &
  Combinatorial auction &
  N/S &
  Ethereum \\ \hline
\multirow{24}{*}{\textbf{Others}} &
  \multirow{3}{*}{\begin{minipage}{0.5in}\textbf{Data Management}\end{minipage}} 
   &
  \cite{xiong_anti-collusion_2020} &
  2020 &
  Big data trading and auction &
  FPSB auction &
  N/S &
  Ethereum \\ \cline{3-8} 
 &
   &
  \cite{an_truthful_2019} &
  2019 &
  Crowdsensed data trading &
  Reverse auction &
  N/S &
  Ethereum \\ \cline{3-8} 
 &
   &
  \cite{chen_secure_2019} &
  2019 &
  Data trading in IoV &
  Iterative double auction &
  Permissioned &
  Ethereum \\ \cline{2-8} 
 &
  \multirow{5}{*}{\begin{minipage}{0.6in}\textbf{Stock Exchange}\end{minipage}} 
   &
  \cite{al-shaibani_consortium_2020} &
  2020 &
  Decentralized stock exchange &
  Double auction &
  Permissioned &
  Ethereum \\ \cline{3-8} 
 &
   &
  \cite{pop_decentralizing_2018} &
  2018 &
  Decentralized stock exchange &
  Double auction &
  N/S &
  Ethereum \\ \cline{3-8} 
 &
  &
  \cite{vishnia_auditchain_2020} &
  2020 &
  Financial trade auditing &
  Periodic double auction &
  Permissioned &
  AuditChain \\ \cline{3-8} 
 &
   &
  \cite{halevi_initial_2019} &
  2019 &
  Secure and efficient IPOs &
  Sealed-bid uniform price auction &
  Permissioned &
  Hyperledger Fabric \\ \cline{2-8} 
 &
  \multirow{3}{*}{\begin{minipage}{0.3in}\textbf{Crowd-sourcing}\end{minipage}} &
  \cite{kadadha_abcrowd_2020} &
  2020 &
  Decentralized spatial crowdsourcing &
  Optimized VCG auction &
  N/S &
  Ethereum \\ \cline{3-8} 
 &
   &
  \cite{hassija_bitfund_2020} &
  2020 &
  Decentralized crowdfunding platform &
  Ascending-price progressive auction &
  N/S &
  Ethereum \\ \cline{2-8} 
 &
  \multirow{4}{*}{\begin{minipage}{0.5in}\textbf{Supply Chain}\end{minipage}} 
   &
  \cite{gupta_bitcom_2020} &
  2020 &
  Decentralized supply chain management &
  Double auction &
  Hybrid &
  Prototype \\ \cline{3-8} 
 &
 &
  \cite{martins_fostering_2020} &
  2020 &
  Customer bargaining and e-procurement &
  Reverse auction &
  Permissionless &
  Ethereum \\ \cline{3-8} 
 &
   &
  \cite{koirala_supply_2019} &
  2019 &
  Multi-attribute carrier procurement &
  Reverse auction &
  N/S &
  Ethereum \\ \cline{3-8} 
 &
   &
  \cite{shwetha2021auction} &
  2021 &
  Food supply chain management &
  English auction &
  N/S &
  Ethereum \\ \cline{2-8} 
 &
  \multirow{2}{*}{\begin{minipage}{0.5in}\textbf{Human Resource}\end{minipage}} &
  \cite{liu_blockchain-based_2020} \cite{liyuan_e2_2019} &
  2019 &
  Education and employment verification &
  VCG auction &
  N/S &
  Simulation \\ \cline{3-8} 
 &
   &
  \cite{ward_establishing_nodate} &
  2018 &
  Employee recognition programs reward &
  N/S &
  N/S &
  Ethereum \\ \cline{2-8} 
 &
  \multirow{5}{*}{\textbf{N/A}} &
  \cite{ramanan_baffle_2020} &
  2020 &
  Decentralized federated learning &
  Scoring and bidding mechanism &
  N/S &
  Ethereum \\ \cline{3-8} 
 &
  &
  \cite{fan_hybrid_2020} &
  2020 &
  Federated learning resource trading &
  Reverse auction &
  Hybrid &
  Ethereum \& FISCO-BCOS \\ \cline{3-8} 
 &
&
  \cite{cheng_auction-based_2020} &
  2020 &
  IoT collaboration &
  Reverse auction &
  Permissioned &
  Prototype \\ \cline{3-8} 
 &
   &
  \cite{seike_blockchain-based_2018} &
  2018 &
  Code ownership management system &
  Vickrey auction &
  Permissionless &
  Ethereum \\ \hline
\end{longtable}
}
\twocolumn

%% file: table/Table_Auction4Blockchain2.tex
\begin{table*}[!htp]
\setlength{\extrarowheight}{0.1pt}
\centering
\caption{Summary of Auction-Based Solutions for Blockchain Enhancement}
\label{tab:auction4blockchain}
\scalebox{0.95}{
\begin{tabularx}{\linewidth}{|m{1.5cm}|m{0.5cm}|m{0.5cm}|m{3cm}|m{3.2cm}|m{6.8cm}|}

\hline
  \textbf{Classification} &
  \textbf{Ref.} &
  \textbf{Year} &
  \textbf{Addressed Issue} &
  \textbf{Auction Model} &
  \textbf{Main Contributions} \\ \hline
\multirow{30}{*}{\begin{minipage}{0.6in}\textbf{Mining Task \\ Offloading}\end{minipage}} &
  \cite{jiao_social_2017} &
  2017 &
  Edge resource offloading &
  VCG auction &
  An auction model that offloads mobile blockchain mining tasks to edge providers and maximizes social welfare. \\ \cline{2-6} 
 &
  \cite{xia_etra_2018} &
  2018 &
  Edge resource offloading &
  VCG auction &
  A three-stage auction model that optimizes edge resource allocation for mobile blockchains. \\ \cline{2-6} 
 &
  \cite{gao_dynamic_2019} &
  2019 &
  Edge resource offloading &
  Optimized Vickrey auction &
  A dynamic auction model for allocating edge resources while maximizing the revenue of all mobile users. \\ \cline{2-6} 
 &
  \cite{li_double_2021} &
  2021 &
  Cloud/Edge resource offloading &
  Combinatorial double auction &
  An efficient resource allocation model for computation offloading in mobile blockchains. \\ \cline{2-6} 
 &
  \cite{xu_hierarchical_2020} &
  2020 &
  Cloud/Edge resource offloading &
  Hierarchical combinatorial auction &
  A hierarchical combinatorial auction model that enhances resource allocation for mobile blockchain. \\ \cline{2-6} 
 &
  \cite{li_resource_2019} &
  2019 &
  Cloud/Edge resource offloading &
  Hierarchical combinatorial auction &
  A hierarchical combinatorial auction model in which both edge and cloud computing resources are considered. \\ \cline{2-6} 
 &
  \cite{guo_blockchain_2020} &
  2020 &
  Edge resource offloading &
  Double auction &
  Non-mining devices and edge clouds can be selected to construct CMNs to offload mining tasks. \\ \cline{2-6} 
 &
  \cite{jiao_auction_2019} &
  2019 &
  Cloud/Edge resource offloading &
  \begin{tabular}[b]{@{}l@{}}Constant-demand and multi-\\demand auction algorithms\end{tabular} 
  &
  Miners with different demand situations are considered, and two efficient auction mechanisms are proposed. \\ \cline{2-6} 
 &
  \cite{guo_differential_2021} &
  2021 &
  Edge resource offloading &
  Multi-item double auction  &
  A privacy-preserving auction model that offloads limited edge servers to blockchain-based IoT devices. \\ \cline{2-6} 
 &
  \cite{liu_efficient_2019} &
  2019 &
  Edge resource offloading &
  Combinatorial double auction &
  A combinatorial double auction mechanism that offloads the mining process from miners to edge servers. \\ \cline{2-6} 
 &
  \cite{jameel_efficient_2020} &
  2020 &
  Mining task offloading in V2X networks &
  First-price auction &
  An efficient auction solution for offloading mining tasks in cellular V2X networks. \\ \cline{2-6} 
 &
  \cite{luong_optimal_2018} &
  2018 &
  Edge resource offloading &
  Deep learning-based optimal auction &
  A deep learning-based optimal auction for edge resource allocation in mobile blockchains. \\ \cline{2-6} 
 &
  \cite{luong_machine-learning-based_2020} &
  2020 &
  Fog resource offloading &
  Deep learning-based optimal auction &
  A deep learning-based optimal auction for fog resource allocation in blockchain networks. \\ \cline{2-6} 
 &
  \cite{sun_joint_2020} &
  2020 &
  Edge resource offloading &
  Double auction &
  Multi-task cross-server resource allocation in mobile edge computing is achieved through auction models. \\ \cline{2-6} 
 &
  \cite{liu_smart_2020} &
  2020 &
  Edge resource offloading &
  Double auction &
  A smart contract-based mobile blockchain computation offloading model using long-term double auctions. \\ \hline
\multirow{18}{*}{\begin{minipage}{0.6in}\textbf{Transaction Fee Mechanism Design}\end{minipage}} &
  \cite{huberman_monopoly_2017} &
  2017 &
  Bitcoin transaction fee mechanism &
  Priority auction with VCG mechanism &
  The Bitcoin protocol, despite the absence of an auctioneer, implicitly includes a priority auction. \\ \cline{2-6} 
 &
  \cite{dimitri_transaction_2019} &
  2019 &
  Bitcoin transaction fee mechanism &
  Auction game &
  The transaction mechanism is modeled as an auction game, where miners sell their space and users bid for such space. \\ \cline{2-6} 
 &
  \cite{daian_flash_2020} &
  2020 &
  Ethereum transaction fee mechanism &
  Priority gas auction &
  Bots in DEXes engage in priority gas auctions to competitively bid up transaction fees to obtain priority orders. \\ \cline{2-6} 
 &
  \cite{roughgarden_transaction_2020} &
  2020 &
  Ethereum transaction fee mechanism &
  GFP auction &
  The transaction fee mechanism of Ethereum has always been a GFP auction, as indicated in EIP-1559. \\ \cline{2-6} 
  &
  \cite{basu_towards_2019} &
  2019 &
  Fee market for cryptocurrencies &
  GSP auction &
  An alternative transaction fee mechanism for cryptocurrencies inspired by the GSP auction. \\ \cline{2-6} 
 &
  \cite{li_novel_2019} \cite{li_novel_2020} &
  2020 &
  Bitcoin transaction fee mechanism &
  GSP auction &
  A novel GSP auction mechanism that deals with the problems caused by the GFP mechanism. \\ \cline{2-6} 
 &
  \cite{yan_dynamic_2020} &
  2020 &
  Bitcoin transaction fee mechanism &
  GSP auction &
  A time-dependent dynamic game model for the Bitcoin transaction fee market under the GSP mechanism. \\ \cline{2-6} 
 &
  \cite{lavi_redesigning_2019} &
  2019 &
  Bitcoin transaction fee mechanism &
  Monopolistic auction &
  Monopolistic auctions are immune to malicious auctioneers and can solve issues in the GFP mechanism. \\ \cline{2-6} 
 &
  \cite{yao_incentive_2018} &
  2018 &
  Bitcoin transaction fee mechanism &
  Monopolistic auction &
  The monopolistic auction mechanism is nearly truthful for any i.i.d. distribution as the number of users grows large. \\ \hline
\multirow{12}{*}{\begin{minipage}{0.6in}\textbf{Miner Selection \& \\ Reward Distribution}\end{minipage}} 
  &
  \cite{saad_e-pos_2021} &
  2021 &
  Miner selection in PoS consensus &
  Blind block auction &
  An enhanced version of PoS called e-PoS is proposed to resist centralization and improve fairness. \\ \cline{2-6} 
  &
  \cite{ai_abc_2020} &
  2020 &
  Miner selection in new consensus &
  Continuous double auction &
  A new auction-based consensus mechanism called ABC is proposed for blockchain. \\ \cline{2-6} 
  &
  \cite{devi_using_2020} &
  2020 &
  Miner selection in blockchain-based IoV &
  Multi-attribute auction &
  An auction-based mechanism for miner selection to encourage miners to participate in block validation in IoV. \\ \cline{2-6} 
 &
  \cite{amin_secured_2020} &
  2020 &
  Miner selection in the ming pool &
  FPSB auction &
  A discretionary mining mechanism for the IOTA blockchain in which miners are nominated through auctions. \\ \cline{2-6} 
 &
  \cite{nadendla_difficulty_2020} &
  2020 &
  Mining cost and allocation function &
  All-pay auction &
  Mining is modeled as an all-pay auction to analyze the mining allocation function of the blockchain. \\ \cline{2-6} 
 &
  \cite{liu_auction_2019} &
  2019 &
  Reward distribution for pool mining &
  Uniform price auction &
  An auction-based reward distribution method that improves miners' enthusiasm and the stability of the mining pool. \\ \hline
\multirow{9}{*}{\begin{minipage}{0.6in}\textbf{Token Sale \& Exchange}\end{minipage}} 
 &
  \cite{zhang_optimum_2020} &
  2020 &
  Cross-chain atomic swap &
  Uniform price auction &
  An atomic swap mechanism with a uniform price auction can save costs, but the optimal result collection is NP-Hard. \\ \cline{2-6} 
 &
  \cite{liu2021aucswap} &
  2020 &
  Cross-chain asset transfer &
  Vickrey auction &
  An efficient cross-chain asset transfer protocol using atomic swap technology and the Vickrey auction model. \\ \cline{2-6} 
&
  \cite{black_atomic_2019} &
  2019 &
  Cross-chain atomic swap &
  Bidding process &
  A competitive bidding process for the liquidation of collateral when defaults occur for atomic loans.
  \\ \cline{2-6} 
 &
  \cite{friedenbach_freimarkets_nodate_2013} &
  2013 &
  Cross-chain atomic swap &
  English, Dutch, and double auction &
  The Freimarkets protocol adds primitives to Bitcoin in order to implement non-currency financial constructs. \\ \cline{2-6} 
 &
  \cite{walther_multi-token_2018} &
  2018 &
  Cross-chain atomic swap &
  Uniform price multi-batch auction &
  A trading platform using Ethereum Plasma in which auctions are implemented among different ERC-20 tokens. \\ \hline
\multirow{4}{*}{\textbf{Others}} 
 &
  \cite{wang_incentivizing_2021} &
  2020 &
  New blockchain incentive mechanism &
  FPSB auction &
  A new relay payment scheme that uses an auction model to solve the relay incentive problem in the blockchain. \\ \cline{2-6} 
 &
  \cite{johnson_ens_nodate} &
  2019 &
  Ethereum naming service &
  Vickrey auction &
  ENS is an auction-based naming system built on top of Ethereum, as defined in EIP-162. \\ \hline
  \end{tabularx}%
}
\end{table*}

%% file: table/Table_Privacy3.tex
\begin{table*}[!htp]
\setlength{\extrarowheight}{0.5pt}
\setlength{\tabcolsep}{4pt}
\centering
\caption{Summary of Privacy Protection Techniques Used in Blockchain-Based Auction Models}
\label{tab:my-table2}
\begin{threeparttable}

\scalebox{0.95}{
\begin{tabularx}{\linewidth}{|m{0.55cm}|m{10cm}|m{0.45cm}|m{0.4cm}|m{0.4cm}|m{0.4cm}|m{0.4cm}|m{0.4cm}|m{0.4cm}|m{0.4cm}|m{0.4cm}|m{0.4cm}|m{0.4cm}|}

\hline
\multirow{2}{*}{\textbf{Ref.}} &
  \multicolumn{1}{c|}{\multirow{2}{*}{\textbf{Main Contributions}}} &
  \multicolumn{10}{c|}{\textbf{Privacy Protection Techniques}\tnote{*}} \\ \cline{3-12} 
 &
   &
  \scriptsize\textbf{MPC} &
  \scriptsize\textbf{ZKP} &
  \scriptsize\textbf{CS} &
  \scriptsize\textbf{TEE} &
  \scriptsize\textbf{AE} &
  \scriptsize\textbf{DS} &
  \scriptsize\textbf{HE} &
  \scriptsize\textbf{Mix} &
  \scriptsize\textbf{DP} &
  \scriptsize\textbf{PB} \\ \hline
  \cite{kosba_hawk_2016} &
  Hawk can help users write private smart contracts without having to implement cryptography. Hawk compiler automatically builds cryptographic protocols for users. &
  \Checkmark &
  \Checkmark &
   &
  \Checkmark &
  \Checkmark &
  \Checkmark &
   &
   &
   &
   \\ \hline
  \cite{blass_strain_2018} &
  Strain is a secure auction protocol based on blockchain. It has a slightly weaker adversary model than traditional MPC and can achieve constant latency. &
  \Checkmark &
  \Checkmark &
   &
   &
   &
   &
   &
   &
   &
   \\ \hline
  \cite{blass_borealis_2020} &
  BOREALIS is an efficient model for sealed-bid auctions on the blockchain. It performs the secure comparison of integer bids among participants using ZKP. &
  \Checkmark &
  \Checkmark &
   &
   &
  \Checkmark &
   &
  \Checkmark &
   &
   &
   \\ \hline
\cite{ma_fully_2019} &
  A secure auction model with affordable computations using an insulated integer comparison protocol, which is more efficient than traditional MPC/ZKP solutions. &
  
  \Checkmark &
   &
  \Checkmark &
   &
   &
   &
   &
   &
   &
   \\ \hline
\cite{galal_verifiable_2018} &
  A smart contract protocol for verifiable sealed-bid auctions on the Ethereum blockchain. Different cryptographic primitives are used during the auction process. &
   &
  \Checkmark &
  \Checkmark &
   &
   &
   &
  \Checkmark &
   &
   &
   \\ \hline
\cite{galal_succinctly_2018} &
  A smart contract protocol for succinctly verifiable sealed-bid auctions on the Ethereum blockchain with various cryptographic primitives. &
  \Checkmark &
  \Checkmark &
  \Checkmark &
   &
  \Checkmark &
   &
   &
   &
   &
   \\ \hline
  \cite{galal_trustee_2019} &
  Trustee is an Ethereum-based Vickrey auction model that protects bids' privacy at a low cost. It consists of a smart contract, an Intel SGX enclave, and a relay scheme. &
   &
   &
   &
  \Checkmark &
  \Checkmark &
   &
   &
   &
   &
   \\ \hline
\cite{benhamouda_supporting_2019} &
  A framework that integrates secure MPC protocols into the blockchain architecture rather than allowing separate nodes to run secure MPC protocols off-chain. &
  \Checkmark &
   &
   &
   &
   &
   &
   &
   &
   &
   \\ \hline
 \cite{liu2020blockchain} &
  A blockchain-based fair and secure MPC model for double auctions. In particular, a more efficient protocol for secure two-party comparison is designed. &
  \Checkmark &
   &
   &
   &
   &
   &
   &
   &
   &
   \\ \hline
\cite{tso_distributed_2019} &
  A smart contract-based electronic voting and bidding system that integrates cryptographic techniques such as the Paillier cryptosystem and homomorphic encryption. &
   &
   &
   &
   &
  \Checkmark &
   &
  \Checkmark &
   &
   &
   \\ \hline
\cite{doweck_multi-party_nodate} &
  A protocol called Time-Capsule that solves the multi-party timed commitments problem for blockchain-based auction applications. &
  
  \Checkmark &
  \Checkmark &
  \Checkmark &
   &
   &
   &
   &
   &
   &
   \\ \hline
  \cite{yuan_shadoweth_2018} &
  ShadowEth is a solution for public blockchains that utilizes hardware enclaves to secure smart contracts while maintaining their integrity and availability. &
   &
   &
   &
  \Checkmark &
   &
  \Checkmark &
   &
   &
   &
   \\ \hline
\cite{enkhtaivan_fair_2019} &
  A blockchain-based anonymous English auction scheme, in which group signatures are used to provide anonymity for bidders and TEE is used to store the secret keys. &
   &
   &
   &
  \Checkmark &
   &
  \Checkmark &
   &
   &
   &
   \\ \hline
  \cite{nguyen_blockchain-based_2020} \cite{nguyen_trustless_2019} &
  An iterative double auction protocol using smart contract and state channel technologies that minimizes blockchain transactions. &
  \Checkmark &
   &
   &
   &
   &
  \Checkmark &
   &
   &
   &
   \\ \hline
\cite{sarfaraz_tree_2021} &
  A public bidding system that integrates ECC and dynamic accumulators in a tree-structured blockchain to protect privacy and achieve great efficiency. &
   &
   &
   &
   &
  \Checkmark &
  \Checkmark &
   &
   &
   &
   \\ \hline
\cite{xiong_anonymous_2019} &
  An anonymous auction model that uses a permissioned blockchain and blind signatures. Specifically, bids are encrypted using a timed-release AE method. &
   &
   &
   &
   &
  \Checkmark &
  \Checkmark &
   &
   &
   &
  \Checkmark \\ \hline
  \cite{krol_pastrami_2020} &
  PASTRAMI makes bidders accountable on the Ethereum blockchain by utilizing threshold blind signatures and commitment schemes to ensure strong privacy guarantees. &
   &
   &
  \Checkmark &
   &
   &
  \Checkmark &
   &
   &
   &
   \\ \hline
\cite{desai_hybrid_2019} &
  A hybrid blockchain-based auction architecture, in which a private blockchain is used to publish sensitive bids and a public blockchain is used to make the auction accountable. &
   &
   &
  \Checkmark &
   &
   &
   &
   &
   &
   &
   \Checkmark
   \\ \hline
  \cite{david_fast_nodate} &
  FAST is an efficient sealed-bid auction protocol on the blockchain. In FAST, fairness is guaranteed (i.e., everybody gets the final output or nobody), and cheaters are punished. &
  \Checkmark &
  \Checkmark &
  \Checkmark &
   &
  \Checkmark &
  \Checkmark &
   &
   &
   &
   \\ \hline
\cite{brandenburger_blockchain_2018} \cite{brandenburger_trusted_2019} &
  A smart contract execution architecture for Hyperledger Fabric that can handle rollback attacks in Intel SGX. &
   &
   &
   &
  \Checkmark &
  \Checkmark &
   &
   &
   &
   &
   \Checkmark
   \\ \hline
  \cite{damle_practical_2019} &
  TPACAS is a secure blockchain-based auction protocol for combinatorial auctions. It introduces a privacy-preserving comparison technique to compare two integers. &
  \Checkmark &
  \Checkmark &
  \Checkmark &
   &
   &
   &
   &
   &
   &
   \\ \hline
  \cite{bunz_zether_2020} &
  Zether is a low-cost privacy-preserving cryptocurrency that encrypts account balances and enforces money transmission using cryptographic proofs. &
   &
  \Checkmark &
  \Checkmark &
   &
  \Checkmark &
   &
  \Checkmark &
   &
   &
   \\ \hline
  \cite{lafourcade_auctionity_2018} \cite{lafourcade_security_2019} &
  Auctionity is a blockchain-based English auction protocol built on Ethereum, in which ECDSA and non-fungible tokens are leveraged to enhance security and privacy. &
   &
   &
   &
   &
  \Checkmark &
  \Checkmark &
   &
   &
   & 
   \\ \hline
\cite{shu_blockchain_2017} &
  A protocol for secure English auction on the blockchain, in which ECDSA, AE, and permissioned blockchains are integrated. &
   &
   &
   &
   &
  \Checkmark &
  \Checkmark &
   &
   &
   &
   \Checkmark
   \\ \hline
\cite{tu_blockchain-based_2020} &
  A blockchain-based spectrum sharing platform that protects users' anonymity by using AE and DP technologies during the bidding process. &
   &
   &
   &
   &
  \Checkmark &
   &
   &
   &
  \Checkmark &
   \\ \hline
\cite{wang_secure_2020} &
  A blockchain-based secure spectrum trading platform that combines Intel SGX, the Paillier cryptosystem, and the Pedersen commitment. &
   &
   &
  \Checkmark &
  \Checkmark &
  \Checkmark &
   &
  \Checkmark &
   &
   &
   \\ \hline
\cite{zhang_privacy_2019} &
  Pseudonyms and pseudonym certificates are issued using blind signatures to enhance the privacy of user identities on the blockchain. &
   &
   &
   &
   &
   &
  \Checkmark &
   &
   &
   &
   \\ \hline
\cite{yu_smart_2019} &
  A secure auction framework built on permissioned blockchains and cryptographic primitives that protects against collusion attacks in auctions. &
   &
   &
   &
   &
  \Checkmark &
   &
  \Checkmark &
  \Checkmark &
   &
  \Checkmark \\ \hline
  \cite{brenzikofer_privacy-preserving_2019} \cite{ableitner_quartierstrom_2019} &
  Quartierstrom utilizes a coin mixing protocol and account-based on-chain smart contracts to create a secure P2P energy marketplace. &
   &
   &
   &
  \Checkmark &
  \Checkmark &
   &
   &
  \Checkmark &
   &
  \Checkmark \\ \hline
  \cite{hassan_deal_2020} &
  A decentralized auction model for energy trading in microgrids, where DP and permissioned blockchains are used to protect bid privacy. &
   &
   &
   &
   &
  \Checkmark &
   &
   &
   &
  \Checkmark &
  \Checkmark \\ \hline
  \cite{sonnino_asterisk_2019} &
  AStERISK is a secure auction framework that uses the mixing function of the Coconut contract to protect auction privacy. &
   &
  \Checkmark &
  \Checkmark &
   &
   &
   &
   &
  \Checkmark &
   &
   \\ \hline
  \cite{laszka_providing_2017} &
    PETra is a blockchain-based microgrid trading platform that uses mixing services to protect energy transaction privacy. &
   &
   &
   &
   &
   &
   &
   &
  \Checkmark &
   &
   \\ \hline
  \cite{liu_blockchain-based_2020} \cite{liyuan_e2_2019} &
  E\textsuperscript{2}C-Chain is a secure blockchain system that protects all users' private information with ZKP in the employment and skill certification process. &
   &
  \Checkmark &
   &
   &
  \Checkmark &
   &
   &
   &
   &
   \\ \hline
  \cite{sun_lottery_2020} \cite{sun_bit_2020} &
  An auction model for quantum blockchains aims to achieve bid privacy, posterior privacy, bid binding, decentralization, and unconditional security. &
   &
   &
  \Checkmark &
   &
   &
   &
   &
   &
   &
   \\ \hline
\end{tabularx}%
}
\begin{tablenotes}
\item[*] \scriptsize Abbreviations: Multi-Party Computation (MPC), Zero-Knowledge Proof (ZKP), Commitment Scheme (CS), Trusted Execution Environment (TEE), Asymmetric Encryption (AE), Digital signature (DS), Homomorphic Encryption (HE), Tumbler/Mixing (MiX), Differential Privacy (DP), Permissioned Blockchain (PB).
\end{tablenotes}
\end{threeparttable}
\end{table*}

%% file: table/Table_Economic.tex
\begin{table*}[!htp]
\setlength{\extrarowheight}{3pt}
\setlength{\tabcolsep}{2.7pt}
\centering
\caption{Summary of Auction Design Properties in Integrated Blockchain-Auction Models}
\label{tab:mechanism}
\begin{threeparttable}

\begin{tabularx}{\linewidth}{|m{0.9cm}|m{0.6cm}|m{13cm}|m{0.4cm}|m{0.4cm}|m{0.4cm}|m{0.4cm}|m{0.4cm}|}

\hline
\multicolumn{1}{|c|}{\multirow{2}{*}{\textbf{Group}}} &
  \multicolumn{1}{c|}{\multirow{2}{*}{\textbf{Ref.}}} &
  \multicolumn{1}{c|}{\multirow{2}{*}{\textbf{Auction Mechanism Design Objectives}}} &
  \multicolumn{5}{c|}{\textbf{Design Properties}\tnote{*}} \\ \cline{4-8} 
\multicolumn{1}{|c|}{} &
  \multicolumn{1}{c|}{} &
  \multicolumn{1}{c|}{} &
  \multicolumn{1}{c|}{\textbf{IR}} &
  \multicolumn{1}{c|}{\textbf{IC}} &
  \multicolumn{1}{c|}{\textbf{BB}} &
  \multicolumn{1}{c|}{\textbf{EE}} &
  \multicolumn{1}{c|}{\textbf{CE}} \\ \hline

\multirow{20}{*}{\begin{minipage}{0.5in}\textbf{Double Auction}\end{minipage}} &
  \cite{ai_abc_2020} &
  A new blockchain consensus-incentive mechanism using a continuous double auction model. &
  \fullcirc &
  \fullcirc &
   &
   &
   \\ \cline{2-8} 
 &
  \cite{xia_bayesian_2020} &
  A Bayesian game-based optimal auction scheme to distribute electricity resources in V2V networks. &
  \fullcirc &
  \emptycirc &
  \fullcirc &
  \fullcirc &
  \fullcirc \\ \cline{2-8} 
 &
  \cite{tu_blockchain-based_2020} &
  A blockchain-based double auction model for spectrum sharing that satisfies different economic properties. &
  \fullcirc &
  \fullcirc &
  \fullcirc &
  \halfcirc &
  \fullcirc \\ \cline{2-8} 
 &
  \cite{chen_secure_2019} &
  An iterative double auction mechanism for IoV data trading aimed at maximizing social welfare. &
  \fullcirc &
  \fullcirc &
  \halfcirc &
  \fullcirc &
   \\ \cline{2-8} 
 &
  \cite{nguyen_blockchain-based_2020} &
  A general iterative double auction model that converges to a Nash equilibrium and maximizes social welfare. &
  \fullcirc &
  \fullcirc &
  \emptycirc &
  \fullcirc &
   \\ \cline{2-8} 
 &
  \cite{guo_differential_2021} &
  A truthful and effective online multi-item double auction mechanism for mobile blockchains. &
  \fullcirc &
  \fullcirc &
  \halfcirc &
  \halfcirc &
  \fullcirc \\ \cline{2-8} 
 &
  \cite{alashery_blockchain-enabled_2020} &
  A novel VCG pricing rule that compensates the balanced budget attribute for the VCG mechanism. &
  \fullcirc &
  \fullcirc &
  \halfcirc &
  \fullcirc &
   \\ \cline{2-8} 
 &
  \cite{li_double_2021} &
  A combinatorial double auction model for computation offloading in mobile blockchains. &
  \fullcirc &
  \fullcirc &
  \fullcirc &
   &
  \fullcirc \\ \cline{2-8} 
 &
  \cite{xu_hierarchical_2020} &
  A hierarchical combinatorial auction model to achieve computing resource allocation for mobile blockchains. &
  \fullcirc &
  \fullcirc &
   &
  \halfcirc &
  \emptycirc \\ \cline{2-8} 
 &
  \cite{xia_etra_2018} &
  A three-stage VCG auction model to achieve resource allocation for mobile blockchains. &
  \fullcirc &
  \fullcirc &
   &
  \fullcirc &
  \fullcirc \\ \cline{2-8} 
 &
  \cite{choubey_energytradingrank_2019} &
  A truthful double auction model to incentivize EVs to participate in the V2V energy trading. &
  \fullcirc &
  \fullcirc &
   &
  \fullcirc &
   \\ \cline{2-8} 
 &
  \cite{liu_efficient_2019} &
  An efficient combinatorial double auction model for mining task assignment with two greedy algorithms. &
  \fullcirc &
  \fullcirc &
  \fullcirc &
  \halfcirc &
  \fullcirc \\ \cline{2-8} 
 &
  \cite{li_resource_2019} &
  An efficient and truthful hierarchical combinatorial auction model for mobile blockchain resource allocation. &
   &
  \fullcirc &
   &
  \halfcirc &
   \\ \cline{2-8} 
 &
  \cite{liu_smart_2020} &
  A long-term auction model for mobile blockchains that satisfies several economic properties. &
  \fullcirc &
  \fullcirc &
  \fullcirc &
   &
  \fullcirc \\ \cline{2-8} 
 &
  \cite{zavodovski_decloud_2019} &
  A truthful double auction model for edge clouds using McAfee’s mechanism with near best social welfare. &
  \fullcirc &
  \fullcirc &
  \fullcirc &
  \halfcirc &
   \\ \cline{2-8} 
 &
  \cite{sun_joint_2020} &
  Two auction models for resource allocation in blockchain-based mobile edge computing.  &
  \fullcirc &
  \fullcirc &
  \fullcirc &
   &
   \\ \cline{2-8} 
 &
  \cite{sun_blockchain-enhanced_2020} &
  A double auction model for V2V energy trading where EVs will bid truthfully based on their private value. &
  \fullcirc &
  \fullcirc &
  \halfcirc &
   &
   \\ \cline{2-8} 
 &
  \begin{tabular}[c]{@{}l@{}}\cite{afraz_distributed_2019} \\ \cite{afraz_sharing_2018}\end{tabular} &
  \multirow{2}{*}{A double auction model satisfies the crucial economic properties of a market while achieving great efficiency.} &
  \multirow{2}{*}{\fullcirc} &
  \multirow{2}{*}{\fullcirc} &
  \multirow{2}{*}{\halfcirc} &
  \multirow{2}{*}{\fullcirc} &
   \\ \cline{2-8} 
 &
  \multirow{2}{*}{\cite{guo_double_2020}} &
  \multirow{2}{*}{\begin{minipage}{4.8in}Two auction algorithms, namely a truthful mechanism for charging and an efficient mechanism for charging, are designed for charging scheduling among EVs.\end{minipage}} &
  \fullcirc &
  \fullcirc &
  \fullcirc &
   &
  \fullcirc \\ \cline{4-8} 
 &
   &
   &
  \fullcirc &
  \emptycirc &
  \fullcirc &
   &
  \fullcirc \\ \hline
\multirow{27}{*}{\begin{minipage}{0.5in}\textbf{Single-Sided Auction}\end{minipage}} &
  \cite{jiao_social_2017} &
  An auction model in edge computing resource allocation for mobile blockchains that maximizes social welfare. &
  \fullcirc &
  \fullcirc &
   &
  \fullcirc &
  \fullcirc \\ \cline{2-8} 
 &
  \cite{gu_cloud_2018} &
  A decentralized cloud storage resource trading model using the VCG auction mechanism. &
   &
  \fullcirc &
   &
   &
   \\ \cline{2-8} 
 &
  \begin{tabular}[c]{@{}l@{}}\cite{liyuan_e2_2019} \\ \cite{liu_blockchain-based_2020}\end{tabular} &
  \multirow{2}{*}{A novel decentralized framework for educational background investigation using the VCG mechanism.} &
  \multirow{2}{*}{\fullcirc} &
  \multirow{2}{*}{\fullcirc} &
   &
  \multirow{2}{*}{\fullcirc} &
  \multirow{2}{*}{\fullcirc} \\ \cline{2-8} 
 &
  \cite{krol_pastrami_2020} &
  An efficient Vicrey-Dutch multi-item auction algorithm that satisfies several economic properties. &
  \fullcirc &
  \fullcirc &
  \fullcirc &
  \fullcirc &
   \\ \cline{2-8} 
 &
  \cite{hassan_deal_2020} &
  A VCG auction model for energy trading that maximizes revenue and ensures the truthful bidding. &
   &
  \fullcirc &
   &
  \fullcirc &
   \\ \cline{2-8} 
 &
  \cite{kadadha_abcrowd_2020} &
  A crowdsourcing platform that motivates workers to bid truthfully through an optimized VCG auction. &
  \fullcirc &
  \fullcirc &
   &
  \fullcirc &
  \fullcirc \\ \cline{2-8} 
 &
  \cite{gu_decentralized_2018} &
  A decentralized cloud storage transaction mechanism based on the reverse VCG auction. &
   &
  \fullcirc &
   &
  \fullcirc &
   \\ \cline{2-8} 
 &
  \cite{chatzopoulos_flopcoin_2018} &
  A computation offloading framework using a truthful auction strategy and a P2P reputation exchange scheme. &
  \fullcirc &
  \fullcirc &
   &
   &
   \\ \cline{2-8} 
 &
  \cite{fan_hybrid_2020} &
  An auction-based resource trading system that encourages more edge nodes to join in the FL model training. &
  \fullcirc &
  \fullcirc &
   &
   &
  \fullcirc \\ \cline{2-8} 
 &
  \cite{guo_reliable_2021} &
  A truthful auction in IoV that motivates vehicles to undertake the tasks issued by traffic administrations. &
  
  \fullcirc &
  \fullcirc &
   &
   &
  \fullcirc \\ \cline{2-8} 
 &
  \cite{an_truthful_2019} &
  A truthful crowdsensing data trading framework based on the reverse auction and blockchain. &
  \fullcirc &
  \fullcirc &
   &
   &
   \\ \cline{2-8} 
 &
  \cite{khan_trusted_2020} &
  A reputation-based truthful auction method for handling interactions between UAV operators and business agents. &
   &
  \fullcirc &
   &
   &
   \\ \cline{2-8} 
 &
  \cite{chen_blockchain_2020} &
  A Vickrey auction model that offloads users from a macrocell base station to small cell access points. &
   &
  \fullcirc &
   &
  \fullcirc &
   \\ \cline{2-8} 
 &
  \cite{chen_safe_2020} &
  A secure and fair auction framework that can achieve high economic efficiency. &
  \fullcirc &
  \fullcirc &
   &
  \fullcirc &
  \fullcirc \\ \cline{2-8} 
 &
  \cite{sonnino_asterisk_2019} &
  An auction framework that automatically determines the best price for cloud services. &
  \fullcirc &
  \fullcirc &
  \fullcirc &
   &
  \fullcirc \\ \cline{2-8} 
 &
  \cite{chatzopoulos_privacy_2018} &
  A truthful and cost-optimal auction model that reduces payments from crowdsensing providers to mobile users. &
  \fullcirc &
  \fullcirc &
   &
   &
   \\ \cline{2-8} 
 &
  \cite{damle_practical_2019} &
  A truthful and secure combinatorial auction solution that focuses on single-minded bidders. &
  \fullcirc &
  \fullcirc &
   &
   &
  \fullcirc \\ \cline{2-8} 
 &
  \cite{wu_cream_2019} &
  A collusion resistance auction solution that maintains social welfare at an acceptable level. &
   &
  \fullcirc &
   &
  \halfcirc &
   \\ \cline{2-8} 
 &
  \cite{wang_smart_2019} &
  An auction model that selects cost-effective service providers and (nearly) maximizes service requesters’ utility. &
   &
  \fullcirc &
   &
  \halfcirc &
   \\ \cline{2-8} 
 &
  \cite{lavi_redesigning_2019} &
  The new monopolistic auction-based Bitcoin fee market mechanism is proved approximately truthful. &
   &
  \halfcirc &
   &
   &
   \\ \cline{2-8} 
 &
  \cite{yao_incentive_2018} &
  The monopolistic auction is nearly truthful for any i.i.d. distribution as the number of users grows large. &
   &
  \halfcirc &
   &
   &
   \\ \cline{2-8} 
 &
  \cite{roughgarden_transaction_2020} &
  EIP-1559 mechanism is truthful for myopic miners and users (except in periods of rapidly increasing demand). &
  \fullcirc &
  \halfcirc &
   &
   &
   \\ \cline{2-8} 
 &
  \cite{lee_trustful_2020} &
  A truthful service allocation model for IoV that uses a VCG auction mechanism. &
  \fullcirc &
  \fullcirc &
   &
   &
  \fullcirc \\ \cline{2-8} 
 &
  \begin{tabular}[c]{@{}l@{}}\cite{luong_optimal_2018} \\ \cite{luong_machine-learning-based_2020}\end{tabular} &
  \multirow{2}{*}{A deep learning-based optimal auction model for blockchain mining tasks offloading.} &
  \multirow{2}{*}{\fullcirc} &
  \multirow{2}{*}{\fullcirc} &
   &
   &
   \\ \cline{2-8} 
 &
  \multirow{3}{*}{\cite{jiao_auction_2019}} &
  \multirow{3}{*}{\begin{minipage}{4.8in}An auction-based market model for blockchain mining tasks offloading, in which two bidding scenarios (the constant demand and the multiple demands) are considered. Accordingly, three different auctions that satisfy different economic attributes are designed.\end{minipage}} &
  \fullcirc &
  \fullcirc &
   &
  \halfcirc &
  \fullcirc \\ \cline{4-8} 
 &
   &
   &
  \fullcirc &
  \emptycirc &
   &
  \halfcirc &
  \fullcirc \\ \cline{4-8} 
 &
   &
   &
  \fullcirc &
  \fullcirc &
   &
  \halfcirc &
  \fullcirc \\ \hline
  
\end{tabularx}%
\begin{tablenotes}
\item[*] \scriptsize Abbreviations: Individual Rationality (IR), Incentive Compatibility (IC), Budget Balance (BB), Economic Efficiency (EE), Computational Efficiency (CE).
\item[*] \scriptsize Notes: Filled (or half-filled) circles indicate that the economic properties are (partially) proven or addressed, while empty circles mean that economic properties are not satisfied. Empty cells represent properties not mentioned in the paper.

\end{tablenotes}
\end{threeparttable}
\end{table*}